\PassOptionsToPackage{usenames,dvipsnames}{xcolor}
\documentclass[acmsmall,screen, nonacm]{acmart}
\AtBeginDocument{%
  }

\def\islong{1} % enable appendix by default

\usepackage{rcompcerto}
\usepackage{lstcoq}

\begin{document}

\title{End-to-end Compositional Verification of Program Safety through Verified and Verifying Compilation}
%% \title{RustCert: A Verified and Verifying Compiler for a Sequential Subset of Rust}

% \affiliation{%

\author{Jinhua Wu}
%%                                         %% can be repeated if necessary
\orcid{0000-0001-5812-053X}             %% \orcid is optional
\affiliation{
  %% \position{Position1}
  \department{John Hopcroft Center for Computer Science, School of Electronic Information and Electrical Engineering}              %% \department is recommended
  \institution{Shanghai Jiao Tong University}            %% \institution is required
  %% \streetaddress{Street1 Address1}
  %% \city{City1}
  %% \state{State1}
  %% \postcode{Post-Code1}
  \country{China}                    %% \country is recommended
}
\email{jinhua.wu@sjtu.edu.cn}          %% \email is recommended

\author{Yuting Wang}
\authornote{Corresponding author}          %% \authornote is optional;
%%                                         %% can be repeated if necessary
\orcid{0000-0003-3990-2418}             %% \orcid is optional
\affiliation{
  %% \position{Position1}
  \department{John Hopcroft Center for Computer Science, School of Electronic Information and Electrical Engineering}              %% \department is recommended
  \institution{Shanghai Jiao Tong University}            %% \institution is required
  %% \streetaddress{Street1 Address1}
  %% \city{City1}
  %% \state{State1}
  %% \postcode{Post-Code1}
  \country{China}                    %% \country is recommended
}
\email{yuting.wang@sjtu.edu.cn}          %% \email is recommended

\author{Liukun Yu}
%%                                         %% can be repeated if necessary
\orcid{0009-0007-7071-7225}             %% \orcid is optional
\affiliation{
  %% \position{Position1}
  \department{School of Electronic Information and Electrical Engineering}              %% \department is recommended
  \institution{Shanghai Jiao Tong University}            %% \institution is required
  %% \streetaddress{Street1 Address1}
  %% \city{City1}
  %% \state{State1}
  %% \postcode{Post-Code1}
  \country{China}                    %% \country is recommended
}
\email{yu_liukun@outlook.com}          %% \email is recommended

\author{Linglong Meng}
%%                                         %% can be repeated if necessary
\orcid{0009-0007-2962-3323}             %% \orcid is optional
\affiliation{
  %% \position{Position1}
%  \department{John Hopcroft Center for Computer Science, School of Electronic Information and Electrical Engineering}              %% \department is recommended
  \institution{University of Minnesota}            %% \institution is required
  %% \streetaddress{Street1 Address1}
  %% \city{City1}
  %% \state{State1}
  %% \postcode{Post-Code1}
  \country{USA}                    %% \country is recommended
}
\email{meng0181@umn.edu}          %% \email is recommended


\begin{abstract}

Safety properties (e.g., absence of undefined behaviors) are critical
for correct operation of programs.
Those properties are usually verified at the source level (e.g., by
separation logics) and preserved to the target by verified compilers
(e.g., CompCert), thereby achieving end-to-end verification of safety.
However, modern safe programming languages like Rust pose new
problems in achieving end-to-end safety.
Because not all functionalities can be implemented in the safe
language, mixing safe and unsafe modules is needed. 
Therefore, verified compilation must preserve a modular notion of
safety which can be composed at the target level.
Furthermore, certain classes of errors (e.g., memory errors) are
automatically excluded by verifying compilation (e.g., borrow checking)
for modules written in safe languages. As a result, verified
compilation needs to cooperate with verifying compilation to ensure
end-to-end safety.

To address the above problems, we propose a language-agnostic approach
to end-to-end compositional verification of program safety. Our
approach provides a modular and generic definition of safety
called \emph{open safety} based on program semantics described as open
labeled transition systems (LTS). By exploiting the compatible
interfaces of open LTS, open safety is composable at the boundary of
modules. Furthermore, it can be modularly preserved by verified
compilers supporting compositional verification. Those properties
enable separate verification of safety for heterogeneous modules and
composition of the safety results at the target level.
Open safety can be generalized to \emph{partial} safety (i.e., only a
certain class of errors can occur).  By this we formalized the
correctness of verifying compilation as derivation of total safety from partial safety.
We demonstrate how our framework can combine verified and verifying
compilation by developing a verified compiler for an ownership
language (called Owlang) inspired by Rust.
We evaluate our approach on the compositional safety verification
using a hash map implemented by Owlang and C.

\end{abstract}


% \keywords{}


\maketitle
% \vspace{-0.4cm}

% !TEX root = main.tex
\section{Introduction}\label{sec:intro}

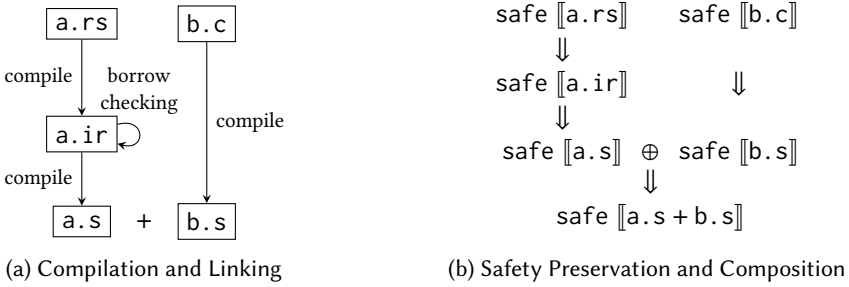
\begin{figure}[t]
\begin{subfigure}[b]{.4\textwidth}
\centering
\begin{tikzpicture}
    \node [draw] (ars) at (-5,0) {$\code{a.rs}$};
    \node [draw, below = 1 of ars] (air) {$\code{a.ir}$};
    \node [draw, below = 0.7 of air] (as) {$\code{a.s}$};

    \draw[-stealth] (ars) -- node[midway, left] {\footnotesize compile} (air);
    \draw[-stealth] (air) -- node[midway, left] {\footnotesize compile} (as);
    \draw[-stealth] (air) to[loop right, looseness=4]
    node[midway, above = 0.1] {\footnotesize
      \begin{tabular}{c}
        borrow\\
        checking
      \end{tabular}}
    (air);
    % \draw[->] (air) to[out=60, in=-60, looseness=4] (air);

    \node [right = 0.8 cm of ars, draw] (bc) {$\code{b.c}$};
    \path let \p1 = (bc) in let \p2 = (as) in
        node [draw] (bs) at (\x1, \y2) {$\code{b.s}$};

    \draw[-stealth] (bc) -- node[midway, right] {\footnotesize compile} (bs);

    % \node [right = 0.6 cm of bs, draw] (cs) {$\code{c.s}$};

    \path (as) -- node[] {$+$} (bs);
    % \path (bs) -- node[] {$+$} (cs);

\end{tikzpicture}
\caption{Compilation and Linking}
\label{sfig:comp-link}
% \vspace{-0.2cm}
\end{subfigure}
\begin{subfigure}[b]{.55\textwidth}
\centering
\begin{tikzpicture}
    \node [] (ars) at (-5,0) {$\safe{\sem{\code{a.rs}}}$};
    \node [below = 0.3 of ars] (air) {$\safe{\sem{\code{a.ir}}}$};
    \node [below = 0.3 of air] (as) {$\safe{\sem{\code{a.s}}}$};

    \path (ars) -- node[sloped] {$\imply$} (air);
    \path (air) -- node[sloped] {$\imply$} (as);

    \node [right = 0.4 cm of ars] (bc) {$\safe{\sem{\code{b.c}}}$};
    \path let \p1 = (bc) in let \p2 = (as) in
        node (bs) at (\x1, \y2) {$\safe{\sem{\code{b.s}}}$};

    \path (bc) -- node[sloped] {$\imply$} (bs);

    % \node [right = 0.4 cm of bs] (cs) {$\safe{\sem{\code{c.s}}}$};

    \path (as) -- node[sloped] {$\semlink$} (bs);
    % \path (bs) -- node[sloped] {$\semlink$} (cs);

    \node [fit = (as) (bs), inner sep = -0.0pt] (tgtsem) {};
    \node [below = 0.25 cm of tgtsem]
    (tgtsyn) { $\safe{\sem{\code{a.s} + \code{b.s}}}$};
    \path (tgtsem) -- node[sloped]
          {$\imply$} ($(tgtsyn.north) + (0, 0.1cm)$);

\end{tikzpicture}    
\caption{Safety Preservation and Composition}
\label{sfig:comp-safe}
% \vspace{-0.2cm}
\end{subfigure}
\caption{Verifying Safety for a Program Mixing Rust and C Code}
\label{fig:safety-exm}
% \vspace{-0.5cm}
\end{figure}

An ultimate goal of formal verification is to ensure executable code
running on hardware or OS does not go wrong
because of program errors (e.g., memory error, division by zero), i.e.,
the executable code is \emph{safe}.
Instead of directly targeting executable code, a large number of
safety verification techniques (e.g., separation logic~\cite{reynolds02}, type systems~\cite{pfpl})
work on source programs as they are easier to reason about with richer structures.
In the end, the gap between safety guarantees for source and target
could be bridged by a verified compiler which propagates
source-level safety properties down to target code.
A typical example is the Verified Software Toolchain
(VST)~\cite{appel11:vst, appel2014program}, which is a higher-order separation logic for proving safety of C code and
relies on CompCert\cite{compcert}---the state-of-the-art verified C
compiler---to propagate this safety down to assembly code.

As software industry has been long plagued by security attacks caused
by memory errors (e.g., invalid pointers, 
use-after-free)~\cite{usgov-safe-roadmap}, there is an increasing trend to steer away from
programming languages inherently not memory safe like C and
C++~\cite{whitehouse-tr}, and to embrace languages like Rust with
memory safety guaranteed by safety checking of its compiler
(e.g., borrow checking)~\cite{rust}.
Still, programming tasks demanding manipulation of low-level memory
(e.g., pointer arithmetic) cannot be implemented in safe Rust because
they are rejected by borrow checking. Instead, they need to be
implemented in the unsafe fragment of Rust, or in unsafe languages
like C.
As a result, modern software is increasingly composed of modules written in
both safe and unsafe languages. A prominent example is the Linux
kernel which now contains both C and Rust
components~\cite{rust-for-linux}.

In this paper, we are concerned with verifying the safety of target
code compiled from source programs mixing safe and unsafe code.
\figref{fig:safety-exm} depicts a typical example. The module
\code{a.rs} is implemented in safe Rust. It is compiled to assembly
code \code{a.s} going through borrow checking in the MIR intermediate language. 
The module \code{b.c} is implemented in C because it needs to manipulate low-level
memory.
By encapsulating \code{b.c} under a safe interface, \code{b.c} could
be invoked by \code{a.rs} like a regular Rust module.
After the source modules are compiled to assembly, they are linked
into a complete program \code{a.s + b.s}.
Our goal is to verify the safety of \code{a.s + b.s}. Following the
approach of VST described in the beginning paragraph, an obvious plan
is depicted in~\figref{sfig:comp-safe}.
First, we individually verify the safety of source code \code{a.rs}
and \code{b.c} (e.g., by exploiting separation logics for Rust and
C). We then utilize verified compilers to propagate safety of
\code{a.rs} and \code{b.c} down to assembly level. Finally, if the
safety assumptions between \code{a.rs} and \code{b.c} are correct, we
should be able to prove that the target code \code{a.s + b.s} is safe
from the individual safety of \code{a.s} and \code{b.s}.

If the above plan works, we get an approach to \emph{end-to-end}
verification of safety guarantees \emph{compositionally}, i.e.,
verification is carried out separately in source modules, and compiled
and composed to form safety of the final target. This approach could
greatly reduce the efforts for verifying safety for heterogeneous
programs by exploiting off-the-shelf verification tools for different
languages (e.g. VST and RustBelt~\cite{RustBelt})
and verified optimizing compilers (e.g., CompCert).

Despite the aforementioned potential, it is not obvious at all how the
above approach could be concretely developed. In particular, it is not
even clear what the definition of safety in~\figref{sfig:comp-safe}
could be for it to be preserved by realistic optimizing compilers. Below we
analyze existing approaches to safety verification to bring out
the problems.

\subsection{Problems}
\label{ssec:problems}

\subsubsection{Preservation of Safety by Verified Compilation of Heterogeneous Modules}
\label{sssec:problem-one}
A major goal of the above approach is to support preservation of
safety for modules written in various safe and unsafe languages by
verified compilers. 
On the surface, this is easy to achieve by using existing safety
definitions in source-level verification tools, and by exploiting
semantics-preserving compilers supporting \emph{verified compositional
  compilation} (VCC)~\cite{stewart15,compcertm,compcerto,cascompcert,wang2019,dscal15,direct-refinement}, i.e., separate compilation of
heterogeneous modules and composition of compilation correctness.
However, it turns out that either obvious safety definitions are not
compatible with VCC, or they require a daunting amount of changes to
verified compilers which defeat our goal of using compilers \emph{as
  they are}.
Below, we discuss the problems of combining common safety verification
techniques, i.e., separation logics and type systems, with the
state-of-the-art of VCC~\cite{compcerto,direct-refinement}, i.e., extensions of CompCert based on
simulation of open modules originating from
Compositional CompCert~\cite{stewart15}.

With separation logics, the standard safety definition is
\emph{partial correctness} in Hoare logics~\cite{hoare69}.
A module $M$ is partially correct, if there exists a Hoare triple
$\htriple{P}{M}{Q}$, indicating that if $M$ start with a state
satisfying the precondition $P$, then it can always take the next step
(i.e., does not crash or go wrong) as long as its execution has not
reach an end. Furthermore, if $M$ terminates then it must terminate in
a state satisfying $Q$.

A Hoare triple $\htriple{P}{M}{Q}$ is a property about the module $M$
\emph{as a whole}. As such, semantics-preserving compilers for
\emph{complete} programs (e.g., the original CompCert) preserves
partial correctness.
For example, VST proves that if a C program $M$ is partially correct,
then the target code $M'$ compiled from $M$ by CompCert is also
partially correct.

However, partial correctness cannot be preserved by the
state-of-the-art compilers supporting VCC for \emph{open modules},
i.e., modules that may invoke other modules. To see that, consider a
Hoare triple $\htriple{P}{M}{Q}$ for the open module $M$. It asserts
that every state $S$ reachable from the initial state of $M$ must be
safe. Preserving this triple demands preservation of execution traces
from initial states of $M$ to $S$ by compilation. Suppose such an
execution trace include invocation of a module in the context of $M$,
then it cannot be preserved by compiler correctness theorems for VCC,
as they do not make any assumption about execution of modules living
in the environment. Therefore, it is impossible to propagate partial
correctness of open modules from source to target via VCC (a more
fundamental explanation is that there are inherent conflicts between the
``big-step'' nature of partial correctness and ``small-step''
simulations used in VCC, which we will discuss in~\secref{sec:ideas}).
As an evidence, VST defines a variant of partial
correctness named safeN~\cite{appel2014program}. However, it is only
used at C level and is not preserved by CompCert.
As a result, partial correctness cannot be applied to our example
in~\figref{fig:safety-exm} where \code{a.rs} invokes unsafe
implementation in \code{b.c} through its interface. %% and their
%% safety is required to be preserved modularly by compilation.
  
Besides separation logics, another class of mainstream technique for verifying safety is type systems~\cite{pfpl}. A well-defined type system
ensures \emph{1)} every well-typed program cannot get stuck or go
wrong in execution, also known as the \emph{progress property}, and
\emph{2)} execution does not break well-typedness of programs.
Then, a program $P$ is safe if it is well-typed.
Since types can be given to sub-components of complete
programs---including modules, functions, expressions, etc.---safety of
open modules can be easily defined and preserved by type-preserving compilers.
In the end, safe modules 
can be composed to form complete programs via typing rules at the
target level.

However, adopting well-typedness as safety in~\figref{sfig:comp-safe}
comes with a high cost.
Although there is prior work on \emph{type-preserving
compilation}~\cite{type-preserve-fj, morrisett98}, integrating those
techniques in semantics-preserving compilers require developing type
systems for every intermediate language and proving that types of
programs are preserved by every compiler pass.
For example, CompCert has 9 intermediate languages with little type
information and 20 passes.
Making it type-preserving would be a project for multiple years.

In summary, we face the following question: \textbf{What is a modular
  definition of safety that is sufficiently strong so that it can be
  preserved by verified compilation, and sufficiently lightweight so
  that it does not require intrusive changes to existing verified
  compilers?}

\subsubsection{Combining Safety Checking by Verifying Compilation with Verified Compilation}
\label{sssec:problem-two}

Modern systems programming languages relies on a combination of
language features and safety checking in compilers to discharge safety
errors. For example, to rule out memory errors, Rust introduces
complex ownership and borrowing mechanisms so that any Rust program
that can go through \emph{borrow checking} is (supposedly) memory
safe. In other words, certain classes of safety properties do
not need to be verified at the source-level, they are automatically
established by compilers.

The idea of using compilers to discharge program errors and to
formally prove their absence is known as \emph{verifying
  compilation}~\cite{tony-verifying}. The most prominent techniques of
verifying compilation originates from proof-carrying code
(PCC)~\cite{necula:pcc}. In the initial version of PCC, the compiler
produces target code along with a set of verification conditions
(VC)~\cite{necula98}. Upon successfully discharging VC, the target code is
guaranteed to be safe. Later, foundational proof-carrying code
proposes a semantic type system such that any well-typed term is
safe~\cite{appel01:fpcc}. Hence verifying compilation can be implemented as type
checking. Recent advances along this line include semantic type
systems supporting multiple languages~\cite{patterson-pldi22, real-abi}.

However, the existing techniques for verifying compilation are not
directly applicable to our example in~\figref{fig:safety-exm}. In our
example, borrow checking can only guarantee the checked program
\code{a.ir} is \emph{partially} safe. That is, it only ensures that
there is no memory error in \code{a.ir}. However, there can still be
other errors such as division by zero in \code{a.ir}. We must rely on
additional verification to discharge the remaining errors, and rely on
verified compilation to propagate safety to \code{a.s}. On the other
hand, existing verifying compilation techniques build 
systems to check that programs are \emph{totally} safe, and they usually
prove this total safety directly for target code without
cooperating with verified compilers.

In summary, we face the following questions: \textbf{How could we
  describe the partial safety of a program, i.e., it is absent of a
  particular class of error and may contain other errors? Furthermore,
  how does verifying compilation for this partial safety cooperate
  with verified compilation to ensure the safety of final target
  program?}

\subsection{Our Contributions}

We answer the above questions by introducing a safety definition
called \emph{open safety} which supports uniform description of
partial and total safety for open modules, and cooperation between verified
and verifying compilation. With open safety, we give a concrete
implementation of the approach to end-to-end and compositional
verification of safety properties described in~\figref{fig:safety-exm}
based on the state-of-the-art verified compiler support VCC, i.e.,
CompCertO~\cite{compcerto,direct-refinement,compcertoc}.
Our technical contributions are summarized as follows:
\begin{itemize}
\item
  We define open safety as a ``small-step'' variant of partial correctness on
  top of small-step operational semantics described as open labeled
  transition systems (LTS).
  It allows pre- and post-conditions on the boundary of modules
  implemented in arbitrary languages, thereby enabling compositional
  safety properties for heterogeneous modules.
  Note that open safety is sound (adequate) w.r.t. partial
  correctness, in that it implies partial correctness when applied to
  complete programs.
  By building in a state invariant for LTS, open safety can be
  modularly preserved by verified compilation by proving preservation
  of invariants under simulations.
  Moreover, because state invariants are \emph{existentially
  quantified}, i.e., they are not visible outside of
  individual modules, proving preservation of open safety does not
  require any changes to the compilers or their correctness proofs.
  The above features make open safety a satisfying answer to the
  question at the end of~\secref{sssec:problem-one}.
  We demonstrate its effectiveness in this regard by proving that the
  entire compilation chain of CompCertO preserves open safety, from C
  modules all the way down to assembly code.

\item We generalize open safety to be parameterized over a class of
  safety errors to describe \emph{partial safety},
  i.e., a module is partially safe if it may only contain those
  errors. The generalized open safety enjoys all the previously
  mentioned benefits for verified compilation. Moreover, in the same
  framework, it gives a formal treatment of verifying compilation,
  i.e, verifying compilation is formalized as derivation of total
  safety from partial safety. This gives a satisfying answer to the
  question at the end of~\secref{sssec:problem-two}.

\item To show that open safety can be checked and preserved by
  compilers combining verified and verifying compilation, we develop a
  compiler for a language called Owlang (\emph{Ownership Language}) which can be viewed as a
  subset of core Rust. Like Rust, Owlang supports automatic heap
  management and ownership transfer of heap values. However, it does
  not support references or borrowing. We implement a compiler from
  Owlang to assembly by combining a frontend from Owlang to Clight
  with CompCert. In particular, the frontend contains a verifying pass
  called \emph{ownership checking} (which is a fragment of the Rust
  borrow checker~\cite{movecheck}) for ensuring memory safety under
  ownership transfer of heap values.
  We formally proved \emph{1)} the semantics preservation of the
  \emph{full} Owlang compiler, \emph{2)} open safety is end-to-end
  preserved by the compiler, and \emph{3)} the ownership checking pass
  is correct, i.e., there can be no \emph{temporal} memory error after
  ownership checking (note that, similar to borrow checking, it does
  not prevent spatial memory errors such as index-out-of-range).

\item To demonstrate the effectiveness of our approach, we carry out a
  concrete exercise for proving end-to-end safety following the
  pattern in~\figref{fig:safety-exm}. We implement a hash map
  implemented partly in Owlang and partly in C. We then verify the
  partial safety of the Owlang module and the total safety of the C
  module, respectively. By compiling the C module using CompCertO and
  the Owlang module using our Owlang compiler, we prove the total
  safety of the linked assembly code. Finally, by the soundness of
  open safety, we prove the regular partial correctness for final
  assembly code.

\end{itemize}

%% Note that this work focuses on integration of open safety with VCC,
%% safety verification for source programs are carried out manually. We
%% conjecture that open safety can be integrated with separation logics
%% as it closely mirrors partial correctness except for containing
%% internal state invariants (it implies partial correctness for closed
%% programs).

The formal development of this work is fully formalized in the Rocq
theorem prover.
% (see the supplementary materials). We also provide a
% technical report in the supplementary materials to explain some
% technical details.
% 
In the rest of the paper, we first present the key ideas of our
approach in~\secref{sec:ideas}. We then introduce necessary technical
background~\secref{sec:back-challenges}. We present the formal
framework of open safety in~\secref{sec:safety} and the Owlang
compiler in~\secref{sec:compiler}. We discuss our application to hash
map in~\secref{sec:app}. Finally, we evaluate our work and discuss
related work in~\secref{sec:eval-related}, and conclude
in~\secref{sec:conclusion}.

\section{Key Ideas}\label{sec:ideas}

\begin{figure}[t]
    \begin{subfigure}{.56\textwidth}
\begin{lstlisting}[language=C, firstline=3]
/* buks.c */
typedef void* List_ptr; // bucket type
typedef List_ptr* HashMap; // buckets array
typedef unsigned int uint;
#define N 10 // number of buckets
// external functions implemented in Owlang
extern List_ptr find_process(List_ptr, int);
extern uint hash(int, uint);

List_ptr* find_bucket(HashMap hmap, int k){
  uint index = hash(k, N);
  return &(hmap[index]);
}

void hmap_process(HashMap hmap, int k){
  List_ptr* buk = find_bucket(hmap, k);
  if(*buk != NULL)//bucket is initialized
    *buk = find_process(*buk, k);
}

int* process(int* v){
    *v ^= 42; return v;
}
\end{lstlisting}
        % \vspace{-0.2cm}
        \caption{Array of Buckets in C}
        \label{sfig:buckets}
        % \vspace{-0.2cm}
    \end{subfigure}
    \begin{subfigure}{.43\textwidth}
\begin{lstlisting}[language=Rust, style=colouredRustFootnote]
/* list.rs (the same suffix as Rust) */
struct Node { key: i32, val: Box<i32>, 
    next: Box<List> }
enum List { Nil, Cons(Node) }
// external function implemented in C
extern fn process(v: Box<i32>) -> Box<i32>  

fn hash(k: i32, range: u32) -> u32 {
    return k % range; }

fn find_process(l: Box<List>, k: i32) 
-> Box<List> { 
 match *l { 
  List::Nil => return l, //value not found
  List::Cons(node) => { 
    if k == node.key { // value found
       node.val = process(node.val); 
    } else { // find recursively
        node.next = find_process(
            node.next, k); 
    } // return the ownership
    *l = List::Cons(node); return l; 
} } }
\end{lstlisting}
    % \vspace{-0.2cm}
    \caption{Linked List in Owlang}
    \label{sfig:list}
    % \vspace{-0.2cm}
    \end{subfigure}
\caption{An Example of Hash Map Implemented in Owlang and C}
\label{fig:hash_map}
% \vspace{-0.6cm}
\end{figure}

% We should mention the ownership transfer in list.rs?

We use a running example in~\figref{fig:hash_map} to illustrate our
key ideas. It presents a snippet of the implementation of our hash map
composed of a C module \code{buks.c} and a Owlang module \code{list.rs}
(Owlang's syntax closely follows Rust).
The two modules are compiled
separately and linked together at assembly level. Our goal is to prove
the safety of the target code following the pattern
in~\figref{fig:safety-exm}.

The main function in~\figref{fig:hash_map} is \code{hmap\_process}.
Given a hash map \code{hmap} and a key \code{k}, it is responsible for
encrypting the value associated with \code{k} in \code{hmap}. The
encryption function \code{process} is provided at the C side. For
simplicity of illustration, it is an XOR operation with a fixed
encryption key value 42.
Our hash map is implemented as an array of buckets where each bucket
stores a pointer to a linked list of key-value pairs.
The bucket array (of type \code{HashMap}) is allocated by
\code{malloc} (not shown here) and managed at the C side, while the
linked list (of type \code{enum List}) is implemented as an algebraic
data type in Owlang and \emph{automatically} managed by ownership semantics
like in Rust.
At the C side, a bucket pointing to a linked list has type \code{void*} (i.e., \code{List\_ptr}), indicating that linked lists are opaque
data structures to C.

Looking more closely at the implementation, we notice that
\code{hmap\_process} first obtains the pointer to the bucket of
\code{k} by calling \code{find\_bucket}.
It then invokes the external function \code{find\_process} in Owlang
which recursively traverses the linked list until it either finds the
corresponding value (then calls back to \code{process} at C side) or
returns without finding the value.
Notice that, in \code{list.rs}, the \code{match} command at line 13
takes over the ownership of the linked list stored in \code{l}. After
processing the linked list, we need to return its ownership back to
\code{l}, as shown at line 22.
This signifies the main difference between Owlang and Rust, i.e., it
only implements Rust's ownership semantics, not its reference or
borrowing semantics. Hence, borrowing needs to be implemented as explicit
ownership transfers which is a bit cumbersome.
Despite that, Owlang's ownership semantics and compiler work very much
like Rust, therefore sufficient for illustrating our approach
in~\figref{fig:safety-exm}.
%
% Notably, the safety of evaluating \code{*buk} in \code{hmap\_process}
% relies on the in-bound index returned from \code{hash} at line 11. The
% safety of \code{\code{k\%range}} at line 9 of~\figref{sfig:list}
% relies on the non-zero \code{range} 
%
% We use CompCert to compile \code{buks.c} and the Owlang compiler
% (\secref{sec:compiler}) to compile \code{list.rs}, and finally link
% them into a single assembly program.

Our running example illustrates a typical scenario in programs
combining safe and unsafe code, i.e., responsibility of managing
resources is divided between safe and unsafe languages. Data
structures that can be naturally described in safe languages are
managed by safe code and automatically checked by verifying
compilation, while the remaining data structures (e.g., array) is
written in the unsafe language encapsulated under safe interfaces.
For example, the linked list is implemented in the Owlang with the
benefit that all memory operations (e.g., dereference of \code{l} at
line 13 and 22 in~\figref{sfig:list}) are ensured safe by ownership checking of the Owlang
compiler. Array is implemented in C as it is difficult to check the
safety of array accesses statically (e.g., line 12
in~\figref{sfig:buckets}).
The safe and unsafe modules may interact with each other in
complicated ways. In our example, the two modules \emph{mutually}
invoke each other, creating a cyclic dependency between them, which
significantly complicates verification of their safety.

Finally, note that the Owlang compiler can only guarantee partial
safety. That is, it can detect all temporal memory errors, but no
more.
For example, \code{hash} requires that its second argument
\code{range} is not zero to prevent division by zero at line 9 in~\figref{sfig:list}, and it
must ensure that the return value is less than \code{range} so that
array access at \code{buks.c} is safe. 
Those safety properties must be captured or checked using other
methods, such as pre- and post-conditions in program logics.

All the above complexity needs to be dealt with to realize the
approach in~\figref{fig:safety-exm}. In the following subsections, we
successively discuss the main challenges and the corresponding key
ideas to solving them following the same structure
of~\figref{ssec:problems}. We then conclude with an overview of our
approach.

\subsection{Challenge: Preserving Modular Safety by Verified Compilation}\label{ssec:challeng1}

First and foremost, we need to define safety for each of the modules
in~\figref{fig:hash_map} and utilize verified compilers to preserve
them to target code. As we have discussed in the introduction, VST has
successfully utilized CompCert to preserved partial correctness for
\emph{complete} programs. However, the same approach does not work for
open modules. Below we analyze the fundamental problem behind
this phenomenon.

The state-of-the-art verified compiler, i.e., CompCert, gives
small-step operational semantics to its languages. Such a semantics is
formalized as an LTS describing a transition relation $s \to s'$,
meaning that execution in one-step changes the program state from $s$
to $s'$, and a set of initial states $I$.  We write $s \nstep s'$ to
denote a state transition in multiple steps. The correctness of a
compiler pass is formalized as \emph{backward simulation} between
LTS. Given the source LTS $L_s$ and target LTS $L_t$, the backward
simulation $\csims{L_t}{L_s}$ maintains an invariant $R$ (binary
relation) between source and target states such that the initial
states of $L_t$ and $L_s$ are related by $R$, and the following
properties hold:
\begin{itemize}
\item \textbf{Small-step Simulation:} For any source state $s_s$ and
  target state $s_t$, if $(s_s, s_t) \in R$ and $s_t \to s_t'$ for
  some $s_t'$, then $s_s \nstep s_s'$ and $(s_s', s_t') \in R$
  for some $s_s'$.

\item \textbf{Forward progress:} For any $(s_s, s_t) \in R$, if there
  is some $s_s'$ s.t. $s_s \to s_s'$, then there exists some $s_t'$
  s.t. $s_t \to s_t'$.
\end{itemize}
The first property denotes that any one-step transition in the target
is simulated by zero or more steps in the source. The second property
denotes that if the source does not get stuck at some state $s_s$,
then the target will also not in the related state $s_t$, i.e., at any
point of execution a safe source state is related to a safe target state.
A critical property of backward simulation is its transitivity:
\begin{itemize}
\item {\bf Transitivity of Backward Simulation:} $\forall\; L_1\; L_2\; L_3, \csims{L_1}{L_2} \imply \csims{L_2}{L_3} \imply \csims{L_1}{L_3}$.
\end{itemize}
With this property, correctness of multiple compiler passes can be
combined to form the correctness of the whole compiler.

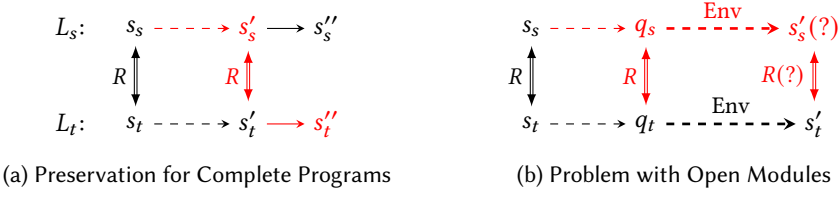
\begin{figure}[t]
\begin{subfigure}{.4\textwidth}
\centering
\begin{tikzpicture}
    
  %% Source States
  \node (s1p) {$s_s$};
  \node[right = 1 of s1p, red] (s2p) {$s_s'$};
  \node[right = 0.5 of s2p] (s3p) {$s_s''$};
  %% Target States
  \node[below = 0.8 of s1p] (s1) {$s_t$};
  \path let \p1 = (s2p) in let \p2 = (s1) in
    node (s2) at (\x1, \y2) {$s_t'$};
  \path let \p1 = (s3p) in let \p2 = (s1) in
    node[red] (s3) at (\x1, \y2) {$s_t''$};
  %% Invariants
  \draw [double, latex-latex] (s1p) -- node[left] {\small $R$} (s1);
  \draw [double, latex-latex, red] (s2p) -- node[left] {\small $R$} (s2);
  %% Source Transitions
  \draw[-stealth, dashed, red] (s1p) -- (s2p);
  \draw[-stealth] (s2p) -- (s3p);
  %% Target Transitions
  \draw[-stealth, dashed] (s1) -- (s2);
  \draw[-stealth, red] (s2) -- (s3);

  %% Text
  \node[left = 0.3 of s1p] (slts) {$L_s$:};
  \path let \p1 = (slts) in let \p2 = (s1) in
    node at (\x1, \y2) {$L_t$:};
  
\end{tikzpicture}
\caption{Preservation for Complete Programs}
\label{sfig:pres-complete}
% \vspace{-0.2cm}
\end{subfigure}
\begin{subfigure}{.5\textwidth}
\centering
\begin{tikzpicture}
    
  %% Source States
  \node (s1p) {$s_s$};
  \node[right = 1 of s1p, red] (q1p) {$q_s$};
  \node[right = 1.5 of q1p, red] (s2p) {$s_s'(?)$};
  %% Target States
  \node[below = 0.8 of s1p] (s1) {$s_t$};
  \path let \p1 = (q1p) in let \p2 = (s1) in
    node (q1) at (\x1, \y2) {$q_t$};
  \path let \p1 = (s2p) in let \p2 = (s1) in
    node (s2) at (\x1, \y2) {$s_t'$};
  %% Invariants
  \draw [double, latex-latex] (s1p) -- node[left] {\small $R$} (s1);
  \draw [double, latex-latex, red] (q1p) -- node[left] {\small $R$} (q1);
  \draw [double, latex-latex, red] (s2p) -- node[left] {\small $R(?)$} (s2);
  %% %% Source Transitions
  \draw[-stealth, dashed, red] (s1p) -- (q1p);
  \draw [-stealth, line width = 1, dashed, red] (q1p) -- node[sloped,
    above] {\small Env} (s2p);
  %% Target Transitions
  \draw[-stealth, dashed] (s1) -- (q1);
  \draw [-stealth, line width = 1, dashed] (q1) -- node[sloped,
    above] {\small Env} (s2);
\end{tikzpicture}
\caption{Problem with Open Modules}
\label{sfig:pres-open}
% \vspace{-0.2cm}
\end{subfigure}

\caption{Preservation of Partial Correctness via Backward Simulation}
\label{fig:pres-partial-correct}
% \vspace{-0.3cm}
\end{figure}

VST is a separation logic for C programs. It operates on
\code{Clight}, a deterministic subset of C defined in CompCert.  A
Hoare triple $\htriple{P}{M}{Q}$ in VST has standard interpretation in
separation logics. For simplicity of the discussion, let us focus on a
complete program $M$, i.e., program with a \code{main} function whose
pre-condition is trivially true (i.e., $P = \top$). We further assume
that $M$ does not terminate. In this case, $\htriple{\top}{M}{Q}$ is
equivalent to that $M$ is \emph{reachable safe}, i.e., any 
state reachable from any initial state in $I$ cannot get stuck. This is formally defined on the semantics
$L$ of $M$:
\begin{itemize}
\item
  $\csafe{L} \;:=\; \forall\; s \in I,\; s', s \nstep s' \imply \exists\; s'', s' \to s''.$
  % \quad (\mbox{$\to$ is the transition relation of $\sem{M}$})$.
\end{itemize}

Partial Correctness is preserved by backward simulation as follows:
\begin{itemize}
\item 
{\bf Preservation of Safety:} $\forall\; L_t\; L_s, \csims{L_t}{L_s} \imply \csafe{L_s} \imply \csafe{L_t}$.
\end{itemize}
The key steps for proving this property are illustrated
in~\figref{sfig:pres-complete} where conclusions (marked in red)
are derived from assumptions (marked in black). To prove
$\csafe{L_s}$, assuming $s_t \nstep s_t'$, we need to prove $s_t' \to
s_t''$ for some $s_t''$. By small-step simulation, we know $s_s \to s_s'$
and $(s_s', s_t') \in R$ for some $s_s'$. By $\csafe{L_s}$, we know
$s_s' \to s_s''$ for some $s_s''$. Finally, by forward progress, we
conclude $s_t' \to s_t''$ for some $s_t''$.
Therefore, in VST, partial correctness for complete programs can be
preserved by the correctness of CompCert.

However, the above proof breaks for open modules. To see that, first
notice that extensions to CompCert supporting open modules (i.e.,
VCC)~\cite{stewart15,compcertm,compcerto,cascompcert,wang2019,dscal15,direct-refinement}
also use small-step simulations which we call \emph{open
simulations}. A key property of open simulations is that they make no
assumption about simulations in the environment. In particular, if a
source module and its compiled target call into the environment, there
is no guarantee about the execution of environment whatsoever (e.g., if they will eventually return). Instead, simulations for open modules
can resume after calling into environment \emph{only if we can provide
some source and target states after the call returns and reestablish the invariant $R$}. By
definition of small-step simulations, this assumption can indeed be
satisfied when the simulation for environment is explicitly
provided.

The above change makes small-step simulation for open modules not
compatible with preservation of $\kwd{reach\_safe}$. If we try to
replay the previous proof, i.e., by assuming $\csims{L_t}{L_s}$ and
$\csafe{L_s}$ to prove $\csafe{L_t}$ for open modules $L_t$ and $L_s$,
we get into problems as depicted in~\figref{sfig:pres-open}. Here, we
assume that the target execution starts from $s_t$, makes a call $q_t$
to the environment, and reaches $s_t'$ after the call returns. As an
example, \code{hmap\_process} may call into the external function
\code{find\_process} in~\figref{fig:safety-exm}. The goal is to show
that $s_t'$ cannot get stuck like in~\figref{sfig:pres-complete}. For
this, we need to show existence of some $s_s'$ such that $(s_s', s_t')
\in R$. However, this is impossible: although we can prove the source
execution reaches some $q_s$ matching $q_t$ by simulating internal
steps, we have no idea if the call $q_s$ into environment will ever
return, since $\csims{L_t}{L_s}$ does not make any assumption about
environment as discussed in the last paragraph.

In conclusion, preservation of partial correctness fails for open
modules because open simulations are small-step, they make no
assumption about how the big-step execution of environment is
simulated, instead relying on assumptions that such execution may
eventually return with reestablished internal invariants. This is
inherently in conflict with the big-step nature of partial
correctness. One solution is to build the assumption about simulation in environment execution into open simulations. This is a
known approach which effectively changes open simulations into
\emph{logical relations}~\cite{hur09, hur11}. It is well-known logical
relations are not transitive in general~\cite{Ahmed06esop}, indicating this change would make it
difficult to support multi-pass compilers like CompCert. Because we
would like to not break verified compilers, we have to find a
suitable alternative to partial correctness.

\subsection{Key Idea: Open Safety}

The above challenge comes from the misalignment between ``big-step''
partial correctness and ``small-step'' simulations. We solve this
challenge by presenting a ``small-step'' variant of partial
correctness such that is composable at the boundary of modules and can
be modularly preserved by the state-of-the-art verified compilers. We
call it \emph{open safety}. The key idea underlying \emph{open safety}
is to maintain internally an invariant for program states at any step
of execution. Instead of stating safety as ``any reachable state does
not get stuck'', we define safety as ``any state satisfying the
invariant does not get stuck''. This change makes safety no longer
dependent on big-step execution history, but only on the current
state. As such, open safety aligns perfectly with the simulation
property of small-step simulations, thereby enables preservation of
safety under them. Below we discuss the design of open safety more
concretely by building it into the framework of
CompCertO~\cite{compcerto,direct-refinement}, and we show how it can
be preserved by CompCertO to target code.

In CompCertO, semantics of modules are defined as \emph{open
labeled transition systems} (open LTS) which communicate with the
environment through \emph{language interfaces}. We write $L: A\arrli
B$ to denote that the LTS $L$ has incoming interface $B$ (accepting
function calls from environment) and outgoing interface $A$
(initiating calls into environment).
For example, $\sem{\code{buks.c}}:\langi{C}\arrli\langi{C}$ denotes
that $\sem{\code{buks.c}}$ communicates with the environment using
C-style function calls and returns.
Compiler correctness theorems for open modules are defined as
\emph{open simulations} between open LTS with \emph{rely-guarantee}
conditions. In this paper, we make use of \emph{open backward
  simulation} $\osims{\scname{R} \arrli \scname{S}}{L_t}{L_s}$ between
the source LTS $L_s$ and target one $L_t$ where $\scname{R}$ and
$\scname{S}$ are \emph{simulation conventions} describing the
rely-guarantee conditions for outgoing and incoming calls,
respectively. For simplicity of discussion, we assume that all the
incoming and outgoing calls have the same language interface and use a
single simulation convention $\scname{R}$. We write backward
simulation as $\osims{\scname{R}}{L_t}{L_s}$ in this case.

We define assertions on language interfaces, referred to as
\emph{safety interfaces}, to capture the pre- and post-conditions of
partial correctness. Later, we show that simulation conventions and
encapsulation for composing safe and unsafe code can also be described
as safety interfaces. Throughout the paper, we denote safety
interfaces using blackboard bold letters (e.g., $\scname{P}$ ,
$\scname{Q}$ and $\scname{I}$). We write the following notation to
represent \emph{open safety} of an LTS $L$:
\begin{align*}
    \mbox{\bf Open Safety:}
    & \quad
    \osafe{L}{\scname{P}}{\scname{Q}}
\end{align*}
Here, $\scname{P}$ specifies the pre- and post-conditions for the
external functions of $L$ and $\scname{Q}$ specifies those for the
internal functions of $L$ which may be invoked by the environment.
Open safety is \emph{composable} under semantic linking of two LTS, as
long as they use complementary safety interfaces
(see~\secref{ssec:safe-compose}).

Roughly speaking, a safety interface $\scname{P}$ denotes a
pre-condition $P'$ and a post-condition $P''$. Intuitively,
$\osafe{L}{\scname{P}}{\scname{Q}}$ may be interpreted as the Hoare
triple $\htriple{Q'}{L}{Q''}$ which depends on an assumption
$\htriple{P'}{L_e}{P''}$ where $L_e$ represents the environment of
$L$. However, this intuition is not entirely accurate as
$\osafe{L}{\scname{P}}{\scname{Q}}$ is stronger than Hoare triples in
the following aspects. First, the LTS $L$ is an abstract notation
agnostic of any particular language while Hoare triples are usually
defined against concrete languages. Second, the small-step nature of
$L$ makes it agnostic of its environment, which may even calls back to
$L$. Therefore, it is very easy to encode mutually recursive safety
requirements in $\osafe{L}{\scname{P}}{\scname{Q}}$. Finally and most
importantly, an invisible (existentially quantified) invariant $\inv$
is maintained during the execution of $L$ s.t. the following
properties hold:
\begin{itemize}
\item {\bf Invariant under Transition:} $\forall\;s\; s', s \in \inv
  \imply s \to s' \imply s' \in \inv$;
\item {\bf Progress under Invariant:} $\forall\; s, s \in \inv \imply
  s \in \mathit{progress}$
\end{itemize}
The first property make sure $\inv$ holds throughout execution. The
second property make sure that, in any state satisfying $\inv$,
execution either ends safely or can still make progress (denoted by $s
\in \mathit{progress}$). Together they enforce the safety of $L$ in a
small-step fashion.

It is worth noting that the above properties about $\inv$ exactly
mirrors type preservation and progress properties in type systems.
Therefore, this invariant can be viewed as the ``type'' over states.
On the other hand, unlike types, our invariants are not visible
outside of a module as they are \emph{existentially quantified} (that
is why it does not appear in the notation
$\osafe{L}{\scname{P}}{\scname{Q}}$). This has two benefits. First,
invariants can be established without relying on any concrete type
systems. Second, open safety for modules with different internal
invariants remain composable. Therefore, the introduction of invariant
is quite lightweight and non-intrusive.

The small-step nature of open safety makes it preservable under open
simulations:
\[
\mbox{\bf Safety Preservation:}
\quad
\osafe{L_s}{\scname{P}}{\scname{Q}} \imply
\osims{\scname{R}}{L_t}{L_s} \imply 
\osafe{L_t}{\scname{P}\compcc\scname{R}}{\scname{Q}\compcc\scname{R}}\]
The simulation convention (i.e., $\scname{R}$) is encoded into the
target safety interfaces via the $\compcc$ operator
(\secref{ssec:safe-pre}).
Intuitively, $\scname{R}$ captures the calling convention of the
compiler, which helps translating safety interface for the source
module into the target. For example, the target safety may depend on
that callee-saved registers are unchanged by external calls, which is
captured in $\scname{R}$.

Finally, we would like to point out that open safety is closely
related to existing safety definitions: the only significant addition
is internal invariants which already takes the form of types in
type systems and is almost always constructed (but hidden in the final
result) when proving partial correctness. Therefore, although we do
not connect with any source-level verification tools in this work, we
conjecture this connection would be quite nature.

\subsection{Challenge: Mixing Safety Checking with Safety Preservation}\label{ssec:ideas-mix-safe-check}

Verifying compilation (e.g., borrow/ownership checking) checks a
certain type of errors (e.g., temporal memory errors), making a
checked program \emph{partially safe}. To preserve this partial safety
by semantics-preserving compilers, a naive approach is to prove that
if the source is absent of this type of errors, then so is the
target. However, this approach does not work in an optimizing
compiler. Consider the following snippet of C code:
\begin{lstlisting}[language=C]
  int x = 10; int *p;   x/0; *p = 10;
\end{lstlisting}
Although \code{p} is an uninitialized pointer, this code does not
contain any memory error as the execution crashes at \code{x/0} before
dereferencing \code{p}.
Since division by zero is an undefined behavior, an optimizing
compiler may compile \code{x/0} into arbitrary code, say a
\code{Nop}. As a result, the optimized code now executes \code{*p =
  10} which causes a memory error.

This example shows that even if a source program is absent of a
certain class of errors, compiler optimizations may reintroduce them.
As discussed in~\secref{sssec:problem-two}, existing approaches to
formalizing verifying compilation does not deal with compilation or
optimization on partially checked programs. Therefore, we need a way
to formalize partial safety that can cooperate with verified
compilation.

\subsection{Key Idea: Open Safety Parameterized by Errors}

To solve the above challenge, we generalize open safety to account for
safety guarantees that \emph{exclude} certain safety properties. We
refer to it as \emph{open partial safety} or just partial safety.

The key idea is to relax the ``make progress'' requirement in open
safety. Specifically, if a module is partially safe, its execution can
either makes progress in a normal way or enters states that violate
given safety properties, e.g., states whose next steps may cause
memory errors by accessing freed memory.
Formally, we use $\opsafep{L}{\scname{P}}{\scname{Q}}{E}$ to denote
partial safety, where $E$ is a state predicate that specifies error
states, i.e., states that violate certain safety properties. For
example, we write $\opsafep{L}{\scname{P}}{\scname{Q}}{\memerr}$ to
represent that $L$ may end execution in states that are \emph{not}
memory safe. The old total safety is thus defined as
$\opsafep{L}{\scname{P}}{\scname{Q}}{\emptyset}$; we often omit
$\emptyset$ for simplicity.
The incremental verification is achieved by first verifying the
partial safety $\opsafep{\sem{M}}{\scname{P}}{\scname{Q}}{E}$ for the
source modules $M$, and then using verifying compilation to ensure
that $M$ cannot reach any state in $E$. This establishes the
correctness theorem for verifying compilation as follows.
\[
\mbox{\bf Correctness of Verifying Compilation:}
\app
\opsafep{\sem{M}}{\scname{P}}{\scname{Q}}{E} \imply
    \checkM{M}\; \imply
    \osafe{\sem{M}}{\scname{P}\compsymb \scname{I}}{\scname{Q}\compsymb \scname{I}}
\]
Here, $(\_\compsymb\_)$ is the conjunction of safety interfaces. We
write $\checkM{M}$ to indicate that $M$ passes the verifying pass.
After verifying compilation, the error states predicate $E$ is ruled
out from the checked module. Moreover, the checked module
automatically satisfies a safety interface $\scname{I}$, which serves
as the safe encapsulation that specifies how this module can safely
interact with other modules, e.g., how modules written in safe Rust
interact with unsafe modules.
Note that the above conclusion can be generalized to partial safety.
For instance, we may establish
$\opsafep{\sem{M}}{\scname{P}\compsymb \scname{I}}{\scname{Q}\compsymb
  \scname{I}}{E_1}$ where $E_1\subset E$, deferring the responsibility
of ruling out $E_1$ to subsequent verifying passes.

\subsection{An Overview of Our Framework}

%% The figure of the general framework
\begin{figure}[t]
\newcommand{\latexarrow}{\tikz[baseline=-1.7ex] \draw[double, -latex] (0,0) -- (0,-0.35);}
\begin{tikzpicture}
    %%%% Source %%%%
    % \node (srcv) {Source Level Verification};
    \node (srcus0) {Source (Unsafe Language)};
    \node [right=0.2cm of srcus0] (srcs0) {Source (Safe Language)};
    \node [below=0.8cm of srcus0] (srcus) {\greentxt{Source}};
    \node [below=0.8cm of srcs0] (srcs) {\redtxt{Source}};

    %%%%% IR and safety checking %%%%%%%
    \node [below=0.4cm of srcs] (ir1) {\redtxt{IR}};
    \node [below=0.4cm of ir1] (ir2) {\greentxt{IR}};

    %%%% Target %%%%
    \node [below=0.4cm of ir2] (tgts) {\greentxt{Target}};
    \path let \p1 = (tgts) in let \p2 = (srcus) in
        node (tgtus) at (\x2, \y1) {\greentxt{Target}};
    \path (tgtus) -- node[sloped] {$\semlink$} (tgts);
    \node [fit = (tgtus) (tgts), inner sep = -0.0pt, draw, rectangle, dashed, ForestGreen] (tgtsem) {};
    \node [below=0.6cm of tgtsem, draw, rectangle, dashed, ForestGreen] (tgtsyn) {\greentxt{Target} $\color{black}\synlink$ \greentxt{Target} };
    \path (tgtsem) -- node [sloped] (sem2syn) {$\imply$} (tgtsyn); 
    % \node [fit = (tgtus) (tgts), inner sep = -0.0pt, draw, rectangle, dashed, ForestGreen] (tgtsem) {};

    %%% Arrows %%%%
    \path (srcus0) -- node [sloped] {$\imply$} (srcus); 
    \path (srcs0) -- node [sloped] {$\imply$} (srcs); 
    \draw [-stealth] (srcus) -- (tgtus);
    \draw [-stealth] (srcs) -- node[right] {\prooflb{(b)}} (ir1);
    \draw [double, -latex] (ir1) --
    node[left] {\small Safety Checking}
    node[right] {\prooflb{(c)}} (ir2);
    \draw [-stealth] (ir2) -- node[right] {\prooflb{(b)}} (tgts);

    %%% Text in the left %%%%
    \path (srcus0) -- (srcus) coordinate[midway] (srcv0);
    \node [left=0cm of srcv0] (srcv) {\small
      \begin{tabular}{c}
        Source-level\\
        Verification
      \end{tabular}};
    \node [right=0cm of srcv0] {\prooflb{(a)}};
    \path (srcus) -- (tgtus) coordinate[midway] (comp0);
    \node [left=0cm of comp0] (comp) {\small Compilation};
    \node [right=0cm of comp0] (comp) {\prooflb{(b)}};
      
    \path (srcs0) -- (srcs) coordinate[midway] (srcv1);
    \node [right=0cm of srcv1] {\prooflb{(a)}};

    %%%% Safety Lifting %%%%
    \node [right=0.4cm of srcs] (srcs1) {\greentxt{Source}};
    \path (srcs) -- node [sloped] {$\imply$} (srcs1);
    \draw [-implies,double equal sign distance] (ir2) -| node[left, pos = 0.75] {\small Lifting} node [right, pos=0.75] {\prooflb{(e)}}  (srcs1);

    \path (tgtsem) -- (tgtsyn) coordinate[midway] (compose);
    \node [left = 0 of compose] {\small Safety Composition};
    \node [right = 0 of compose] {\prooflb{(d)}};
    
\end{tikzpicture}
\caption{The Framework (\greentxt{Green}: total safety, \redtxt{Red}: partial safety)}
\label{fig:e2e-safety-diag}
% \vspace{-0.3cm}
\end{figure}
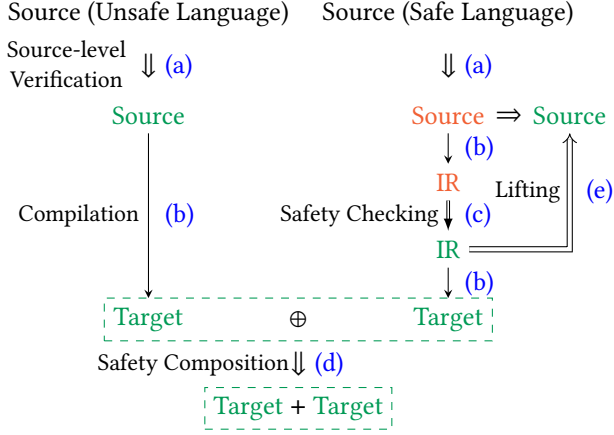

% We first introduce our framework for the end-to-end compositional
% verification of program safety in~\secref{ssec:framework} and then
% illustrate how our framework supports~\figref{fig:mtv-exm} with a
% concrete example in~\secref{ssec:running-example}.

%% introduce the end-to-end safety framework
% divide into three principles: composition, preservation and support safety checking and discuss how does we design the open safety

The complete framework is depicted in~\figref{fig:e2e-safety-diag} with
{\color{blue}blue} labels for each step of verification. From the top
to bottom, we have two source modules written in unsafe and safe
languages, respectively. Their safety is verified at the source level
\refplb{(a)}.
%  by possibly
% different techniques (e.g., program logics, model checking or manual
% proofs). 
For the module in unsafe language, it is verified to be totally safe
(colored in \greentxt{green}). The total safety requires that the
execution of module never gets stuck, i.e., has no undefined
behaviors. 
For the module in safe language, it is verified to be partially safe
(colored in \redtxt{red}). The partial safety is defined by
\emph{allowing} the modules to step to the states that violate certain
safety properties (e.g., it may step to memory error states). These
violation will be proved impossible by the safety checking passes.
The two source modules are compiled to IRs and then to the same target
(e.g., assembly) for linking. The compilation \refplb{(b)} in our
framework preserves the total and partial safety from Source to IR and
then to Target. The safety checking \refplb{(c)} performed on the IR
of the module in safe language ensures that the partially safe module is
totally safe.
%  and then its total safety is preserved to the target.
%
Moreover, the total safety verified in \greentxt{IR} can be propagated to
the source module \refplb{(e)}, which is useful for further refinement
proofs at the source-level.
Finally, the safety composition \refplb{(d)} ensures
that the syntactically linked target module is totally safe.
If this target is a complete program, we can derive regular partial
correctness from total safety.

An an example, the proof of our running example
in~\figref{fig:hash_map} proceeds as follows. We first prove that
\code{buks.c} is totally safe and \code{list.rs} is partially safe.
We then prove the total safety of \code{list.ir} by ownership
checking. We then prove the total safety of the target assembly
modules \code{list.s} and \code{buks.s} by semantics preservation of
CompCertO and the Owlang compiler. We apply horizontal
compositionality and the safety preservation under the syntactic
linking to prove \code{list.s + buks.s} is totally safe, from which we
finally derive the partial correctness of \code{list.s +
  buks.s}. We shall discuss the details in~\secref{sec:app}.

\section{Background}\label{sec:back-challenges}

\subsection{Closed Semantics and Reachability Safety}\label{ssec:back-pre-reach-safe}

In CompCert~\cite{leroy09, Leroy-backend}, the semantics of a whole
program, which we refer to as \emph{closed semantics}, can be
represented by $L=\clts{S}{I}{\to}{F}$, consisting of a set of states
$S$, initial states $I\subseteq S$, an internal transition relation
$\to \subseteq S \times \mathcal{E} \times S$ and final states
$F\subseteq S\times \kwd{int}$ that return integers.
We often omit the \emph{event} ($\mathcal{E}$) of the internal
transition in closed semantics and the following modular semantics.

In closed semantics, \emph{reachability safety} is a predicate defined on
LTS as follows:
\begin{definition}[Reachability Safety]\label{def:reach-safe}
   An LTS $L$ is reachable safe, denoted as $\csafe{L}$, if: 
   \[\forall\app s\app s',\app s\in I \imply s \nstep s' \imply (\exists\app s'',\app s' \to s'') \vee (\exists\app r,\app (s', r)\in F)\]
\end{definition}
%
% An LTS is reachability safe if every state reachable (denoted by
% $\to^*$) from the initial state can make progress, i.e., it either can
% take a step or is in the final state associated with a return value.
Here, $\nstep$ denotes the multi-steps transition. In this paper, we
also write $\csafes{s}$ to denote that the state $s$ is reachable
safe, i.e., all reachable states from $s$ either can take a next step
or it is in the final state.
In the above definition, we do not specify the postcondition for the
return value $r$, which is different from the definition of partial
correctness as mentioned in~\secref{ssec:problems}.
We will see in~\secref{ssec:open-safe-adequacy} that proving a whole
program is reachable safe also requires the proofs of partial
correctness for this program, so in this paper reachability safety is
the final goal of our verification.

% Note that reachability safety is often derived from the partial
% correctness of a whole program, by proving that its \code{main}
% function satisfies the pre- and post-condition.

\subsection{Semantics for Open Modules}
We use \emph{open labeled transition systems} (open LTS) from
CompCertO~\cite{compcerto} to define semantics of modules. The open LTS
extends CompCert’s LTS by characterizing modules' interaction with
environments through explicit transition relations that are
parameterized by \emph{language interfaces}.

A language interface $A = \linterface{A^q}{A^r}$ consists of a set of
queries $A^q$ and replies $A^r$. The language interface used in the C
module is
$\cli = \linterface{\kval \times \ksig_{\cli} \times \kval^* \times
  \kmem}{\kval \times \kmem}$. Here $\kval$, $\kmem$ and
$\ksig_{\cli}$ are the types of values, memory and the function
signature of the callee.
%
% The Owlang interface is $\rli = \linterface{\kval \times \ksig_{\rli}
% \times \kval^* \times \kmem}{\kval \times \kmem}$ which is almost the
% same as the C one except that the signature $\ksig_{\rli}$ contains
% Owl types for the arguments and return value.
%
The queries of $\cli$ have the form $\cquery{v_f}{\sig}{\vec{v}}{m}$
where $v_f$ is the function pointer points to the callee, $\sig$ is
the signature, $\vec{v}$ is the list of arguments and $m$ is the
memory. The replies have the form $\creply{v'}{m'}$ containing the
return value and the memory.
The assembly interface is $\asmli = \linterface{\kregset \times
\kmem}{\kregset \times \kmem}$ where its queries and replies take the
form $\asmquery{\regset}{m}$ containing the registers set and memory.

The open LTS, denoted by $L : A \arrli B$, is a tuple
$\myolts{S}{I}{\to}{F}{X}{Y}$ where $A$ ($B$) is the language
interface for outgoing (incoming) calls and returns.
The transition relation $I \subseteq B^q \times S$ receives an
incoming query to initialize the LTS;
$\to \subseteq S \times \mathcal{E} \times S$ is the internal
transition relation where $\mathcal{E}$ is the event type that is
often omitted in our discussion; $F \subseteq S \times B^r$ finalizes
the LTS with an outgoing reply; $X \subseteq S \times A^q$ models
outgoing queries (i.e., external calls) from the external states;
$Y \subseteq S \times A^r \times S$ defines how an external state
continues execution by selecting a resumed state based on incoming
replies (i.e., the return values of external calls).
To facilitate the following discussion, we first formalize the
\emph{progress property} for an internal state:
\begin{definition}[Progress Property]
    Given $L: A \arrli B$ and a state $s\in S$, $\progress{s}$ holds
    if: 
    \[(\exists\app s',\app s \to s') \vee (\exists\app r,\app (s, r)\in F) \vee (\exists\app q,\app (s, q)\in X)\]
\end{definition}
The first and the second clauses in the progress property are the same
as the conclusion in~\defref{def:reach-safe}, while the third clause
ensures that external states can make progress by emitting some query.
We can then define reachable safe states as
$\csafes{s} = \forall\app s', s \nstep s' \imply \progress{s'}$.

\paragraph{Semantics Linking.}\label{ssec:back-compose-lts}
Two open LTS with compatible interfaces, i.e., $L_1, L_2: A \arrli A$,
can be composed as $L_1 \semlink L_2 : A \arrli A$.
In the composed LTS, the internal state
$s = s_{i_1} :: \ldots :: s_{i_n}$ is a stack of states from $L_1$ and
$L_2$ (i.e., $s_{i_k} \in S_1$ or $s_{i_k} \in S_2$) to simulate the
mutual invocation between $L_1$ and $L_2$.
The key part of the semantic linking is to merge $X$ ($Y$) of one LTS
to $I$ ($F$) of another into an internal step of the composed
semantics.
Therefore, safety composition must ensure that the conditions required
by one module for making external calls (i.e., $X$) satisfy the
assumptions expected by the other module’s initial transition (i.e.,
$I$), and similarly for returns from one's $F$ to $Y$ of another.

\subsection{Open Backward Simulations}
\label{ssec:back-simulation}

% How to divide the challenges into two cases

% A natural extension of reachability safety to open LTS is to define it
% as partial correctness in the style of program logics. In this
% setting, safety conditions are specified over the language interfaces,
% capturing pre- and post-conditions for both incoming calls to the
% module and its outgoing calls to the environment.
% %
% To enable the preservation of this safety predicate, we must refine
% its design in two key aspects: (1) defining safety conditions that can
% capture the rely-guarantee conditions imposed by compilers, and (2)
% identifying an internal safety representation that can be preserved
% through simulation of open LTS. Both aspects present non-trivial
% design challenges.

% The separate compilation of the running example illustrates how
% compiler assumptions about the environment can affect the safety of
% the compiled code.
% %
% In the optimized version $M_1'$ shown
% in~\figref{sfig:back-run-div-asm}, the division at line 13 is no
% longer guarded, as the compiler has eliminated the check
% $\code{d != 0}$ as shown in \figref{sfig:back-run-div-opt}.
% %
% Its safety is no longer ensured by $M_1'$ alone but depends on the
% interference of environments (i.e., the call to \code{div}). If $M_1'$
% is linked with the a modified version of $M_2$ that assigns zero to
% \code{*b}, the resulting program would exhibit division by zero.
%

% The above compilation is considered correct in CompCertO because the
% interference is captured as rely-guarantee conditions within the
% \emph{open simulation}~\cite{compcerto}.

In CompCertO~\cite{compcerto}, simulations between open LTS is defined
as \emph{open simulations}, where the simulation relation is extended
to a \emph{Kripke relation} used to describe the evolution of program
states.
A Kripke relation $R : W \to \pset{S}{S \subseteq A \times B}$ is a
family of relations indexed by a \emph{Kripke world} $W$ and we define
$\krtype{W}{A}{B} = W \to \pset{S}{S \subseteq A \times B}$. The
evolution of Kripke worlds is described by the accessibility relation,
which is denoted as $w \accsymb w'$ for some $w$ and $w'$.
At the boundary of open LTS, the simulation is expressed by
\emph{simulation conventions} which relate source and target language
interfaces $A_1$ and $A_2$. The simulation convention is defined as a
tuple
$\scname{R} = \safeface{W}{\scname{R}^q :
  \krtype{W}{A^q_1}{A^q_2}}{\scname{R}^r : \krtype{W}{A^r_1}{A^r_2}}$
which we write as $\scname{R}: \sctype{A_1}{A_2}$.
For some optimization passes, CompCertO uses unary simulation
conventions to prove their correctness by establishing invariants on
the interfaces. As we will see, these unary simulation conventions are
well-suited for our safety definition.

\begin{figure}
  \hspace{-1.1cm}
  \centering
  \begin{subfigure}[b]{0.2\textwidth}
    \centering
  \begin{tikzpicture}
    \node (q1) {\small $q_s$};
    \node [below = 0.3cm of q1] (w1) {};
    \node [below = 0.3cm of w1] (q2) {\small $q_t$};
    \node [right = 0.4cm of q1, red] (s1) {\small $s_s$};
    \path let \p1 = (s1) in let \p2 = (w1) in 
    node (w2) at (\x1, \y2) {};
    \path let \p1 = (s1) in let \p2 = (q2) in 
    node (s2) at (\x1, \y2) {\small $s_t$};

    \draw [-stealth,red] (q1) -- node[above]{\small $I_s$} (s1);
    \draw [double,latex-latex] (q1) -- node[sloped,midway,below]{\tiny $\scname{R}_B^q(w_B)$}(q2);
    \draw [-stealth] (q2) -- node[above]{\small $I_t$} (s2);
    \draw [double, latex-latex, red] (s1) -- node[sloped,midway,above, red]{\tiny $R(w_B)$}(s2);
    %\path (w1) -- node {\small $=$} (w2);    
  \end{tikzpicture}
\caption{Initial states}
\label{sfig:initial-osim}
\end{subfigure}
\begin{subfigure}[b]{0.2\textwidth}
\centering
  \begin{tikzpicture}
    \node (s1) {\small $s_s$};
    \node [below = 0.3cm of s1] (w1) {};
    \node [below = 0.3cm of w1] (s2) {\small $s_t$};
    \node [right = 0.4cm of s1, red] (s1p) {\small $s_s'$};
    \path let \p1 = (s1p) in let \p2 = (s2) in 
    node (s2p) at (\x1, \y2) {\small $s_t'$};
    \path let \p1 = (s1p) in let \p2 = (w1) in 
     node (w2) at (\x1, \y2) {};

     \draw [-stealth,red, dashed] (s1) -- (s1p);
     \draw [-stealth] (s2) --(s2p);
     \draw [double,latex-latex] (s1) -- node[sloped,midway,below]{\tiny $R(w_B)$}(s2);
     \draw [double,latex-latex,red] (s1p) -- node[sloped,midway,above,red]{\tiny $R(w_B)$}(s2p);
     %\path (w1) -- node {\small $=$} (w2);
  \end{tikzpicture}
\caption{Internal steps}
\label{sfig:internal-osim}
\end{subfigure}
\begin{subfigure}[b]{0.33\textwidth}
    \centering
    \begin{tikzpicture}
    \node (s1) {\small $s_s$};
    \node [below = 0.3cm of s1] (w1) {};
    \node [below = 0.3cm of w1] (s2) {\small $s_t$};
    
    \node [right = 0.4cm of s1, red] (q1) {\small $q_s$};
    \path let \p1 = (q1) in let \p2 = (s2) in 
    node (q2) at (\x1, \y2) {\small $q_t$};
    \path let \p1 = (q1) in let \p2 = (w1) in 
    node (w2) at (\x1, \y2) {};

    \draw [-stealth, red] (s1) -- node[above]{\small $X_s$} (q1);
    \draw [-stealth] (s2) -- node[above]{\small $X_t$} (q2);
    \draw [double,latex-latex] (s1) -- node[sloped,midway,below]{\tiny $R(w_B)$}(s2);
    \draw [double,latex-latex,red] (q1) -- node[sloped,midway,above]{\tiny $\scname{R}_A^q(w_A)$}(q2);
    %\path (w1) -- node {\tiny $w_\kp \accsymb_i \get{w_A}$} (w2);

    \node [right = 1.5cm of q1] (r1) {\small $r_s$};
    \path let \p1 = (r1) in let \p2 = (q2) in 
    node (r2) at (\x1, \y2) {\small $r_t$};
    \path let \p1 = (r1) in let \p2 = (w2) in 
    node (w3) at (\x1, \y2) {};

    \draw[-stealth, line width=1, dashed] (q1) -- node[above] {\tiny External} (r1);
    \draw[-stealth, line width=1, dashed] (q2) -- node[above] {\tiny External} (r2);
    \draw [double,latex-latex] (r1) -- node[sloped,midway,below]{\tiny $\scname{R}_A^r({w_A})$}(r2);
    %\path (w2) -- node {\tiny $\accsymb$} (w3);

    \node [right = 0.4cm of r1, red] (s1p) {\small $s_s'$};
    \path let \p1 = (s1p) in let \p2 = (r2) in 
    node (s2p) at (\x1, \y2) {\small $s_t'$};
    \path let \p1 = (s1p) in let \p2 = (w3) in 
    node (w4) at (\x1, \y2) {};

    \draw[-stealth,red] (r1) -- node[above] {\small $Y_s$} (s1p);
    \draw[-stealth] (r2) -- node[above] {\small $Y_t$} (s2p);
    \draw[double,latex-latex,red] (s1p) -- node[sloped,midway,above]{\tiny $R(w_B)$}(s2p);
    %\path (w3) -- node {\small $=$} (w4);
\end{tikzpicture}
\caption{External calls}
\label{sfig:external-osim}
\end{subfigure}
\begin{subfigure}[b]{0.2\textwidth}
  \centering
  \begin{tikzpicture}
    \node (s1) {\small $s_1$};
    \node [below = 0.3cm of s1] (w1) {};
    \node [below = 0.3cm of w1] (s2) {\small $s_t$};
    \node [right = 0.4cm of s1, red] (r1) {\small $r_s$};
    \path let \p1 = (r1) in let \p2 = (s2) in 
    node (r2) at (\x1, \y2) {\small $r_t$};
    \path let \p1 = (r1) in let \p2 = (w1) in 
    node (w2) at (\x1, \y2) {};

     \draw [-stealth,red] (s1) -- node[above]{\small $F_s$} (r1);
     \draw [-stealth] (s2) -- node[above]{\small $F_t$} (r2);
     \draw [double,latex-latex] (s1) -- node[sloped,midway,below]{\tiny $R(w_B)$}(s2);
     \draw [double,latex-latex,red] (r1) -- node[sloped,midway,above]{\tiny $\scname{R}_B^r(w_B)$}(r2);
     % \path (w1) -- node{\small $\get{w_B} \accsymb_e {w_\kp}' \land w_\kp \accsymb_i {w_\kp}'$} (w2);
     %\path (w1) -- node{\tiny $\accsymb$} (w2);
     %\path (w1) -- node{\small $\accsymb_e \land \accsymb_i $} (w2);
  \end{tikzpicture}
\caption{Final states}
\label{sfig:final-osim}
\end{subfigure}
\caption{Simulation Diagrams for Open Backward Simulation}
\label{fig:open-bsim}
% \end{figure}

% \begin{figure}
%   % \hspace{-1.1cm}
%   \centering
%
\begin{subfigure}[b]{0.2\textwidth}
  \centering
  \begin{tikzpicture}
    \node (q1) {\small $q_s$};
    \node [below = 0.3cm of q1] (w1) {};
    \node [below = 0.3cm of w1] (q2) {\small $q_t$};
    \node [right = 0.4cm of q1] (s1) {\small $s_s$};
    \path let \p1 = (s1) in let \p2 = (w1) in 
    node (w2) at (\x1, \y2) {};
    \path let \p1 = (s1) in let \p2 = (q2) in 
    node [red] (s2) at (\x1, \y2) {\small $s_t$};

    \draw [-stealth] (q1) -- node[above]{\small $I_s$} (s1);
    \draw [double,latex-latex] (q1) -- node[sloped,midway,below]{\tiny $\scname{R}_B^q(w_B)$}(q2);
    \draw [-stealth, red] (q2) -- node[above]{\small $I_t$} (s2);
    % \draw [latex-latex, dashed] (s1) -- node[sloped,midway,above]{\tiny $R(w_B)$}(s2);
    %\path (w1) -- node {\small $=$} (w2);    
  \end{tikzpicture}
\caption{Initial progress}
\label{sfig:initial-osim-progress}
\end{subfigure}
\begin{subfigure}[b]{0.3\textwidth}
\centering
  \begin{tikzpicture}
    \node (s1) {\small $s_s$};
    \node [below = 0.3cm of s1] (w1) {};
    \node [below = 0.3cm of w1] (s2) {\small $s_t$};
    \node [right = 0.8cm of s1] (s1p) {\small $s_s'$};
    \node [right = 0.8cm of s1p] (s1pp) {};    
    \path let \p1 = (s1p) in let \p2 = (s2) in 
    node (s2p) at (\x1, \y2) {};
    \path let \p1 = (s1p) in let \p2 = (w1) in 
     node (w2) at (\x1, \y2) {};
     
     \draw [-stealth, dashed] (s1) -- (s1p);
     \draw [-stealth] (s1p) -- node [above] {\tiny progress} (s1pp);
     \draw [-stealth,red] (s2) -- node [above] {\tiny progress} (s2p);
     \draw [double,latex-latex] (s1) -- node[sloped,midway,below]{\tiny $R(w_B)$}(s2);
     % \draw [latex-latex,dashed] (s1p) -- node[sloped,midway,above]{\tiny $R(w_B)$}(s2p);
     %\path (w1) -- node {\small $=$} (w2);
  \end{tikzpicture}
\caption{Internal progress}
\label{sfig:internal-osim-progress}
\end{subfigure}
\begin{subfigure}[b]{0.33\textwidth}
    \centering
    \begin{tikzpicture}
    \node (s1) {\small $s_s$};
    \node [below = 0.3cm of s1] (w1) {};
    \node [below = 0.3cm of w1] (s2) {\small $s_t$};
    
    % \node [right = 0.4cm of s1] (q1) {\small $q_s$};
    % \path let \p1 = (q1) in let \p2 = (s2) in 
    % node (q2) at (\x1, \y2) {\small $q_t$};
    % \path let \p1 = (q1) in let \p2 = (w1) in 
    % node (w2) at (\x1, \y2) {};

    % \draw [-stealth, dashed] (s1) -- node[above]{\small $X_s$} (q1);
    % \draw [-stealth] (s2) -- node[above]{\small $X_t$} (q2);
    \draw [double,latex-latex] (s1) -- node[sloped,midway,below]{\tiny $R(w_B)$}(s2);
    % \draw [double,latex-latex,dashed] (q1) -- node[sloped,midway,above]{\tiny $\scname{R}_A^q(w_A)$}(q2);
    %\path (w1) -- node {\tiny $w_\kp \accsymb_i \get{w_A}$} (w2);

    \node [right = 1.5cm of s1] (r1) {\small $r_s$};
    \path let \p1 = (r1) in let \p2 = (q2) in 
    node (r2) at (\x1, \y2) {\small $r_t$};
    \path let \p1 = (r1) in let \p2 = (w2) in 
    node (w3) at (\x1, \y2) {};

    \draw[-stealth, line width=1, dashed] (s1) -- node[above] {\tiny External} (r1);
    \draw[-stealth, line width=1, dashed] (s2) -- node[above] {\tiny External} (r2);
    \draw [double,latex-latex] (r1) -- node[sloped,midway,below]{\tiny $\scname{R}_A^r({w_A})$}(r2);
    %\path (w2) -- node {\tiny $\accsymb$} (w3);

    \node [right = 0.4cm of r1] (s1p) {\small $s_s'$};
    \path let \p1 = (s1p) in let \p2 = (r2) in 
    node [red] (s2p) at (\x1, \y2) {\small $s_t'$};
    \path let \p1 = (s1p) in let \p2 = (w3) in 
    node (w4) at (\x1, \y2) {};

    \draw[-stealth] (r1) -- node[above] {\small $Y_s$} (s1p);
    \draw[-stealth,red] (r2) -- node[above] {\small $Y_t$} (s2p);
    % \draw[latex-latex,dashed] (s1p) -- node[sloped,midway,above]{\tiny $R(w_B)$}(s2p);
    %\path (w3) -- node {\small $=$} (w4);
\end{tikzpicture}
\caption{External progress}
\label{sfig:external-osim-progress}
\end{subfigure}
\caption{Forward Progress Property of Open Backward Simulation}
\label{fig:open-bsim-progress}
\end{figure}

The compilation correctness of modules is defined as \emph{open
  backward simulation} denoted as
$\osim{\scname{R}_A}{\scname{R}_B}{L_t}{L_s}$ (or $\osims{\scname{R}_A}{L_t}{L_s}$ if $\scname{R}_A  = \scname{R}_B$).
% where the target LTS
% ($L_t$) simulates the source LTS ($L_s$).
%
It is formally defined as follows
% (we shall write
% $\osims{\scname{R}}{L_t}{L_s}$ to denote
% $\osim{\scname{R}}{\scname{R}}{L_t}{L_s}$)
\footnote{For simplicity, we
  omit the requirements of $\csafes{s_s}$ in (2), (3) and (4), as
  these requirements can be simply derived in the proof of safety
  preservation by the safety of $L_s$.}:
\begin{definition}[Open Backward Simulation]\label{def:open-bsim}
  Given $L_s: A_s \arrli B_s$, $L_t: A_t \arrli B_t$, $\scname{R}_A :
  \sctype{A_s}{A_t}$ and $\scname{R}_B : \sctype{B_s}{B_t}$,
  $\osim{\scname{R}_A}{\scname{R}_B}{L_t}{L_s}$ holds if there is some
  Kripke relation $R \in \krtype{W_B}{S_s}{S_t}$ that satisfies:
  \begin{tabbing}
    \quad\=\quad(1)\=\quad\=\kill
    % \>(1) $\forall\app q_1\app q_2,\app (q_1, q_2) \in \scname{R}_B^q(w_B) \imply (q_1 \in D_1 \iff q_2 \in D_2)$\\
    \>(1) $\forall\app w_B\app q_s\app q_t\app s_t,\app (q_s, q_t) \in \scname{R}_B^q(w_B) \imply (q_t, s_t) \in I_t \imply
    \exists\app s_s, (s_s, s_t) \in R(w_B) \land (q_s, s_s) \in I_s.$\\
    \>(2) $\forall\app w_B\app s_s\app s_t,\app (s_s, s_t) \in R(w_B) \imply \stept{s_t}{s_t'}  \imply
    \exists\app s_s',\app s_s \nstep s_s'\land (s_s', s_t') \in R(w_B).$\\
    \>(3) $\forall\app w_B\app s_s\app s_t\app q_t,\app (s_s, s_t) \in R(w_B) \imply (s_t, q_t) \in X_t \imply$\\
    \>\>$\exists w_A\app q_s,\app (q_s, q_t) \in \scname{R}_A^q(w_A) \land (s_s, q_s) \in X_s\; \land$\\
    \>\>\>$\forall\app r_s\app r_t\app s_t', (r_s, r_t) \in \scname{R}_A^r(w_A) \imply (s_t, r_t, s_t') \in Y_t \imply \exists\app s_s',  (s_s, r_s, s_s') \in Y_s \land (s_s', s_t') \in R(w_B).$\\
    \>(4) $\forall\app w_B\app s_s\app s_t\app r_t,\app (s_s, s_t) \in R(w_B) \imply (s_t, r_t) \in F_t \imply 
    \exists\app r_s, (s_s, r_s) \in F_s\land (r_s, r_t) \in \scname{R}_B^r(w_B).$\\
    \> (5) The below forward progress property holds.
    % (or $\Progress{L_s, L_t}{R}$)
    % $\forall\app w_B\app s_s\app s_t,\app (s_s, s_t) \in R(w_B) \imply \csafes{s_s} \imply$ \\
    % \>\> $(\exists\app s_t',\app s_t \to s_t') \vee (\exists\app r_t,\app (s_t, r_t)\in F_t) \vee (\exists\app q_t,\app (s_t, q_t)\in X_t).$
  \end{tabbing}
\end{definition}
\begin{definition}[Forward Progress Property]\label{def:progress-prop}
  Given $L_s: A_s \arrli B_s$, $L_t: A_t \arrli B_t$,
  $\scname{R}_A : \sctype{A_s}{A_t}$,
  $\scname{R}_B : \sctype{B_s}{B_t}$, and the simulation relation $R$,
  the forward progress property holds if:
    \begin{tabbing}
        \quad\=\quad(1)\=\quad\=\kill
        \>(1) $\forall\app w_B\app q_s\app q_t\app s_s,\app (q_s, q_t) \in \scname{R}_B^q(w_B) \imply (q_s, s_s) \in I_s \imply
    \exists\app s_t,\app (q_t, s_t) \in I_t.$\\
    \>(2) $\forall\app w_B\app s_s\app s_t,\app (s_s, s_t) \in R(w_B) \imply \csafes{s_s} \imply \progress{s_t}$\\
    % \>\> $(\exists\app s_t',\app s_t \to s_t') \vee (\exists\app r_t,\app (s_t, r_t)\in F_t) \vee (\exists\app q_t,\app (s_t, q_t)\in X_t).$\\
        \>(3) $\forall\app w_B\app w_A \app r_s\app r_t\app s_s\app s_s'\app s_t, (s_s, s_t)\in R(w_B) \imply (r_s, r_t) \in \scname{R}_A^r(w_A) \imply (s_s, r_s, s_s') \in Y_s \imply$ \\
        \>\> $\exists \app s_t', (s_t, r_t, s_t') \in Y_t.$
    \end{tabbing}
\end{definition}
Properties (1) to (4) describe how the source LTS simulates the target
one, each of whose simulation diagram is depicted
in~\figref{fig:open-bsim} (in the same order). 
% of open backward simulation are illustrated
% in~\figref{fig:open-bsim}, which describe how source LTS simulates
% the target one.
%
Property (5) is the forward progress property introduced
in~\secref{ssec:challeng1} but is extended for the open semantics. The
three clauses of forward progress properties are depicted in
~\figref{fig:open-bsim-progress}.

% illustrated in~\figref{fig:open-bsim-progress}, which
% ensures that if source can make progress then the target also can do.
% %
% Notably, the forward progress property is also defined in the original
% CompCert, where it is used to prove behavior refinement.

In~\figref{fig:open-bsim} and~\figref{fig:open-bsim-progress}, black
arrows and symbols represent assumptions and red ones represent
conclusions.
In the open backward simulation, property (1) requires $R(w_B)$ holds for
the initial states initialized by matched queries; (2) requires
$R(w_B)$ to be preserved during the execution;
(3) requires that related external states emit related queries to the
environment, and if the environment return related replies, then
$R(w_B)$ can be reestablished;
(4) requires that the final
replies are related; (5) requires that when the source and target are
matched--either by matching queries/replies or by matching internal
states--the safety of the source LTS can be transferred to the
target.

\section{End-to-end Safety for Verified and Verifying Compilation}\label{sec:safety}

\subsection{Definition of Open Safety}\label{ssec:safe-def}

% Safety in closed programs is often defined as reachable safety, i.e.,
% all states reachable from the initial state cannot get stuck. It
% becomes subtle to define reachable safety in open semantics, as it is
% unclear what is the meaning of reachable state when the state relies
% on the replies from the environment. One may think that we can define
% reachable states as there is a legal history from the initial state to
% the current state, where the legal history is defined as all the
% interaction with the environment in the history satisfies the safety
% interfaces.
% %
% However, to the best of our knowledge, it is hard to prove (or
% requires lots of unreasonable assumptions) that it is preservable
% under the open simulation due to the incompatibility between inductive
% defined reachability and non-inductive defined open semantics and open
% simulation.
% %
% We follow the way of proving type soundness by type preservation and
% progress, to define open safety as \emph{invariant preservation and
% progress} along with the satisfaction of safety interfaces at the
% modules boundary.

We first introduce \emph{safety interfaces}, which 
are the unary simulation conventions in CompCertO. These
conventions, originally used to prove compiler correctness, are
well-suited for expressing the boundary assertions required by our
open safety definition.
Specifically, given a language interface $A$, the safety interface for
$A$ is a tuple
$\scname{I}_A = \simconv{W}{\scname{I}_A^q :
  \krtypeunary{W}{A^q}}{\scname{I}_A^r : \krtypeunary{W}{A^r}}$ which
we write as $\scname{I}$ if $A$ can be inferred from the context.
In our safety definition, the component $\scname{I}_A^q$ specifies the
preconditions of incoming or outgoing calls, while $\scname{I}_A^r$
specifies the postconditions of returns to or from the environment.
The Kripke relation
$\krtypeunary{W}{A} = W \to \pset{S}{S \subseteq A}$ is essential to
capture the rely-guarantee conditions in verified and verifying
compilation.
The rely-guarantee conditions are similar to propositions in post
conditions of Hoare triples in separation
logic~\cite{marriage-rely-sep} which describes changes to states
w.r.t. pre-conditions.
For example, if a module $L$ emits an external call satisfying
$\scname{I}_A^q(w_A)$ for some Kripke world $w_A$, it can assume that
the return (e.g., the return memory) from the external call must
satisfy $\scname{I}_A^r(w_A)$. Note that the accessibility relation is
often implicit in $\scname{I}_A^r$, i.e.,
$r \in {\scname{I}_A^r}(w) = \exists\app w', w \accsymb w' \land r \in
\scname{I}_A^r(w')$. The accessibility relation $w \accsymb w'$ serves
to enforce invariants across possible worlds, e.g., to ensure that
some memory owned by $L$ is protected, which corresponds to the
frame-preserving update in separation logics
%
% The accessibility $w \accsymb w'$ can be used to
% ensure that some memory owned by $L$ is protected, i.e., like
% frame-preserving update in separation logics.
%
We will revisit this usage in our definition of the safety interface
required by the ownership checking in~\secref{ssec:own-checking}.

\begin{figure}[t]
\begin{subfigure}[t]{.23\textwidth}
\centering
\begin{tikzpicture}
    \node (q1)  {\small $q_B$};
    % \node [left=0.3 of q1] (d) {\small $D$};
    % \path (q1) -- node [sloped] {\small $\ni$} (d);
    \node [right=0.7 of q1] (s1) {\small $s$};
    \draw [-stealth] (q1) -- node [sloped, above] {\small $I$} (s1);
    \node [above=0.5 of q1] (qb) {\tiny $\scname{Q}_B^q(w_B)$};
    \path (q1) -- node [sloped] {\small $\in$} (qb);
    \node [above=0.58 of s1, red] (sinv) {\tiny $\inv(w_B)$};
    \path (s1) -- node [sloped, red] {\small $\in$} (sinv);
    
    \node [below=0.5 of s1, inner sep=0.2] (s1') {\small \color{red} $s'$};
    \draw [-stealth, red] (q1) -- node [below] {\small $I$} (s1');

\end{tikzpicture}
\caption{Initial States}
\label{sifg:diag-safe-init}
\end{subfigure}
\begin{subfigure}[t]{.2\textwidth}
  \centering
\begin{tikzpicture}
    \node (s1) at (q1) {\small $s$};
    \node [above=0.58 of s1] (sinv) {\tiny $\inv(w_B)$};
    \path (s1) -- node [sloped] {\small $\in$} (sinv);
     %%% s2 %%%
    \node [right=0.7 of s1] (s2) {\small $s'$};
    \draw [-stealth] (s1) -- node [above, sloped] {} (s2);
    \path let \p1=(sinv) in let \p2=(s2) in
        node [red] at (\x2, \y1) (sinv2) {\tiny $\inv(w_B)$};
    \path (s2) -- node [sloped, red] {\small $\in$} (sinv2);
    \node [red] at (s1') (s1p) {\phantom{\small \color{red} $s'$}};
    \draw [red, -stealth] (s1) -- node [below, sloped] {\tiny $\mathit{progress \vee E}$} (s1p);
  \end{tikzpicture}
\caption{Internal Steps}
\label{sifg:diag-safe-internal}
\end{subfigure}
\begin{subfigure}[t]{.35\textwidth}
\centering
\begin{tikzpicture}
     %%% s3 %%%
    \node (s3) at (q1) {\small $s$};
    \path let \p1=(sinv2) in let \p2=(s3) in
        node at (\x2, \y1) (sinv3) {\tiny $\inv(w_B)$};
    \path (s3) -- node [sloped] {\small $\in$} (sinv3);
    \node [right=0.5 of s3] (q2) {\small $q_A$};
    \draw [-stealth] (s3) -- node [above, sloped] {\small $X$} (q2);
    \path let \p1=(qb) in let \p2=(q2) in
        node [red] at (\x2,\y1) (qb2) {\tiny $\scname{P}_A^q(w_A)$};
    \path (q2) -- node [sloped, red] {\small $\in$} (qb2);

    % \path let \p1 = (q2) in let \p2 = (s1p) in
    %   node [red] at (\x1, \y2) (s3p) {\small \dots};
    % \draw [dashed, red, -stealth] (s3) -- node [below, sloped] {\tiny $\mathit{progress}$} (s3p);

    %%% rA and s4 %%%%
    \node [right=1 of q2] (r1) {\small $r_A$};
    \draw[-stealth, line width = 1, dashed] (q2) -- (r1);
    \path let \p1 = (qb2) in let \p2 = (r1) in
        node at (\x2, \y1) (rp1) {\tiny $\scname{P}_A^r(w_A)$};
    \path (r1) -- node [sloped] {\small $\in$} (rp1);
    \node [right=0.7 of r1] (s4) {\small $s'$};
    \draw [-stealth] (r1) -- node [sloped, above] {\small $Y$} (s4);
    \path let \p1 = (sinv3) in let \p2 = (s4) in
        node [red] at (\x2, \y1) (sinv4) {\tiny $\inv(w_B)$};
    \path (s4) -- node [sloped, red] {\small $\in$} (sinv4);
    % progress for rA
    \path let \p1 = (s1') in let \p2 = (s4) in
      node [inner sep=0.1] at (\x2, \y1) (s4') {\small \color{red} $s''$};
    \draw [-stealth, red] (r1) -- node [below] {\small $Y$} (s4');  
\end{tikzpicture}
\caption{External Calls and Returns}
\label{sifg:diag-safe-external}
\end{subfigure}
\begin{subfigure}[t]{.2\textwidth}
\centering
\begin{tikzpicture}
    \node (s5) {\small $s$};
    \path let \p1 = (sinv4) in let \p2 = (s5) in
        node at (\x2, \y1) (sinv5) {\tiny $\inv(w_B)$};
    \path (s5) -- node [sloped] {\small $\in$} (sinv5);
    \node [right=0.5 of s5] (r2) {\small $r_B$};
    \draw [-stealth] (s5) -- node [above, sloped] {\small $F$} (r2);
    \path let \p1=(rp1) in let \p2=(r2) in
        node [red] at (\x2,\y1) (rb2) {\tiny $\scname{Q}_B^r(w_B)$};
    \path (r2) -- node [sloped, red] {\small $\in$} (rb2);

    \path let \p1 = (r2) in let \p2 = (s1p) in
      node [red] at ($(\x1, \y2) + (0, 0.1)$) (s5p) {\phantom{small \dots}};
    % \draw (current bounding box.south east) rectangle (current bounding box.north west);
\end{tikzpicture}
\caption{Final States}
\label{sifg:diag-safe-final}
\end{subfigure}
% \vspace{-0.2cm}
\caption{Diagrams for the Properties of Open Safety}
\label{fig:diag-safe}
% \vspace{-0.3cm}
\end{figure}
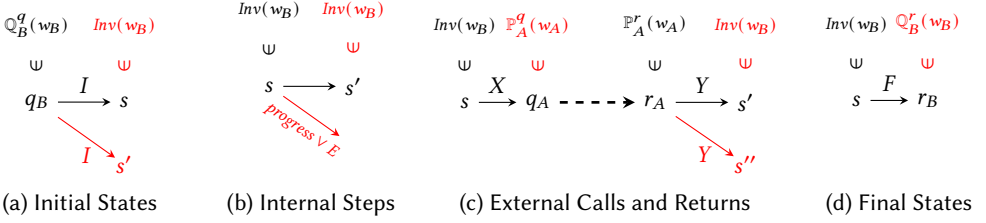

We define open safety that is parametric in a state predicate $E$. By
instantiating the predicate to specific error states, we can obtain
total safety (by setting $E$ to $\emptyset$) and partial safety. The assertions
at the boundaries are specified by the safety interfaces.
%
%Its definition is presented as follows:
%
\begin{definition}[Open Safety]\label{def:osafe}
    Given $L: A \arrli B$, $\scname{P}_A$, $\scname{Q}_B$ and a state predicate $E\subseteq S$, $\opsafep{L}{\scname{P}_A}{\scname{Q}_B}{E}$ holds if there is some invariant $\inv\in \krtypeunary{W_B}{S}$ that satisfies:
    \begin{tabbing}
      \quad\=\quad(1)\=\quad\=\kill
      \>(1) $\forall\app w_B\app q_B, q_B\in \scname{Q}_B^q(w_B) \imply (\exists\app s,\app (q_B, s)\in I) \land (\forall\app s,\app (q_B, s) \in I \imply s\in \inv(w_B)).$\\
      \>(2) $\forall\app w_B\app s\app s',\app s\in \inv(w_B) \imply \stept{s}{s'} \imply  s'\in \inv(w_B).$\\
      \>(3) $\forall\app w_B\app s\app q_A,\app s\in \inv(w_B) \imply (s, q_A) \in X \imply \exists\app w_A,\app q_A\in \scname{P}_A^q(w_A)\app  \land$ \\
      \>\> $\forall\app r_A,\app r_A\in \scname{P}_A^r(w_A) \imply (\exists\app s',\app (s, r_A, s')\in Y) \land (\forall\app s',\app (s, r_A, s') \in Y \imply s'\in \inv(w_B)).$\\
      \>(4) $\forall\app w_B\app s\app r_B,\app s\in \inv(w_B) \imply (s, r_B)\in F \imply r_B\in \scname{Q}_B^r(w_B).$\\
      \>(5) $\forall\app w_B\app s,\app s\in \inv(w_B) \imply \progress{s} \lor s\in E.$
    \end{tabbing}
\end{definition}
%
% Here a progress state is a state that cannot get stuck i.e., it can
% either return to the environment by $F$, emit an outgoing call by
% $X$ or take an internal step.
%
Open safety requires the LTS to be equipped with an invariant, which
we refer to as \emph{safety invariant}. Depending on the verification
techniques, the invariant may take different forms---for instance,
type information, Hoare triples, or state invariant established via
model checking.
The properties of open safety are depicted in~\figref{fig:diag-safe},
where black symbols represent assumptions and red ones represent
conclusions.
For each properties, the open safety requires that: (1) $L$ can enter
an initial state if the query satisfies the precondition
$\scname{Q}_B^q(w_B)$ and every initial state from this query
satisfies the invariant $\inv$; (2) $\inv$ is preserved by each
internal step; (3) if the state satisfying $\inv$ can emit an external
query, then this query satisfies the precondition
$\scname{P}_A^q(w_A)$, and if the environment returns a reply
satisfying the postcondition $\scname{P}_A^r(w_A)$, then $L$ continues
with this reply and all continued states from this reply satisfy
$\inv$; (4) the final reply must satisfy the postcondition
$\scname{Q}_B^r(w_B)$; (5) every state satisfying $\inv$ can make
progress or is in the error states specified by $E$.

The ``small-step'' nature of open safety---maintaining invariant at
each small-step, makes it structurally aligned with the open
simulation (i.e., \defref{def:open-bsim}), where the invariant plays a
role analogous to the simulation relation and both must be preserved
throughout the module's execution.
The open safety also captures the progress property, both at the
module boundaries by safety interfaces, and during its internal
execution by $\progress{s}$, corresponding to the forward progress in
backward simulation.
The cooperation between invariant preservation and progress, just like
the type preservation and progress, ensure that open safety implies
partial correctness.

% small step nature of the open safety, align with open simulation 

\subsubsection{Partial Safety and Correctness of Verifying
  Compilation}\label{sssec:safe-verifying}
The partial safety is an instance of open safety such that
$E\neq \emptyset \wedge E\cap \mathit{progress} = \emptyset$.
% The
% instance of $E$ is often defined as the violation of safety properties
% that are ensured by the verifying compilation or other verification
% methods.
%
Theoretically, verifying partial safety is easier than verifying total
safety, since when the program enters a state in $E$, the make
progress requirement is trivially satisfied.
The error states predicate $E$ in partial safety can be refined by
safety checking when it guarantees that a subset of error states
$E_1 \subset E$ is unreachable. This refinement is considered as the
correctness criterion for this safety checking pass, as shown below:
\begin{definition}[Correctness of Verifying Compilation]\label{def:correct-safe-check}
  Given a module $M$ where $\sem{M}:A \arrli B$, $\scname{P}_A$,
  $\scname{Q}_B$, $E$, and safety interfaces $\scname{I}_A$ and
  $\scname{I}_B$ for the verifying compilation, the safety checking
  $\texttt{Check}$ which rules out $E_1\subseteq E$ is correct if:
    \[ \opsafep{\sem{M}}{\scname{P}_A}{\scname{Q}_B}{E} \imply
        \checkM{M}\; \imply
       \opsafep{\sem{M}}{\scname{P}_A\compsymb \scname{I}_A}{\scname{Q}_B\compsymb \scname{I}_B}{E \backslash E_1}.
    \]
\end{definition}
\noindent Here, $\scname{P}\compsymb \scname{I} = \simconv{W_{\scname{P}} \times W_{\scname{I}}}{\scname{P}^q\compsymb \scname{I}^q}{\scname{P}^r\compsymb \scname{I}^r}$ where for any $q$, $q\in \scname{P}^q\compsymb \scname{I}^q(w_\scname{P}, w_\scname{I})\iff q\in \scname{P}^q(w_\scname{P}) \land q\in \scname{I}^q(w_\scname{I})$ (similarly for $\scname{P}^r\compsymb \scname{I}^r$).
The safety interfaces $\scname{I}_A$ and $\scname{I}_B$ are the
formalized assumptions of this verifying pass, e.g., type interfaces
for type checker and aliasing discipline for the Rust borrow checker.

% %
% The conclusion of total safety in the above definition can be
% generalized to partial safety to support multiple checking passes. For
% example, if the compiler can check the memory safety and the absence
% of integer overflow, we can prove the safety of source module by
% assuming these two properties hold by defining $P(s)= P_1(s) \lor
% P_2(s)$ where $P_1 = \memerr$ and $P_2 = \kwd{int-overflow}$. Assume
% that the compiler first checks the memory safety at the $i$th pass and
% then checks the integer overflow at the $j$th pass. 
% %
% The verification of the $i$th pass is defined as
% $\opsafep{L_i}{\scname{P}_1}{\scname{Q}_1}{P} \imply
% \opsafep{L_i}{\scname{P}_2}{\scname{Q}_2}{P_2}$ where $L_i$ denotes
% the language semantics at the $i$th pass and $\scname{P}_2$
% ($\scname{Q}_2$) is the strengthened interface.
% %
% With the removed assumption $P_1$, the verification of the $j$th pass
% is defined as $\opsafep{L_j}{\scname{P}_2}{\scname{Q}_2}{P_2} \imply
% \osafe{L_j}{\scname{P}_3}{\scname{Q}_3}$, meaning that all the
% assumptions in the source have been verified by the checking. We will
% discuss how do we preserve partial safety from the source to the $i$th
% pass and then to the $j$th pass in~\secref{sssec:partial-safe-pre}.
%

\subsection{Compositionality of Open Safety}\label{ssec:safe-compose}
% $\osafe{L_1}{\scname{P}}{\scname{Q}}$ and
% $\osafe{L_2}{\scname{Q}}{\scname{P}}$.
% %
% Specifically, the guarantee conditions of $L_1$ (i.e., the red safety
% interfaces in~\figref{sifg:diag-safe-external}
% and~\figref{sifg:diag-safe-final}) satisfy the rely conditions of
% $L_2$ (i.e., the black safety interfaces in
% ~\figref{sifg:diag-safe-init} and~\figref{sifg:diag-safe-external}),
% and vice versa.
%
% The safety interfaces of the composed module is the disjoint union
% $\uplus$ of $\scname{P}$ and $\scname{Q}$, which is defined as the
% tuple
%
% We focus on the compositionality of total safety, as safety
% composition occurs at the target level between total safety
% guarantees. In particular, open total safety properties with
% complementary safety interfaces can be composed, as shown below:

We focus on the compositionality of total safety, as safety
composition in practice typically occurs at the target level between
modules that have already been individually verified. In contrast,
composing partial safety would require explicitly combining the error
states of the modules, which complicates the safety checking.
\begin{theorem}[Compositionality of Open Safety]\label{thm:safety-compose}
  Given $L_1, L_2: A\arrli A$, $\scname{P}_A$ and $\scname{Q}_A$,
    \[
    \osafe{L_1}{\scname{P}_A}{\scname{Q}_A} \imply
    \osafe{L_2}{\scname{Q}_A}{\scname{P}_A} \imply
    \osafe{L_1 \semlink L_2}{\scname{P}_A\uplus \scname{Q}_A}{\scname{P}_A\uplus \scname{Q}_A}.
    \]
\end{theorem}
Since there may be mutual invocations between $L_1$ and $L_2$, the
outgoing safety interface of one (e.g., $\scname{Q}_A$ of $L_2$) and
the incoming interface of the other (e.g., $\scname{Q}_A$ of $L_1$)
are often assumed to be identical. More generally, this requirement
can be relaxed: it suffices for a stronger $\scname{Q}_A'$ interface
of $L_2$ to logically entail $\scname{Q}_A$ of $L_1$ by the refinement
of safety interface (see \secref{sec:app}).

The safety interface for $L_1\semlink L_2$ is the disjoint union
$\uplus$ between the interfaces of two modules:
\[\scname{P}\uplus \scname{Q} = 
    \simconv{W_{\scname{P}} + W_{\scname{Q}}}
    {\scname{P}^q \uplus \scname{Q}^q: \krtypeunary{W_{\scname{P}} + W_{\scname{Q}}}{A^q}}
    {\scname{P}^r \uplus \scname{Q}^r: \krtypeunary{W_{\scname{P}} + W_{\scname{Q}}}{A^r}}.
\]
Here $+$ is the notation for building a sum type. For any $q$, 
$q\in \scname{P}^q \uplus \scname{Q}^q(w_{\scname{P}\uplus \scname{Q}}) \iff 
(w_{\scname{P}\uplus \scname{Q}} = \kwd{inl}\app w_{\scname{P}} \imply q\in \scname{P}^q(w_{\scname{P}})) \land 
(w_{\scname{P}\uplus \scname{Q}} = \kwd{inr}\app w_{\scname{Q}} \imply q\in \scname{Q}^q(w_{\scname{Q}}))$
(similarly for ${\scname{P}^r \uplus \scname{Q}^r}$), where \kwd{inl}
and \kwd{inr} are the two constructors for the sum type.
The disjoint union is used to distinguish whether a call targets $L_1$
or $L_2$, so that the corresponding incoming interface can be applied;
the same applies to outgoing calls/returns.
When composing multiple modules, this disjoint union can often be
simplified by using unique function names to distinguish different
modules, thereby avoiding explicit tagging.

For the proof of this compositionality theorem, by the definition of
open safety, the key is to construct a safety invariant for
$L_1\semlink L_2$.
Similar to the disjoint union of safety interfaces at module
boundaries, the internal safety invariant is chosen based on the
currently executing module, i.e., $\inv_1$ (the invariant of
$\osafe{L_1}{\scname{P}_A}{\scname{Q}_A}$) when $L_1$ is active, and
$\inv_2$ when $L_2$ is. Specifically, this internal invariant is
structured as a stack of invariants selected from $\inv_1$ and
$\inv_2$, reflecting the control flow between $L_1$ and $L_2$
(\secref{ssec:back-compose-lts}).
As we can see, the composition does not rely on the internal
definitions of $\inv_1$ and $\inv_2$. This allows results from
different verification techniques to be composed at the level of open
safety, as long as they conform to safety interfaces at boundaries.

\subsection{Preservation of Open Safety through Open Simulation}\label{ssec:safe-pre}

We show that open total safety can be preserved without modifying
existing compiler correctness proofs (i.e., reusing the proof
of~\defref{def:open-bsim}), as illustrated below:
\begin{theorem}[Preservation of Open Total Safety]\label{thm:open-safety-preserve}
  Given $L_s: A_s \arrli B_s$, $L_t: A_t \arrli B_t$, $\scname{R}_A :
  \sctype{A_s}{A_t}$, $\scname{R}_B : \sctype{B_s}{B_t}$, $\scname{P}_{A_s}$ and $\scname{Q}_{B_s}$,
  \[ \osafe{L_s}{\scname{P}_{A_s}}{\scname{Q}_{B_s}} \imply
    \osim{\scname{R}_A}{\scname{R}_B}{L_t}{L_s} \imply
    \osafe{L_t}{\scname{P}_{A_s}\compcc \scname{R}_A}{\scname{Q}_{B_s}\compcc \scname{R}_B}.
  \]
\end{theorem}

We first introduce how to encode the simulation conventions
$\scname{R}$ into the target safety interfaces by the $\compcc$
operator.
Intuitively, if we view $\scname{R}$ as the calling convention of the
compiler, a target query $q_t$ satisfies
$\scname{P} \compcc \scname{R}$ if it can \emph{simulate} a source
query $q_s$ such that they are related by this calling convention and
$q_s$ satisfies $\scname{P}$. Therefore, we define $\compcc$ as
follows:
% %
% We illustrate this using the incoming safety interface (i.e.,
% $\scname{P}\compsymb\rsown\compcc\compra$) of $\semA{\code{list.s}}$
% as shown in~\figref{fig:proof-struct}.
% %
% The meaning of \code{list.s} satisfying
% $\scname{P}\compsymb\rsown\compcc\compra$ is that it can
% \emph{simulates} an Owlang module satisfying
% $\scname{P}\compsymb\rsown$. This simulation is reflected by
% $\compra$: if \code{list.s} receives an incoming query $q_t$ that
% satisfies $\scname{P}\compsymb\rsown\compcc\compra$, then it can
% assume that $q_t$ adheres to the calling convention described by
% $\compra$, that is, there exists a source-level query $q_s$ such that
% $(q_s, q_t)\in \compra^q$.
% %
% Furthermore, $q_s$ must satisfy $\scname{P}\compsymb\rsown$, e.g., the
% argument \code{range} of \code{hash} is not zero, then the
% corresponding register in $q_t$ is also not zero.
% %
% Symmetrically, when \code{list.s} emits a reply $r_t$ to the
% environment, it must ensure that $r_t$ conforms to the calling
% convention, e.g., it does not change the callee-save registers, and
% satisfies the postconditions.
%
%
\begin{definition}[$\compcc$ operator]\label{def:compcc}
  Given source and target interfaces $A_s$ and $A_t$,
  simulation convention $\scname{R} : \sctype{A_s}{A_t}$ such that
  $\scname{R} = \safeface{W_{st}}{\scname{R}^q}{\scname{R}^r}$, and
  source safety interface
  $\scname{P} = \safeface{W_s}{\scname{P}^q}{\scname{P}^r}$:
  \[\scname{P} \compcc \scname{R} :=
    \simconv{W_{s} \times W_{st}} 
    {\scname{P}^q \compcc \scname{R}^q: \krtypeunary{W_{s} \times W_{st}}{A_t^q}}
    {\scname{P}^r \compcc \scname{R}^r: \krtypeunary{W_{s} \times W_{st}}{A_t^r}}
    \quad \text{where}
  \] 
  \[q_t\in \scname{P}^q \compcc \scname{R}^q(w_{s}, w_{st}) \iff 
    \exists\app q_s,\app (q_s, q_t) \in \scname{R}^q(w_{st}) \land q_s\in \scname{P}^q(w_{s})\]
  \[r_t\in \scname{P}^r \compcc \scname{R}^r(w_{s}, w_{st}) \iff 
    \exists\app r_s,\app (r_s, r_t) \in \scname{R}^r(w_{st}) \land r_s\in \scname{P}^r(w_{s}).\]
\end{definition}
%
% Here $\scname{P} \compcc \scname{R}$ requires that if any query
% (reply) satisfies the interface, there exists a source query (reply)
% which satisfies the $\scname{P}$ and it is related to the target query
% (reply) by $\scname{R}$. i.e., it can simulate $\scname{P}$. For
% example, when composing a hand-written assembly and an assembly
% compiled from C, the $\exists$ here requires that the hand-written
% assembly should simulate C function calls i.e., it satisfies the
% calling convention specified by $\scname{R}$.
%
Here, when we consider incoming safety interfaces (dually for outgoing
interfaces), the $\exists$ quantifiers in pre-conditions (i.e.,
$\scname{P}^q \compcc \scname{R}^q$) will appear in the premise of the
open safety judgement, behaving like a $\forall$. It means that the
environment has picked a right (i.e., adhere to the calling
convention) query satisfying $\scname{P}^q$. The $\exists$ quantifiers
in post-conditions (i.e., $\scname{P}^r \compcc \scname{R}^r$) will
appear in the conclusion, meaning that the modules (i.e., the provers)
must choose a right reply satisfying $\scname{P}^r$.
%
% This encoding corresponds to the simulation-based description above.
% maybe need to discuss the exists here can be refined in specific
% verification problems.

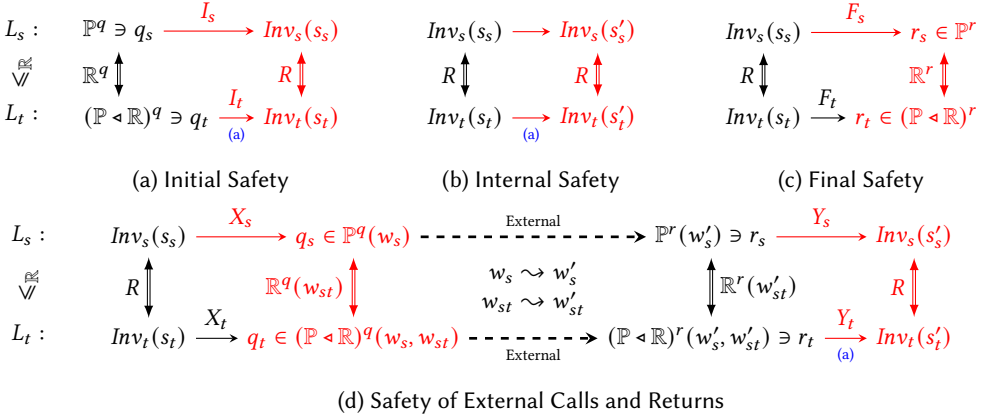
\begin{figure}
\begin{subfigure}[t]{0.05\textwidth}
  \centering
  \begin{tikzpicture}
    \node [] (ls) {\small $L_s:$};
    \node [below = 0.6cm of ls] (lt) {\small $L_t:$};
    \node [below = -0.2cm of lt] {\phantom{\tiny \bluelb{(a)}}};
    \path (ls) -- node [rotate = 90] {$\refinesymb_{\scname{R}}$} (lt);
  \end{tikzpicture}
\end{subfigure}
\begin{subfigure}[t]{.3\textwidth}
\centering
\begin{tikzpicture}
  \node [red] (ss) {\small $\inv_s(s_s)$};
  \node [left = 1.2 of ss] (qs) {\small $\scname{P}^q\ni q_s$};
  \node [below = 0.6 of qs.south west, anchor=north west] (qt)
  {\small $(\scname{P}\compcc\scname{R})^q\ni q_t$};
  \path let \p1 = (ss) in let \p2 = (qt) in
  node [red] (st) at (\x1, \y2) {\small $\inv_t(s_t)$};

  %% Simulation arrows
  \draw [double,latex-latex, red] (ss) -- node[left] {\small $R$} (st);
  \draw [double, latex-latex] (qs) -- node[left] {\small $\scname{R}^q$} (qt.north -| qs);

  %% transition
  \draw[-stealth, red] (qs) -- node[sloped, above] {\small $I_s$} (ss);
  \draw[-stealth, red] (qt) -- node[sloped, above]  {\small $I_t$} node[below] {\tiny \bluelb{(a)}} (st);
  
\end{tikzpicture}
\caption{Initial Safety}
\label{sfig:diag-safe-pre-init}
\end{subfigure}
\begin{subfigure}[t]{.3\textwidth}
\centering
\begin{tikzpicture}
  \node (ss) {\small $\inv_s(s_s)$};
  \node [right = 0.5 of ss, red] (ss1) {\small $\inv_s(s_s')$};
  \node [below = 0.6 of ss] (st) {\small $\inv_t(s_t)$};
  \path let \p1 = (ss1) in let \p2 = (st) in
  node [red] (st1) at (\x1, \y2) {\small $\inv_t(s_t')$};

  %% Simulation arrows
  \draw [double,latex-latex] (ss) -- node[left] {\small $R$} (st);
  \draw [double,latex-latex, red] (ss1) -- node[left] {\small $R$} (st1);

  %% transition
  \draw[-stealth, red] (ss) --  (ss1);
  \draw[-stealth, red] (st) -- node[below] {\tiny \bluelb{(a)}} (st1);
  
\end{tikzpicture}
\caption{Internal Safety}
\label{sfig:diag-safe-pre-internal}
\end{subfigure}
\begin{subfigure}[t]{.3\textwidth}
\centering
\begin{tikzpicture}
  \node (ss) {\small $\inv_s(s_s)$};
  \node [right = 1.2 of ss, red] (rs) {\small $r_s\in \scname{P}^r$};
  \node [below = 0.6 of rs.south east, anchor=north east, red] (rt)
  {\small $ r_t \in (\scname{P}\compcc\scname{R})^r$};
  \path let \p1 = (ss) in let \p2 = (rt) in
  node (st) at (\x1, \y2) {\small $\inv_t(s_t)$};

  %% Simulation arrows
  \draw [double,latex-latex] (ss) -- node[left] {\small $R$} (st);
  \draw [double, latex-latex, red] (rs) -- node[left] {\small $\scname{R}^r$} (rt.north -| rs);

  %% transition
  \draw[-stealth, red] (ss) -- node[sloped, above] {\small $F_s$} (rs);
  \draw[-stealth] (st) -- node[sloped, above] {\small $F_t$} node[below] {\phantom{\tiny \bluelb{(a)}}} (rt);
  
\end{tikzpicture}
\caption{Final Safety}
\label{sfig:diag-safe-pre-final}
\end{subfigure}
\begin{subfigure}[t]{0.05\textwidth}
  \centering
  \begin{tikzpicture}
    \node [] (ls) {\small $L_s:$};
    \node [below = 0.8cm of ls] (lt) {\small $L_t:$};
    \node [below = -0.2cm of lt] {\phantom{\tiny \bluelb{(a)}}};
    \path (ls) -- node [rotate = 90] {$\refinesymb_{\scname{R}}$} (lt);
  \end{tikzpicture}
\end{subfigure}
\begin{subfigure}[t]{.9\textwidth}
\centering
\begin{tikzpicture}
  \node (ss) {\small $\inv_s(s_s)$};
  \node [right = 1.2 of ss, red] (qs) {\small $q_s\in \scname{P}^q(w_s)$};
  \node [below = 0.8 of ss] (st) {\small $\inv_t(s_t)$};
  \path let \p1 = (qs) in let \p2 = (st) in
  node [red] (qt) at (\x1, \y2) {\small $q_t \in (\scname{P}\compcc \scname{R})^q(w_s, w_{st})$};

  \node [right = 3 of qs] (rs) {\small $\scname{P}^r(w_s') \ni r_s$};
  \node [right = 1.2 of rs, red] (ss1) {\small $\inv_s(s_s')$};
  \path let \p1 = (rs) in let \p2 = (st) in
  node (rt) at (\x1, \y2){\small $(\scname{P}\compcc\scname{R})^r(w_s', w_{st}') \ni r_t$};
  \path let \p1 = (ss1) in let \p2 = (st) in
  node [red] (st1) at (\x1, \y2){\small $\inv_t(s_t')$};

  %% Simulation arrows
  \draw [double,latex-latex] (ss) -- node[left] {\small $R$} (st);
  \draw [double, latex-latex, red] (qs) -- node[left] {\small $\scname{R}^q(w_{st})$} (qt);
  \draw [double,latex-latex, red] (ss1) -- node[left] {\small $R$} (st1);
  \draw [double, latex-latex] (rs) -- node[right] {\small $\scname{R}^r(w_{st}')$} (rt);

  %% transition
  \draw[-stealth, red] (ss) -- node[sloped, above] {\small $X_s$} (qs);
  \draw[-stealth] (st) -- node[sloped, above] {\small $X_t$} (qt);
  \draw[-stealth, red] (rs) -- node[sloped, above] {\small $Y_s$} (ss1);
  \draw[-stealth, red] (rt) -- node[sloped, above] {\small $Y_t$} node[below] {\tiny \bluelb{(a)}} (st1);
  \draw [-stealth, dashed, line width = 1] (qs) --
  node [above] {\tiny External}
  node [below = 0.15cm] (acc) {
    \small
    \begin{tabular}{c}
      $w_s \accsymb w_s'$\\
      $w_{st} \accsymb w_{st}'$\\
      \end{tabular}} (rs);
    \draw [-stealth, dashed, line width = 1] (qt) --  node [below] {\tiny External}
 (rt);
    
\end{tikzpicture}
\caption{Safety of External Calls and Returns}
\label{sfig:diag-safe-pre-external}
\end{subfigure}
% \vspace{-0.2cm}
\caption{Preservation of Open Safety through Open Backward Simulation}
\label{fig:safe-pre-proof}
% \vspace{-0.3cm}
\end{figure}

We illustrate the proof structure of the open safety preservation
in~\figref{fig:safe-pre-proof}. Here, each subfigure corresponds to
specific clauses in~\defref{def:osafe}, where
\figref{sfig:diag-safe-pre-internal} includes clause (2) and (5). In
each subfigure, red symbols and arrows represent conclusions (or
intermediate steps) we need to prove, and black symbols and arrows
represent premises (or direct consequences of the premises).

% Among them, the red dashed arrows correspond to ``make
% progress'' requirements that are discharged by applying the forward
% progress property (\defref{def:progress-prop}) to the source-level
% open safety.

% For illustration, we annotate each intermediate proof step with one of
% three labels to indicate the main assumption used in that step: (1)
% refers to the open safety of source LTS (i.e., $L_s$), (2) to the
% forward progress property (\defref{def:progress-prop}), and (3) to the
% open backward simulation (\defref{def:open-bsim}) excluding the part
% of (2).

We define the safety invariant for $L_t$ as $\inv_t$ such that
$\forall\app s_t \in \inv_t \iff \exists\app s_s,\app s_s\in \inv_s
\wedge (s_s, s_t) \in R$ where $R$ is the simulation relation.
Once $s_t\in \inv_t$ (or $\inv_t(s_s)$) holds, we can apply the
forward progress (\figref{fig:open-bsim-progress}) to conclude that
$s_t$ can make progress ($\csafe{s_s}$ is proved by the safety of
$L_s$), thereby proving the steps annotated by \bluelb{(a)}.
To show that the invariant $\inv_t$ is preserved during the execution,
we apply the properties in~\figref{fig:open-bsim}. Now the steps
annotated by \bluelb{(a)} become assumptions.
Consider the proof of~\figref{sfig:diag-safe-pre-internal}.  Given
that $s_t$ steps to $s_t'$, the step simulation ensures that $s_s$
steps to $s_s'$ and $(s_s', s_t') \in R$ holds. By the preservation of
$\inv_s$ and the definition of $\inv_t$, we conclude
$s_t' \in \inv_t$.

% Can remove it
% We use the proof of clause (3) in open safety (as shown
% in~\figref{sfig:diag-safe-pre-external}) to explain the proofs in
% details.
% %
% From left to right: by the definition of $\inv_t$, we obtain
% $\inv_s(s_s)$ and $(s_s, s_t) \in R$. Then, by backward simulation, we
% know that $s_s$ can emit an outgoing query $q_s$ and a world
% $w_s$. The open safety of $L_s$ guarantees that $q_s$ satisfies the
% precondition $\scname{P}^q(w_s)$. By the simulation
% $(q_s, q_t)\in \scname{R}^q(w_{st})$ and the definition of $\compcc$,
% we prove that $q_t$ satisfies the corresponding precondition.
% %
% After the external call returns, we are given that the reply $r_t$
% satisfies the postcondition, which implies
% $(r_s, r_t)\in \scname{R}^r(w_{st}')$. Using clause (3) of
% \defref{def:progress-prop} to derive $Y_t$ and the backward simulation
% to derive $Y_s$, we reestablish the simulation $(s_s', s_t') \in R$,
% from which we conclude $\inv_t(s_t')$.

Although the definition of $\inv_t$ introduces the compiler-level
invariant (i.e., the simulation relation) into the internal invariant
of the target module, this does not affect modular verification in
practice.
% As discussed earlier, verification is performed at the
% source level, where users only need to construct $\inv_s$.
%
As discussed earlier, the composition of modules does not require
knowledge of the internal structure of $\inv_t$. It relies solely on
complementary safety interfaces between modules. Although the
preserved safety interface on the target side may still carry
compiler-specific calling conventions,
% this is the most general case,
% which aims to preserve the source-level safety interface.
%
it is entirely possible to refine it into a native safety interface
(i.e., does not rely on $\compcc$) that encodes only the requirements
imposed by the calling convention on the target program.
% , rather than the full calling convention itself.

Based on the above proof structure, we now revisit how our open safety
addresses the challenge discussed in~\secref{ssec:challeng1}, i.e.,
how to reestablish the simulation relation after external calls.
First, we use $\compcc$ to constrain the environment so that any
target reply it returns corresponds to a source reply.
However, this condition alone is insufficient. If we use partial
correctness to quantify the related replies, e.g., via universal
quantification, they may deviate from what the open simulation
expects, i.e., $\scname{R}^r(w_{st}')$
in~\figref{sfig:diag-safe-pre-external}.
We resolve this with the safety invariant that aligns with the simulation,
to ensure the environment’s replies are consistent with the
requirements of open simulation.

% requirements of open simulation.  To resolve this, our open
% safety formulation introduces a \emph{small-step invariant} that
% ensures the environment’s replies are consistent with the stepwise
% requirements of open simulation.

\begin{figure}
\centering
\begin{tikzpicture}
  \node (l1) {\small Source Partial Safe};
    % $\opsafep{L_s}{\scname{P}}{\scname{Q}}{E}$};
  \node [right = 2.4 of l1](l2)
  {\small IR Partial Safe};
    % $\opsafep{L_{ir}}{\scname{P}\compcc\scname{R}_1}{\scname{Q}\compcc\scname{R}_1}{E}$};
  \node [below = 1 of l2](l2s)
  {\small IR Total Safe};  
  % {\small $\osafe{L_{ir}}{\scname{P}\compcc\scname{R}_1\compsymb \scname{I}}{\scname{Q}\compcc\scname{R}_1\compsymb \scname{I}}$};
  \node [right = 2.4 of l2s](lt)
  {\small Target Total Safe};
  % {\small $\osafe{L_{t}}{\scname{P}\compcc\scname{R}_1\compsymb \scname{I}\compcc\scname{R}_2}{\scname{Q}\compcc\scname{R}_1\compsymb \scname{I}\compcc\scname{R}_2}$};
  \path let \p1 = (l1) in let \p2 = (l2s) in
  node (l1s) at (\x1, \y2)
  {\small Source Total Safe};
  % {\small $\osafe{L_s}{\scname{P}\compsymb \scname{I}}{\scname{Q}\compsymb \scname{I}}$};
  % \path let \p1 = (l2) in let \p2 = (lt) in
  % node (l1s2) at (\x1, \y2)   {\small $\osafe{L_s}{\scname{R}_1\cccomp \scname{P}\compcc\scname{R}_1\compsymb \scname{I}}{\mydots}$};
  \path [draw, -stealth, line width = 1.2](l1) -- node [above] {\small BSim: ${}_{\scname{R}}^E\!\!\bsim$}
  node [below] {\small \thmref{thm:open-partial-safety-preserve}} (l2);
  \path [draw, -stealth, line width = 1.2](l2s) -- node [above] {\small BSim: ${}_{\scname{R}}\!\!\bsim$}
  node [below] {\small \thmref{thm:open-safety-preserve}} (lt);
  \path [draw, -stealth, line width = 1.2](l2) -- node [right] {\small By the proof of \defref{def:correct-safe-check}} (l2s);
  \path [draw, -stealth, line width = 1.2](l2s) -- node [above] {\small FSim: $\fsim_{\scname{R}}$}
  node [below] {\small \thmref{thm:open-safety-preserve}} (l1s);

\end{tikzpicture}
% \vspace{-0.2cm}
\caption{A Road Map of Safety Propagation in Compilation Chain}
\label{fig:safe-summary}
% \vspace{-0.3cm}
\end{figure}
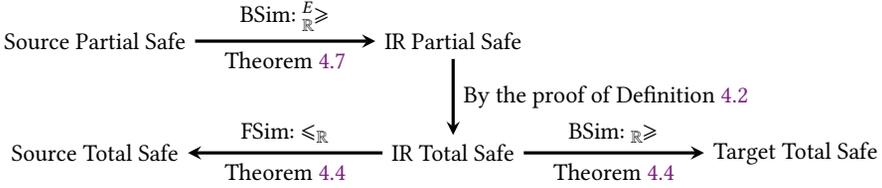

% % \vspace{-0.2cm}
\subsubsection{Preservation of Open Partial Safety}\label{sssec:partial-safe-pre}

% TODO: summarize the simulation we have mentioned: 

% So far, our discussion has focused on the preservation of open total
% safety. However, i
As safety checking (e.g., borrow checking) is usually performed at
intermediate stages, it is essential to preserve partial safety to
these stages, ensuring that the safety checking can indeed verify the
intended properties.
% TODO: discuss the difference from preserving safety properties
The key to preserving partial safety lies in ensuring that the
compiler does not miscompile error states specified by $E$ in the
source-level partial safety, into different kinds of target-level
errors. For instance, a memory error should not be transformed into a
division-by-zero error. Such transformations are not ruled out by
backward simulation.
Note that this problem (i.e., preserving certain errors) is different
from that discussed in~\secref{ssec:ideas-mix-safe-check}, where we
discuss the difficulties of preserving certain safety properties.
% the
% challenge, therefore, is how to extend backward simulation to prevent
% compilers from introducing unintended errors.
 
Our approach mirrors the way partial safety extends total safety. We
extends the forward progress property in backward simulation to ensure
that source-level error states (i.e., $E_s$) are preserved through the
compilation, as shown below:
\begin{definition}[Partial Forward Progress Property]\label{def:partial-progress-prop}
  Given $L_s: A_s \arrli B_s$, $L_t: A_t \arrli B_t$,
  $\scname{R}_A : \sctype{A_s}{A_t}$,
  $\scname{R}_B : \sctype{B_s}{B_t}$, $E_s\subseteq S_s$, $E_t\subseteq S_t$, and $R$,
  the partial forward progress property holds if:
    \begin{tabbing}
      \quad\=\quad(1)\=\quad\=\kill
      \>(1), (2), and (3) are the same as those in~\defref{def:progress-prop}. \\
      \> (4) $\forall\app w_B\app s_s\app s_t,\app (s_s, s_t) \in R(w_B) \imply s_s\in E_s \imply\progress{s_t} \lor s_t\in E_t$.
    \end{tabbing}
\end{definition}
Here, clause (4) indicates that any error state in the source ($E_s$)
can only be compiled to either a corresponding error state in the
target ($E_t$) or a safe state, but never to an unintended error
state. We replace the original forward progress property in open
backward simulation with the above definition, yielding a new backward
simulation, denoted as $\osimE{\scname{R}_A}{\scname{R}_B}{L_t}{L_s}$
or $\osimsE{\scname{R}}{L_t}{L_s}$.
Using this definition, we establish the theorem of partial safety
preservation as follows:
\begin{theorem}[Preservation of Open Partial Safety]\label{thm:open-partial-safety-preserve}
  Given $L_s: A_s \arrli B_s$, $L_t: A_t \arrli B_t$, $\scname{R}_A :
  \sctype{A_s}{A_t}$, $\scname{R}_B : \sctype{B_s}{B_t}$, $\scname{P}_{A_s}$, $\scname{Q}_{B_s}$ $E_s\subseteq S_s$ and $E_t\subseteq S_t$,
  \[ \opsafep{L_s}{\scname{P}_{A_s}}{\scname{Q}_{B_s}}{E_s} \imply
    \osimE{\scname{R}_A}{\scname{R}_B}{L_t}{L_s} \imply
    \opsafep{L_t}{\scname{P}_{A_s}\compcc \scname{R}_A}{\scname{Q}_{B_s}\compcc \scname{R}_B}{E_t}.
  \]
\end{theorem}

\subsubsection{Lifting Total Safety from IR to Source}

% Although our primary concern is with the total safety of the target
% program, and only partial safety needs to be verified at the source
% level, certain properties, such as the behaviors refinement in
% CompCert, require the source program to be totally safe.
%
We show that even when total safety is established at an intermediate
stage, e.g., through safety checking on an intermediate representation
(IR), it can still be propagated back to the source level.
The idea is to apply \thmref{thm:open-safety-preserve} to the forward
simulation (denoted as $\osims{\scname{R}}{L_s}{L_t}$) between source
and target modules.
Specifically, we can reuse the existing forward simulation proofs
established at each compilation pass, and additionally prove the
forward progress property in the reverse direction, that is, if a
target state is not stuck, then the related source state must also be
not stuck.
This condition ensures that the compiler does not compile an erroneous
source state into a safe target state, which would otherwise make the
back-propagation of total safety impossible.

% % \vspace{-0.2cm}
\paragraph{Summary}
We summarize the aforementioned properties of safety verification and
preservation in \figref{fig:safe-summary}, which illustrates how
partial safety at the source level can derive total safety for both the
target and source levels by applying them to forward (FSim) and
backward (BSim) simulations, respectively. Note that we also use
$L_s \; {}_{\scname{R}}\!\!\bsim L_t$ to represent the backward
simulation $L_t \fsim_{\scname{R}} L_s$.
%
% In the next section, we will see its application in a realistic
% verified and verifying compilation.

% We have introduced three kinds of simulations used in the preservation
% of open safety. The first is for the preservation of total safety from
% the source to target. It is proved by
% applying~\thmref{thm:open-safey-preserve} in the backward simulation
% which is flipped from the forward simulation by proving receptiveness
% for the source and determinism for the target as CompCert does. The
% backward simulation contains the progress property which is derived
% from the forward simulation and the safety of the source.
% %
% The second is for the preservation of partial safety from source to
% target. The extra requirement is to prove the property of errors
% preservation in the simulation. The third is for lifting total safety
% from the target to source, which is proved by
% applying~\thmref{thm:open-safey-preserve} in the forward simulation of
% the passes along with the additional progress property.

\subsection{Soundness of Open Safety}\label{ssec:open-safe-adequacy}

If a module $M$ is a complete program, i.e., it has a main entry and
has no external calls, then its open semantics can be transformed into
closed semantics, e.g., by eliminating the transitions of $X$ and
$Y$. Accordingly, its open safety implies the safety under closed
semantics, namely, that all reachable states can make progress. This
is formalized in the following theorem:

\begin{theorem}[Soundness of Open Safety]\label{thm:adequacy}
  Given a module $M$ and a safety interface $\scname{Q}$,
  \[ \osafe{\sem{M}}{\bot}{\scname{Q}} \imply
    \code{init}(M)\app \text{satisfies}\app \scname{Q} \imply
    \csafe{\sem{M}_{\code{closed}}}.
  \]
\end{theorem}
Here, the first premise requires that $M$ satisfies open safety under
the incoming interface $\bot$, meaning that it cannot call external
functions as $\bot$ cannot be satisfied. The second premise ensures
that the initialization of $M$ (e.g., memory initialization) satisfies
$\scname{Q}$ to enable the premise for using open safety. $\csem{\_}$ in the conclusion represents the
closed semantics.

% !TEX root = main.tex

\section{Proving End-to-end Open Safety for an Ownership Language}\label{sec:compiler}

% Rust is a nice example, we choose the ownership subset for simplicity, but it is effective

% In this section, we introduce our verified compiler for the ownership
% language (Owlang), the ownership checking phase and their integration
% within the end-to-end open safety framework. 

% in the ownership language (Owl) which is inspired from
% Rust. We will introduce the key features of Owl and its verified
% compiler frontend from Owl to Clight in~\secref{ssec:own-compiler},
% the ownership checking which ensures memory safety
% in~\secref{ssec:own-checking}, and the theorem of end-to-end safety
% in~\secref{ssec:owl-dr}.

\begin{figure}
\def\bheight{0.8cm}
\begin{tikzpicture}
    %%% languages
    \hblock{\bheight}{draw, minimum width = 1.4cm, rounded corners=0.1cm}  (owl) {Owlang};
    \hblock{\bheight}{draw, minimum width = 1.4cm, rounded corners=0.1cm, right = 1.4cm of owl}  (ir1) {Owl-IR};
    \hblock{\bheight}{draw, minimum width = 1.4cm, rounded corners=0.1cm, right = 1.6cm of ir1}  (ir2) {Owl-IR};
    \hblock{\bheight}{draw, minimum width = 1.4cm, rounded corners=0.1cm, right = 1.6cm of ir2}  (cl) {Clight};
    \hblock{\bheight}{draw, minimum width = 1.4cm, rounded corners=0.1cm, right = 1.5cm of cl}  (asm) {Asm};

    %%% compilation 
    \draw [->] (owl) -- node [above] {\small Lowering} node [below] {$\fsim_{{\rid}}$} (ir1);
    \draw [->] (ir1) -- node [above, align=center] {\small Drop\\ \small Elaboration} node [below] {$\fsim_{{\rinjp}}$} (ir2);
    \draw [->] (ir2) -- node [above, align=center] {\small Clight\\ \small Generation}  node [below] {$\fsim_{{\rinjp}\compsymb \comprc}$} (cl);
    \draw [->] (cl) -- node [above] {\small CompCert} node [below] {$\fsim_{\compca}$} (asm);

    %%% move checking
    \draw[->] (ir1.north west) to[bend left=130, looseness=1.5]
        node[midway, above] {\small Ownership Checking} (ir1.north east);
 \end{tikzpicture}
% \vspace{-0.1cm}
\caption{Structure of the Owlang Compiler}
\label{fig:owl-compiler}
% \vspace{-0.3cm}
\end{figure}
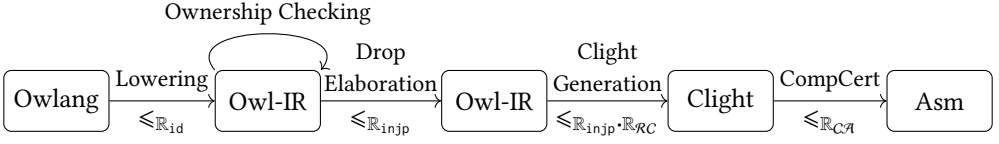

In this section, we present the application of our open safety
approach to a verified and verifying compiler for Owlang.
As shown in~\figref{fig:owl-compiler}, the compiler consists of a
verified frontend from Owlang to Clight and a verified backend from
Clight to (CompCert x86-64) assembly.
The frontend proceeds in three passes. The first pass lowers Owlang
into an intermediate representation called Owl-IR. The second pass
performs Drop Elaboration modeled after~\cite{non-zero-copy},
inserting explicit drop operations (like destructors in C++) for
values that go out of scope. The final pass translates Owl-IR into
Clight.
%%% Technical report
\ifdefined\islong
Detailed explanations of the frontend can be
  found in~\secref{sec:compiler-frontend}.
\else Detailed explanations of the frontend can be
  found in the supplementary materials.  \fi
The backend is reused from the CompCertO optimizing
compiler\footnote{It supports the compilation passes from
  Clight to assembly except for the \kwd{Unusedglob} which removes
  unused module-local definitions, due to the limitation of
  distinguishing module-local and global definitions in CompCertO (see
  Section 7.3 in~\cite{direct-refinement})}.
The Owlang compiler performs ownership checking at the Owl-IR level
after the Lowering pass, which detects illegal uses of variables that
do not have full ownership of their values.

We annotate the simulation conventions used in the proof of each pass
in~\figref{fig:owl-compiler}. Here, $\rid$ is the identity relation
and $\rinjp$ is used for the verification of compilation that changes
the memory structure, e.g., by adding or removing stack allocated
variables.
%%% Technical report
\ifdefined\islong
$\rinjp$ is required in Drop Elaboration as it would insert auxiliary
variables to control the execution of drop operations. For the pass of
Clight Generation, it is required as this pass adds C functions to
implement the drop operations, which would change the structure of
stack.
\fi
$\comprc$ is the foreign function interface between Owlang and C,
primarily relating their types and data layouts, e.g., how \code{enum}
in Owlang is translated to the tagged union in C and how they are
represented in memory.
$\compca$ is the C calling convention adopted from the compiler
correctness of CompCertO~\cite{compcerto, direct-refinement}.

\myparagraph{Compiler Correctness}
We first give the correctness theorem of the Owlang compiler:
% is stated as the open backward
% simulation between Owlang and assembly modules:
%
\begin{theorem}[Compiler Correctness]
  Compilation in the Owlang compiler (denoted as \code{OwlComp}) is
  correct in terms of open backward simulation,
    \[ \forall\app (M_s: \textnormal{Owlang})\app (M_t: \textnormal{Asm}),\app
       \code{OwlComp}(M_s) = M_t \imply 
        \osims{\compra}{\sem{M_t}}{\sem{M_s}}
    \]
\end{theorem}
This theorem is proven by first composing the open forward simulation
of the Owlang compiler frontend and the CompCertO backend, and then
converting the composed forward simulation to backward simulation as
CompCert does, i.e., through proving \emph{receptiveness} for the
source semantics and \emph{determinism} for the target semantics. The
result is
$\osims{\rid \compsymb \rinjp \compsymb (\rinjp \compsymb \comprc)
  \compsymb \compca}{\sem{M_t}}{\sem{M_s}}$.
We then refine
$\rid \compsymb \rinjp \compsymb (\rinjp \compsymb \comprc) \compsymb
\compca$ into $\compra$---the calling convention between Owlang and
assembly, using the techniques of direct
refinement~\cite{direct-refinement}. As a result, $\compra$ directly
relates Owlang and assembly interfaces without mentioning internal
simulation conventions, which simplifies the reasoning about safety
composition across languages (see~\secref{sec:app}).

\begin{figure}
\centering
\begin{tikzpicture}
  \node (l1) {\small $\opsafep{\sem{M_s}}{\scname{P}}{\scname{Q}}{\memerr}$};
  \node [right = 2.2 of l1](l2)
  {\small $\opsafep{\sem{M_{ir}}}{\scname{P}}{\scname{Q}}{\memerr}$};
  \node [below = 1 of l2](l2s)
  {\small $\osafe{\sem{M_{ir}}}{\scname{P}\compsymb\rsown}{\scname{Q}\compsymb\rsown}$};  
  \node [right = 1.8 of l2s](lt)
  {\small $\begin{aligned}
    \sem{M_t}\models \scname{P}\compsymb \rsown\compcc \compra &\\
    \arrli \scname{Q}\compsymb \rsown\compcc \compra&
  \end{aligned}$};

  % {\small $\osafe{L_{t}}{\scname{P}\compcc\scname{R}_1\compsymb \scname{I}\compcc\scname{R}_2}{\scname{Q}\compcc\scname{R}_1\compsymb \scname{I}\compcc\scname{R}_2}$};
  \path let \p1 = (l1) in let \p2 = (l2s) in
  node (l1s) at (\x1, \y2)
  {\small $\osafe{\sem{M_s}}{\scname{P}\compsymb \rsown}{\scname{Q}\compsymb \rsown}$};
  % {\small $\osafe{L_s}{\scname{P}\compsymb \scname{I}}{\scname{Q}\compsymb \scname{I}}$};
  % \path let \p1 = (l2) in let \p2 = (lt) in
  % node (l1s2) at (\x1, \y2)   {\small $\osafe{L_s}{\scname{R}_1\cccomp \scname{P}\compcc\scname{R}_1\compsymb \scname{I}}{\mydots}$};
  \path [draw, -stealth, line width = 1.2](l1) -- node [above] {\small $\prescript{\memerr}{\rid}{\bsim}$}
  node [below] {\small \thmref{thm:open-partial-safety-preserve}} (l2);
  \path [draw, -stealth, line width = 1.2](l2s) -- node [above] {\small $\prescript{}{\compra}{\bsim}$}
  node [below] {\small \thmref{thm:open-safety-preserve}} (lt);
  \path [draw, -stealth, line width = 1.2](l2) -- node [right] {\small By \thmref{thm:own-check-correct}}
  node [left]
  {\small
    \begin{tabular}{c}
    Ownership\\
    Checking
  \end{tabular}} (l2s);
  \path [draw, -stealth, line width = 1.2](l2s) -- node [above] {\small $\fsim_{\rid}$}
  node [below] {\small \thmref{thm:open-safety-preserve}} (l1s);

\end{tikzpicture}
% \vspace{-0.5cm}
\caption{Proof Structure of End-to-end Open Safety of the Owlang Compiler}
\label{fig:owl-comp-safe-proof}
% \vspace{-0.2cm}
\end{figure}
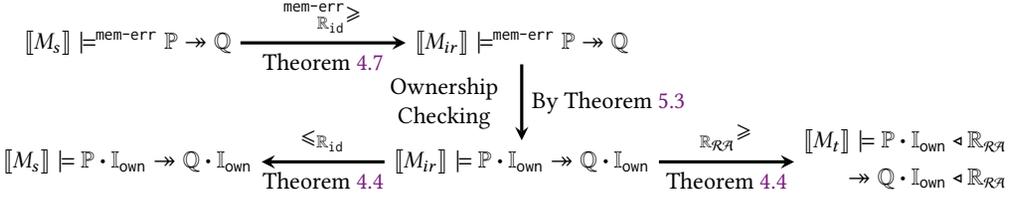

\myparagraph{End-to-end Open Safety}
With ownership checking that ensures memory safety, users only need to
prove partial safety of source Owlang modules. The predicate of error
states in the partial safety is instantiated to \memerr for memory
error states.
We show the theorem of end-to-end safety verification and preservation
of the Owlang compiler as follows:
\begin{theorem}[End-to-end Open Safety of the Owlang Compiler]\label{thm:own-e2e-safe}
  The Owlang compiler $\owlcheckcomp$ verifies memory safety and
  preserves open safety to the assembly,
    \begin{tabbing}
    \quad\=$\forall\app (M_s: \textnormal{Owlang})\app (M_t: \textnormal{Asm})$\=\kill
    \>$\forall\app (M_s: \textnormal{Owlang})\app (M_t: \textnormal{Asm})\app \scname{P}\app \scname{Q}\app,\app 
    \opsafep{\sem{M_s}}{\scname{P}}{\scname{Q}}{\memerr} \imply
       \owlcheckcomp(M_s) = M_t \imply$ \\
    \>\>$\osafe{\sem{M_t}}{\scname{P}\compsymb \rsown\compcc \compra}{\scname{Q}\compsymb \rsown\compcc \compra} \land \osafe{\sem{M_s}}{\scname{P}\compsymb \rsown}{\scname{Q}\compsymb \rsown}.$
    \end{tabbing}
\end{theorem}
The above conclusion consists of two parts: (1) the assembly module
$M_t$ is totally safe and (2) the source Owlang module $M_s$ is also
totally safe, under the safety interface $\rsown$ imposed by the
ownership checking (\defref{def:own-interface}). The proof structure
of this theorem is depicted in~\figref{fig:owl-comp-safe-proof},
following the road map in~\figref{fig:safe-summary}.
Note that $\rid$ can be absorbed by other safety interfaces and is
thus omitted after the safety preservation.
We preserve source-level partial safety to Owl-IR by the extended
backward simulation ($\prescript{\memerr}{\rid}{\bsim}$) of the
Lowering pass. This backward simulation is derived from the original
forward simulation by additionally establishing clause (4)
of~\defref{def:partial-progress-prop}.
The remaining steps follow directly from the established theorems.
% In
% particular, the verification of the ownership checking will be
% discussed in~\secref{ssec:own-checking}.

\subsection{Implementation and Verification of the Ownership Checking}\label{ssec:own-checking}

% Existing approaches to ownership analysis are mostly
% type-based~\cite{wadler1990linear, linear-region, ahmed20073}.
% However, as discussed earlier, verifying total safety solely through a
% type system is impractical for our ownership language, as Owlang
% includes unsafe operations (e.g., division) that are difficult to
% analyze statically.
%
Similar to move checking in the Rust borrow checker~\cite{movecheck},
our ownership checking is implemented as a data-flow analysis based on
the Kildall algorithm from CompCert.
This analysis computes, at each program point, the ownership status of
variables and checks that every use of a variable is justified by
ownership-—that is, only variables that hold full ownership of their
values can be used.

% The verification of this checking pass follows the methodology of
% abstract interpretation.
% %
% We establish a relation between the abstract domain (i.e., the
% ownership status tracked by the analysis) and the concrete domain
% (i.e., the memory state in the Owl-IR semantics). Since the analysis
% enforces that only variables with ownership can be used, this relation
% allows us to infer the validity of the memory resources owned by these
% variables. As a result, all memory accesses through these variables
% are guaranteed to be safe by the ownership checking.

% emphesize that we use data flow analysis instead of type system

\subsubsection{Implementation.}

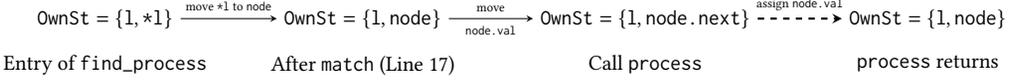
\begin{figure}[t]
\def\bheight{0.6cm}
\def\bstarl{$b_{\code{*l}}$}
\def\bnode{$b_{\code{node}}$}
\newcommand{\defnode}[1]{
  \edef\oenv{oenv#1}
  \edef\bl{bl#1}
  \edef\bltxt{bltxt#1}
  \edef\bsltxt{bsltxt#1}
  \edef\bslcons{bslcons#1}
  \edef\bslk{bslk#1}
  \edef\bslv{bslv#1}
  \edef\bslnext{bslnext#1}
  \edef\bntxt{bntxt#1}
  \edef\bnk{bnk#1}
  \edef\bnv{bnv#1}
  \edef\bnnext{bnnext#1}
  \edef\bv{bv#1}
  \edef\bvtxt{bvtxt#1}
}
\scalebox{0.8}{
\begin{tikzpicture}
  %%%%%%%%%%%%%%%%%%%%%%%%%%%%%%%%%%%%%%%%%%%%%%%%%%%%
  %%%%%%%%%% Sequences of Ownership Status %%%%%%%%%%%
  %%%%%%%%%%%%%%%%%%%%%%%%%%%%%%%%%%%%%%%%%%%%%%%%%%%%

  \node(oenv1) {$\ownst=\{\code{l}, \code{*l}\}$};
  \node[right=1.6cm of oenv1](oenv2) {$\ownst=\{\code{l}, \code{node}\}$};
  \node[right=1.4cm of oenv2](oenv3) {$\ownst=\{\code{l}, \code{node.next}\}$};
  \node[right=1.4cm of oenv3](oenv4) {$\ownst=\{\code{l}, \code{node}\}$};
  \node [below = 0.2 of oenv1] (st1) {Entry of \code{find\_process}};
  \node [below = 0.2 of oenv2] (st2) {After \code{match} (Line 17)};
  \node [below = 0.2 of oenv3] (st3) {Call \code{process}};
  \node [below = 0.2 of oenv4] (st4) {\code{process} returns};
  %%%%%%%%%%%%%%%%%%%%%%%%%%%%%%%%%%%%%%%%%%%%%
  %%%%%%%%%% State Transition Arrows %%%%%%%%%%
  %%%%%%%%%%%%%%%%%%%%%%%%%%%%%%%%%%%%%%%%%%%%%

  % \draw[-stealth] (st1) -- (st2);
  % \draw[-stealth] (st2) -- (st3);
  % \draw[dashed, -stealth] (st3) -- 
  %     node[sloped,above] {\tiny \text{call \code{process} and}} 
  %     node[sloped,below] {\tiny \text{if \code{process} returns}}
  %     (st4);
  \draw[-stealth] (oenv1) -- node[sloped, above] {\tiny \text{move \code{*l} to \code{node}}}(oenv2);
  \draw[-stealth] (oenv2) -- 
    node[sloped, above] {\tiny \text{move}}
    node[sloped, below] {\tiny \text{\code{node.val}}} (oenv3);
  \draw[-stealth, line width=1, dashed] (oenv3) -- 
    node[sloped, above] {\tiny \text{assign} \code{node.val}} (oenv4);

\end{tikzpicture}
}
% \vspace{-0.2cm}
\caption{The Results of Ownership Checking for \code{find\_process}}
\label{fig:own-impl}
% \vspace{-0.3cm}
\end{figure}

We briefly illustrate the workflow of our ownership checking using the
linked list example in~\figref{sfig:list}. We present the ownership
analysis results for the \code{find\_process} function
in~\figref{fig:own-impl}, focusing on the segment from the function
entry to the invocation of \code{process}, and then to the return from
\code{process}.
% FIXME: if we do not remove the compilation section, we should revise
% this paragraph
In~\figref{fig:own-impl}, the \code{OwnSt} set, as computed by the
analysis, represents the ownership status at each program point.
It tracks the variables that have ownership of its value at that
point. When a variable are moved from, then it would be removed from
\code{OwnSt}. When a variable is assigned (or initialized), it would
be added to \code{OwnSt}.

% For simplicity, \code{OwnSt} is an
% over-approximation of the exact ownership status in our analysis, which does not
% affect our discussion.

At the function entry of \code{find\_process}, the analysis assumes
that the input parameter \code{l} has full ownership of its value
(i.e., the linked list). We distinguish between \code{l} and \code{*l}
because our analysis tracks ownership at the level of memory blocks,
i.e., whether a block owns the value stored in it.
After the \code{match} statement, the ownership of \code{*l} is
transferred to \code{node}. When calling \code{process}, part of
\code{node}'s ownership, namely \code{node.val}, is transferred to the
environment. Once the return value of \code{process} is assigned back
to \code{node.val}, the full ownership of \code{node} is restored.

\subsubsection{Verification.}\label{sssec:own-check-verification}

The goal of verifying the ownership checking is to ensure that any
module that passes the ownership checking has no memory error. As
introduced in~\secref{sssec:safe-verifying}, we formalize the correctness of
verifying pass as the implication from partial safety to open
safety.
First, we should define the predicate for memory error state,
which we denote as $\memerr$. The definition of $\memerr$ utilizes the
permission model in the CompCert memory. Specifically, a state belongs
to $\memerr$ if and only if its \emph{next} memory access operation
would violate the permission required by the accessed memory block.
For example, in the CompCert memory, freeing a memory block first
checks whether the block has the \code{Freeable} permission. If it
does, the block is freed by clearing its permission; otherwise, the
memory operation is undefined.

Following~\defref{def:correct-safe-check}, we prove the correctness of
the ownership checking (denoted by \code{OwnCheck}) as shown below:
\begin{theorem}[Correctness of the Ownership Checking]\label{thm:own-check-correct}
    Given an \textnormal{Owl-IR} module $M$, $\scname{P}$ and $\scname{Q}$, $\texttt{OwnCheck}$ is correct in terms of ensuring memory safety:
    \[ \opsafep{\sem{M}}{\scname{P}}{\scname{Q}}{\memerr} \imply
       \owncheckM{M}\; \imply
       \osafe{\sem{M}}{\scname{P}\compsymb \rsown}{\scname{Q}\compsymb \rsown}.
    \]
\end{theorem}
Here, $\rsown$ is the safety interface imposed by the ownership
checking, which we refer to as \emph{ownership interface}.  The
checked module is guaranteed to adhere to $\rsown$ automatically, but
it also relies on the environment to respect $\rsown$ as well;
otherwise, the memory safety guarantee may no longer hold. 

\myparagraph{High-level Proof Structure}
The idea of the verification is to show that there is a safety
invariant for the modules that pass the ownership checking, and this
invariant has no intersection with $\memerr$.
By the definition of open safety, the verification of the above
theorem demands a safety invariant $\inv$ for the checked module.
With the partial safety premise whose safety invariant is
$\inv_p$, we define $\inv$ such that
$s \in \inv \iff s \in \inv_p \wedge s \in \owninv$.
The \emph{ownership invariant} $\owninv$ relates the produced
\code{OwnSt} (i.e., the abstract domain) and the memory state (i.e.,
the concrete domain).
Inside the proof, the preservation of $\inv$ is proved by the
preservation of $\inv_p$ from the premise of partial safety and the
preservation of $\owninv$.
The make progress requirement is established by combining the make
progress requirement of partial safety with the fact that the
ownership invariant $\owninv$ excludes memory error states, i.e.,
$\owninv \cap \memerr = \emptyset$.

To account for interference from the environment, we define the
\emph{ownership interface} $\rsown$ as an abstraction of $\owninv$ at
module boundaries. We prove that: (1) on the outgoing side, any state
satisfying $\owninv$ guarantees that $\rsown$ holds for outgoing
queries and replies; and (2) on the incoming side, any query or reply
satisfying $\rsown$ ensures that $\owninv$ can be established in the
resulting state.

\begin{figure}[t]
\def\bheight{0.6cm}
\def\bstarl{$b_{\code{*l}}$}
\def\bnode{$b_{\code{node}}$}
\newcommand{\defnode}[1]{
  \edef\oenv{oenv#1}
  \edef\bl{bl#1}
  \edef\bltxt{bltxt#1}
  \edef\bsltxt{bsltxt#1}
  \edef\bslcons{bslcons#1}
  \edef\bslk{bslk#1}
  \edef\bslv{bslv#1}
  \edef\bslnext{bslnext#1}
  \edef\bntxt{bntxt#1}
  \edef\bnk{bnk#1}
  \edef\bnv{bnv#1}
  \edef\bnnext{bnnext#1}
  \edef\bv{bv#1}
  \edef\bvtxt{bvtxt#1}
}
\scalebox{0.8}{
\begin{tikzpicture}
  %%%%%%%%%%%%%%%%%%%%%%%%%%%%%%%%%%%%%%%%%%%%%%%%%%%%
  %%%%%%%%%% Sequences of Ownership Status %%%%%%%%%%%
  %%%%%%%%%%%%%%%%%%%%%%%%%%%%%%%%%%%%%%%%%%%%%%%%%%%%

  \node(oenv1) {$\ownst=\{\code{l}, \code{*l}\}$};
  \node[right=1.6cm of oenv1](oenv2) {$\ownst=\{\code{l}, \code{node}\}$};
  \node[right=1.4cm of oenv2](oenv3) {$\ownst=\{\code{l}, \code{node.next}\}$};
  \node[right=1.4cm of oenv3](oenv4) {$\ownst=\{\code{l}, \code{node}\}$};

  %%%%%%%%%%%%%%%%%%%%%%%%%%%%%%%%%%%
  %%%%%%%%%% Memory Blocks %%%%%%%%%%
  %%%%%%%%%%%%%%%%%%%%%%%%%%%%%%%%%%%

  %%%%%%%% The first %%%%%%%%%%%
  \defnode{1}
  %%% block l
  \hblock{\bheight}{draw, minimum width = 0.4cm,
                    below left = 1.2 and -0.7 of \oenv} (\bl) {\bstarl};
  \node[above = 0 of \bl] (\bltxt) {\small $b_\code{l}$};
  %%% block *l
  \hblock{\bheight}{draw, minimum width = 0.4cm, right = 0.3 of \bl} (\bslcons) {$1$};
  \node[above = 0 of \bslcons] (\bsltxt) {\small $b_\code{\kwd{*l}}$};
  \hblock{\bheight}{draw, minimum width = 0.4cm, right = 0 of \bslcons} (\bslk) {$12$};
  \hblock{\bheight}{draw, minimum width = 0.4cm, right = 0 of \bslk} (\bslv) {$b_v$};
  \hblock{\bheight}{draw, minimum width = 0.4cm, right = 0 of \bslv} (\bslnext) {$b_\code{nx}$};
  %%% block node
  \hblock{\bheight}{draw, minimum width = 0.4cm,
                    below = 0.6 of \bl.south west, anchor=north west, pattern = north east lines} (\bnk) {\phantom{$12$}};
  \hblock{\bheight}{draw, minimum width = 0.4cm, right = 0 of \bnk, pattern = north east lines} (\bnv) {\phantom{$b_v$}};
  \node[above = 0 of \bnk.north west, anchor=south west] (\bntxt) {\small $b_\code{\kwd{node}}$};
  \hblock{\bheight}{draw, minimum width = 0.4cm, right = 0 of \bnv, pattern = north east lines} (\bnnext) {\phantom{$b_\code{nx}$}};
  %%% block v
  \hblock{\bheight}{draw, minimum width = 0.4cm, below = 0.6 of \bslnext} (\bv) {$v$};
  \node[above = 0 of \bv] (\bvtxt) {\small $b_v$};
  %%% pointers
  \draw[-stealth,red] (\bl) -- ++(0.5, 0) |- (\bslcons.north west);
  \draw[-stealth,red] (\bslv) |- (\bv.north west);
  %%% state (enter find_process)
  \path let \p1 = (\oenv.west) in let \p2 = (\bl) in
      node [align=left, anchor=west] (st1) at ($(\x1,\y2) + (0, -2.2)$) {Entry State of \\ \code{find\_process}$(b_{\kwd{*l}}, 12)$};

  %%%%%%%% The second %%%%%%%%%%%
  \defnode{2}
  %%% block l
  \hblock{\bheight}{draw, minimum width = 0.4cm,
                    below left = 1.2 and -0.7 of \oenv} (\bl) {\bstarl};
  \node[above = 0 of \bl] (\bltxt) {\small $b_\code{l}$};
  %%% block *l
  \hblock{\bheight}{draw, minimum width = 0.4cm, right = 0.3 of \bl, pattern = north east lines} (\bslcons) {$1$};
  \node[above = 0 of \bslcons] (\bsltxt) {\small $b_\code{\kwd{*l}}$};
  \hblock{\bheight}{draw, minimum width = 0.4cm, right = 0 of \bslcons, pattern = north east lines} (\bslk) {$12$};
  \hblock{\bheight}{draw, minimum width = 0.4cm, right = 0 of \bslk, pattern = north east lines} (\bslv) {$b_v$};
  \hblock{\bheight}{draw, minimum width = 0.4cm, right = 0 of \bslv, pattern = north east lines} (\bslnext) {$b_\code{nx}$};
  %%% block node
  \hblock{\bheight}{draw, minimum width = 0.4cm,
                    below = 0.6 of \bl.south west, anchor=north west} (\bnk) {$12$};
  \hblock{\bheight}{draw, minimum width = 0.4cm, right = 0 of \bnk} (\bnv) {$b_v$};
  \node[above = 0 of \bnk.north west, anchor=south west] (\bntxt) {\small $b_\code{\kwd{node}}$};
  \hblock{\bheight}{draw, minimum width = 0.4cm, right = 0 of \bnv} (\bnnext) {$b_\code{nx}$};
  %%% block v
  \hblock{\bheight}{draw, minimum width = 0.4cm, below = 0.6 of \bslnext} (\bv) {$v$};
  \node[above = 0 of \bv] (\bvtxt) {\small $b_v$};
  %%% pointers
  \draw[-stealth,red] (\bl) -- ++(0.5, 0) |- (\bslcons.north west);
  \draw[-stealth,red] (\bnv) -- ++(0, 0.5) -| (\bv.north west);
  %%% state
  \path let \p1 = (\oenv.west) in let \p2 = (\bl) in
  node [anchor=west](st2) at ($(\x1,\y2) + (0, -2.2)$) {State Before Line 17};

  %%%%%%%% The third %%%%%%%%%%%
  \defnode{3}
  %%% block l
  \hblock{\bheight}{draw, minimum width = 0.4cm,
                    below left = 1.2 and -0.7 of \oenv} (\bl) {\bstarl};
  \node[above = 0 of \bl] (\bltxt) {\small $b_\code{l}$};
  %%% block *l
  \hblock{\bheight}{draw, minimum width = 0.4cm, right = 0.3 of \bl, pattern = north east lines} (\bslcons) {$1$};
  \node[above = 0 of \bslcons] (\bsltxt) {\small $b_\code{\kwd{*l}}$};
  \hblock{\bheight}{draw, minimum width = 0.4cm, right = 0 of \bslcons, pattern = north east lines} (\bslk) {$12$};
  \hblock{\bheight}{draw, minimum width = 0.4cm, right = 0 of \bslk, pattern = north east lines} (\bslv) {$b_v$};
  \hblock{\bheight}{draw, minimum width = 0.4cm, right = 0 of \bslv, pattern = north east lines} (\bslnext) {$b_\code{nx}$};
  %%% block node
  \hblock{\bheight}{draw, minimum width = 0.4cm,
                    below = 0.6 of \bl.south west, anchor=north west} (\bnk) {$12$};
  \hblock{\bheight}{draw, minimum width = 0.4cm, right = 0 of \bnk, pattern = north east lines} (\bnv) {$b_v$};
  \node[above = 0 of \bnk.north west, anchor=south west] (\bntxt) {\small $b_\code{\kwd{node}}$};
  \hblock{\bheight}{draw, minimum width = 0.4cm, right = 0 of \bnv} (\bnnext) {$b_\code{nx}$};
  %%% block v
%   \hblock{\bheight}{draw, minimum width = 0.4cm, below right = 0.6  and 0.4 of \bslnext.south east, anchor=north west} (\bv) {$v$};
  \hblock{\bheight}{draw, minimum width = 0.4cm, below = 0.6 of \bslnext} (\bv) {$v$};
  \node[left = -0.1 of \bv.north west, anchor=east] (\bvtxt) {\small $b_v$};
  %%% pointers
  \draw[-stealth,red] (\bl) -- ++(0.5, 0) |- (\bslcons.north west);
  %%% separation line
  \draw[-,dashed] ($(\bslnext.north east) + (0.1, 0.1)$) -- ++(0, -1) -- ++(-1.1, 0) -- ++(0, -1);
  %%% state
  \path let \p1 = (\oenv.west) in let \p2 = (\bl) in
  node [anchor=west] (st3) at ($(\x1,\y2) + (0, -2.2)$) {Call \code{process} State};

  %%%%%%%% The forth %%%%%%%%%%%
  \defnode{4}
  %%% block l
  \hblock{\bheight}{draw, minimum width = 0.4cm,
                    below left = 1.2 and -0.7 of \oenv} (\bl) {\bstarl};
  \node[above = 0 of \bl] (\bltxt) {\small $b_\code{l}$};
  %%% block *l
  \hblock{\bheight}{draw, minimum width = 0.4cm, right = 0.3 of \bl, pattern = north east lines} (\bslcons) {$1$};
  \node[above = 0 of \bslcons] (\bsltxt) {\small $b_\code{\kwd{*l}}$};
  \hblock{\bheight}{draw, minimum width = 0.4cm, right = 0 of \bslcons, pattern = north east lines} (\bslk) {$12$};
  \hblock{\bheight}{draw, minimum width = 0.4cm, right = 0 of \bslk, pattern = north east lines} (\bslv) {$b_v$};
  \hblock{\bheight}{draw, minimum width = 0.4cm, right = 0 of \bslv, pattern = north east lines} (\bslnext) {$b_\code{nx}$};
  %%% block node
  \hblock{\bheight}{draw, minimum width = 0.4cm,
                    below = 0.6 of \bl.south west, anchor=north west} (\bnk) {$12$};
  \hblock{\bheight}{draw, minimum width = 0.4cm, right = 0 of \bnk} (\bnv) {$b_{v_1}$};
  \node[above = 0 of \bnk.north west, anchor=south west] (\bntxt) {\small $b_\code{\kwd{node}}$};
  \hblock{\bheight}{draw, minimum width = 0.4cm, right = 0 of \bnv} (\bnnext) {$b_\code{nx}$};
  %%% block v
  \hblock{\bheight}{draw, minimum width = 0.4cm, below = 0.6 of \bslnext} (\bv) {$v_1$};
  \node[above = 0 of \bv] (\bvtxt) {\small $b_{v_1}$};
  %%% pointers
  \draw[-stealth,red] (\bl) -- ++(0.5, 0) |- (\bslcons.north west);
  \draw[-stealth,red] (\bnv) -- ++(0, 0.5) -| (\bv.north west);
  %%% state
  \path let \p1 = (\oenv) in let \p2 = (\bl) in
  node (st4) at ($(\x1,\y2) + (0, -2.2)$) {State After Line 17};

  %%%%%%%%%%%%%%%%%%%%%%%%%%%%%%%%%%%%%%%%%%%%%
  %%%%%%%%%% State Transition Arrows %%%%%%%%%%
  %%%%%%%%%%%%%%%%%%%%%%%%%%%%%%%%%%%%%%%%%%%%%

  \draw[-stealth] (st1) -- (st2);
  \draw[-stealth] (st2) -- (st3);
  \draw[dashed, -stealth, line width=1] (st3) -- 
      node[sloped,above] {\tiny \text{call \code{process}$(b_v)$ and}} 
      node[sloped, below] {\tiny \text{if \code{process} returns $b_{v_1}$}}
      (st4);
  \draw[-stealth] (oenv1) -- node[sloped, above] {\tiny \text{move \code{*l} to \code{node}}}(oenv2);
  \draw[-stealth] (oenv2) -- 
    node[sloped, above] {\tiny \text{move}}
    node[sloped, below] {\tiny \text{\code{node.val}}} (oenv3);
  \draw[-stealth, line width=1, dashed] (oenv3) -- 
    node[sloped, above] {\tiny \text{assign} \code{node.val}} (oenv4);

  %%%%%%%%%%%%%%%%%%%%%%%%%%%%%%%%%%%%%%%%%%%%%
  %%%%%%%%%% Invariant Symbols %%%%%%%%%%%%%%%%
  %%%%%%%%%%%%%%%%%%%%%%%%%%%%%%%%%%%%%%%%%%%%%
  \foreach \i in {1,...,4}{
    \draw [double, latex-latex] (oenv\i) -- 
        node[midway, left] {\small $I_\kwd{own}$} +(0, -1);
  };  

  %%%% Rely-Guarantee Symbol %%%%%
  \path let \p1 = (oenv3.east) in let \p2 = (st4.west) in
  node[single arrow,draw] (iown) at ({(\x1+\x2)*.5 - 0.4cm},{(\y1+\y2)*.5}) {$\rsown$};

\end{tikzpicture}
}
% \vspace{-0.2cm}
\caption{Relation Between Ownership Status (Above) and Memory States (Below)}
\label{fig:own-proof}
% \vspace{-0.3cm}
\end{figure}
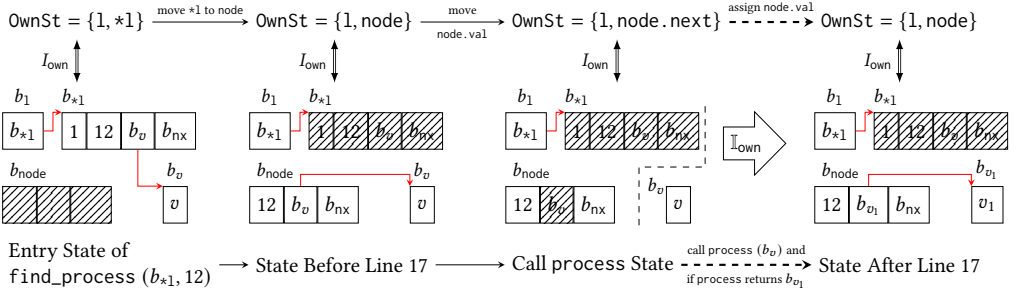

\myparagraph{Ownership Invariant and Ownership Interface}
We illustrate the definition of $\owninv$ and $\rsown$
using~\figref{fig:own-proof}, where the upper part is the transition
of ownership status (the same as~\figref{fig:own-impl}), and the lower
part is the transition of memory states.
The memory is represented as blocks, each of which is denoted as
$b_{\code{x}}$ for the location of some variable \code{x} (except for
$b_{v_1}$ in ~\figref{fig:own-proof}).
A pointer value $(b,o)$ contains a pair of memory block and offset
which is often omitted if it is zero.
We use unshaded memory blocks to indicate that the memory locations
own values, and shaded blocks for those do not.

\paragraph{Semantic Interpretation for the Ownership Invariant.}
%
% According to the above explanation, a memory region is ``bright'' when
% the value stored in that memory region is valid, e.g., it is safe to
% free the memory pointed by this value.
%
Since our goal is to prove $\owninv \cap \memerr = \emptyset$ and the
ownership checking has guaranteed that only variables in $\ownst$ can
be used, $\owninv$ should require that memory locations that may be
accessed by the variable in $\owninv$ are valid. In terms of CompCert
memory, we define ``valid'' as: the memory location has freeable
permission (i.e., the highest permission).
As in ownership type systems like Rust, values are treated as
resources (e.g., memory resources). Therefore, the meaning of ``memory
locations that may be accessed'' should correspond to the memory
resources owned by this variable.
We can thus define $\owninv$ as follows:
\begin{definition}[Ownership Invariant (Simplified)]\label{def:own-inv}
  Given \ownst produced by the ownership checking and memory $m$,
  $\owninv\app \ownst\app m$ holds if there is a resource environment
  $\Omega$ such that:
    \[\forall\app p\app v\app \mfp, \app p\in \ownst \imply \text{load}(m,p)=v\imply \Omega(p)=\mfp \imply \wtval{m}{\mfp}{v}\]
\end{definition}
The resource environment $\Omega$ maps variables to their memory
resources $\mfp$ (i.e., memory footprint). In details, $\mfp$ is the
semantic interpretation of the variable types, which is an algebraic
structure that represents the memory footprint that can be touched by
a value of this type.
The conclusion $\wtval{m}{\mfp}{v}$ requires that the value $v$ loaded
from the memory location of $p$ is semantically well-typed. In simple
terms, \code{wt-val} specifies that the memory blocks in $\mfp$ are
reachable from $v$ and are also freeable (i.e., safe to access).
For example, consider the variable \code{node} at the second state of
\figref{fig:own-proof}.
Since \code{node} is in \ownst and its value is
$[12;b_v;b_{\code{nx}}]$, $\owninv$ ensures that the footprint of
\code{node}, i.e., the blocks $b_v$ and $b_{\code{nx}}$ including its
footprint (i.e., the rest of the list) are freeable. Therefore, the
value $b_v$ passed to the \code{process} is guaranteed not to be a
dangling pointer.

\paragraph{Rely-guarantee Conditions for the Ownership Interface.}
% Now consider the definition of $\rsown$.
At the boundary of Owlang modules, passing values to the environment
can be viewed as transferring their memory footprint. To ensure that
$\owninv$ still holds after the environment returns, we rely that the
environment does not modify the memory outside the transferred
footprint, i.e., memory footprint owned by the module.
For instance, if the environment (i.e., \code{process}) frees the
block $b_{\code{nx}}$, then $\owninv$ cannot be reestablished after
\code{process} returns as \code{node} is in \ownst.
% This rely-guarantee condition represents a form of ownership
% transfer at module-level, which is the core of $\rsown$.
Specifically, we define a Kripke relation called \kown to protect
memory that is not transferred to the environment and then use it to
define $\rsown$:
\begin{definition}[Kripke Relation with Protection on Ownership Memory]
    $\kown = \klr{W_{\kown}}{\accsymb_{\kown}}$ where $W_{\kown} = ((\klist\app \kblock)\times \kmem)$ and $\accsymb_{\kown}$ is the accessibility relation such that
    \begin{tabbing}
    \quad\=$(\mfp, m) \accsymb_{\kown} (\mfp', m') \; \iff \;$\=\kill
    \> $(\mfp, m) \accsymb_{\kown} (\mfp', m') \iff (\forall\app b\app \delta,\app b\notin \mfp \imply (b,\delta) \in\unchangedon{m}{m'})$\\
    \>\> $\land\; (\forall\app b,\app b \notin\mfp \imply b\in \mfp' \imply \neg \kvalidblock(m, b))$
    \end{tabbing}
\end{definition}
\begin{definition}[Ownership Interface (Simplified)]\label{def:own-interface}
    $\rsown = \simconv{W_{\kown}}{\rsown^q}{\rsown^r}$ where
    \begin{tabbing}
      \quad\=$\cquery{v_f}{\sig}{\vec{v}}{m} \in \rsown^q(\mfp, m)$\;\=$\iff$\;\=$\exists\app$\=\kill
      \> $\cquery{v_f}{\sig}{\vec{v}}{m} \in \rsown^q(\mfp, m)$ \>$\iff$ \> $\kwd{norepeat}(\mfp) \land \overrightarrow{\code{wt-val}}\app m\app \mfp\app \vec{v}$\\
      \> $\creply{v}{m'} \in \rsown^r(\mfp, m)$ \>$\iff$ \>  $\exists\app \mfp'\app \mathit{rfp},\app (\mfp,m)\accsymb_{\kown}(\mfp', m') \land \mathit{rfp}\subseteq \mfp' \land\ \wtval{m'}{\mathit{rfp}}{v}$
      % \>\>\>\> $ \land\; \wtval{m'}{\mathit{rfp}}{v}$
    \end{tabbing}
\end{definition}

\noindent Here, the Kripke world of \kown is a pair consisting of a
list of memory blocks representing the memory footprint and the memory
itself. The protection is enforced through the accessibility relation,
where the \kunchangedon predicate ensures that memory outside the
footprint remains unchanged. Additionally, any fresh footprint (i.e.,
memory blocks in $\mfp'$ but not in $\mfp$) is considered invalid in
$m$ as it is allocated by the environment. 
For the ownership interface, its domain is the Owlang interface, whose
queries and replies follow the same format as the C interface, except
that the signatures ($\sig$) use Owlang types instead of C types.
$\rsown^q$ requires that the footprint $\mfp$ contains no duplicates,
and that each argument of $\vec{v}$ satisfies $\code{wt-val}$ with its
footprint in $\mfp$. $\rsown^r$ protects the memory outside $\mfp$
with $\accsymb_{\kown}$ and ensures the return value satisfies
$\code{wt-val}$.

To illustrate the definition of $\rsown$, we use the transfer of
memory footprint during the call to \code{process}
in~\figref{fig:own-proof} as an example.
For the outgoing query, the Kripke world of $\rsown$ is set to
$([b_v], m)$ where $b_v$ is the block pointed to by $\code{node.val}$
and $m$ is the memory state before the call. Since $\owninv$ ensures
that $\wtval{m}{[b_v]}{(b_v, 0)}$ holds (as $\code{node}\in \ownst$
before the call), the query condition $\rsown^q$ is satisfied.
Upon receiving the reply $\creply{(b_{v_1}, 0)}{m}$, the reply
condition $\rsown^r$ guarantees that the memory outside $\mfp$ (i.e.,
$[b_v]$), specifically, the region to the left of the dashed line
in~\figref{fig:own-proof}, remains unchanged. This ensures that
$\owninv$ can be reestablished after returning from the environment.

% !TEX root = main.tex

\section{End-to-end Compositional Verification of Safety for Heterogeneous
Modules}\label{sec:app}

\begin{figure}
\begin{tikzpicture}
    \node [] (llrs) at (0,0) 
        {$\opsafep{\semR{\code{list.rs}}}{\top}{\scname{P}}{\memerr}$ \exmprooflb{(a.1)}};
    \node [below=1.0cm of llrs.south west, anchor=west] (llir0) 
        {$\opsafep{\semR{\code{list.ir}}}
            {\top}
            {\scname{P}}{\memerr}$};
    \node [below=1.0cm of llir0.south west, anchor=west, align=left] (llir) 
        {$\osafe{\semR{\code{list.ir}}}
            {\top\compsymb\rsown}
            {\scname{P}\compsymb\rsown}$};
    \node [below=1.0cm of llir.south west, anchor=west] (lls) 
        {$\osafe{\semA{\code{list.s}}}
            {\top\compsymb\rsown\compcc\compra}
            {\scname{P}\compsymb\rsown\compcc\compra}$};

    %%%% Phantom nodes
    \node [anchor=west] (llrs_pha) at (llrs.west) {\phantom{$\semR{\code{list.rs}}$}};
    \node [anchor=west] (llir0_pha) at (llir0.west) {\phantom{$\semR{\code{list.ir}}$}};
    \node [anchor=west] (llir_pha) at (llir.west) {\phantom{$\semR{\code{list.ir}}$}};
    \node [anchor=west] (lls_pha) at (lls.west) {\phantom{$\semA{\code{list.s}}$}};
    %%%% Owlang compilation
    \draw[-stealth] (llrs_pha) -- (llir0_pha) 
        node[midway, right] {{\small Owlang Compiler} \exmprooflb{(b.1)}} 
        node[midway, above, rotate=90] {$\fsim_{\rid}$};
    \draw[double, -latex] (llir0_pha) -- (llir_pha) 
        node[midway, right] {{\small Ownership Checking} \exmprooflb{(c)}};
    \draw[-stealth] (llir_pha) -- (llir_pha |- lls_pha.north) 
        node[midway, right] {{\small Owlang Compiler} \exmprooflb{(b.2)}} 
        node[midway, above, rotate=90] {$\fsim_{\compra}$};

    \node [right=1.7cm of llrs.east, anchor=west] (bukc)
        {$\osafe{\semC{\code{buks.c}}}
        {\scname{P}\compsymb\rsown\compcc\comprc}
        {\top\compsymb\rsown\compcc\comprc}$ \exmprooflb{(a.2)}};
    \path let \p1 = (bukc.west) in let \p2 = (lls) in
        node [anchor=west] (buks2) at (\x1, \y2) 
        {$\osafe{\semA{\code{buks.s}}}
        {\scname{P}\compsymb\rsown\compcc\compra}
        {\top\compsymb\rsown\compcc\compra}$};
    \node [above=1.4cm of buks2.west, anchor=west] (buks1)
        {$\begin{aligned}
            \semA{\code{buks.s}} \models \scname{P}\compsymb\rsown\compcc\comprc\compcc\compca &\\
            \arrli \top\compsymb\rsown\compcc\comprc\compcc\compca &
        \end{aligned}$};
    
    %%% Phantom nodes
    \node [anchor=west] (bukc_pha) at (bukc.west) {\phantom{$\semC{\code{buks.c}}$}};
    \node [anchor=north west] (buks1_pha) at (buks1.north west) {\phantom{$\semA{\code{buks.s}}$}};
    %%%% CompCert compilation
    \draw[-stealth] (bukc_pha) -- (buks1_pha) 
        node[midway, right] {{\small CompCert} \exmprooflb{(b.3)}} 
        node[midway, above, rotate=90] {$\fsim_{\compca}$};

    \path (buks2) -- node[sloped] {$\semlink$} (lls);
    \node [fit = (buks2) (lls), inner sep = -0.0pt] (tgtsafe) {};
    \node [below = 0.5 cm of tgtsafe] (tgtsyn) 
          {$\osafe{\semA{\code{list.s} + \code{buks.s}}}
          {(\scname{P}\uplus \top)\compsymb\rsown\compcc\compra}
          {(\scname{P}\uplus \top)\compsymb\rsown\compcc\compra}$};
    \node [below = 0.3cm of tgtsyn] (tgtclose)
          {$\csafe{\sem{\code{list.s} + \code{buks.s}}_{\code{closed}}}$};
        
    \path (tgtsafe) -- node[sloped,rotate=180] {$\fsim_{\rid}$} 
        node [right] {\app \exmprooflb{(e)}}(tgtsyn);

    \path (tgtsyn) -- node[sloped] {$\imply$} 
        node [right] {\app \exmprooflb{(f)}}(tgtclose);
        
    \path (buks1) -- node[rotate=90] (inveq) {$\equiv$}
        node [right] {\app \exmprooflb{(d)}} (buks2);

\end{tikzpicture}
% \vspace{-0.5cm}
\caption{Proof Structure of the Hash Map Example}
\label{fig:proof-struct}
% \vspace{-0.3cm}
\end{figure}
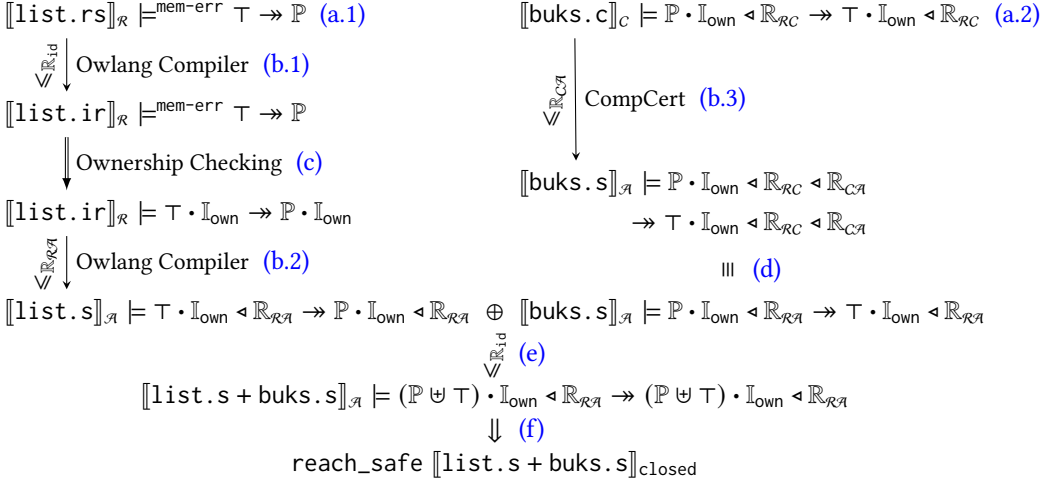

We give a formal account of end-to-end safety verification for
heterogeneous modules based on our approach, using the hash map
example in~\figref{fig:hash_map}.
% TODO: add reference to former high-level proof structure
The proof structure is depicted in~\figref{fig:proof-struct}.
As shown at the bottom, our \textbf{final result} is that the linked
assembly program is safe under the \emph{closed} semantics, which
matches the assembly semantics in the original CompCert.
For open semantics $\sem{\_}_{\_}$, we annotate their corresponding
language interfaces.

% : $\langi{R}$ for Owlang, $\langi{C}$ for C, and
% $\langi{A}$ for assembly.

\myparagraph{Source Level Verification[\refexmplb{(a.1)} to
  \refexmplb{(a.2)}]}
We prove partial safety for \code{list.rs} and total safety for
\code{buks.c}, following~\defref{def:osafe}.
For \refexmplb{(a.1)}, its incoming safety interface $\scname{P}$
mainly specifies the assertions for \code{hash}. Since the outgoing
interface of \code{list.rs} (i.e., the assertions for \code{process})
imposes no specific constraints, we denote it by $\top$ (i.e.,
True). We define $\scname{P}$ as follows:
\begin{definition}[Safety Interface $\scname{P}$]
    $\scname{P} = \simconv
    {\kwd{ident}\times \kwd{symtbl} \times \kwd{int}}
    {\scname{P}^q}{\scname{P}^r}$ where
    \begin{tabbing}
        \quad\=$\scname{P}^q(f, se, r):=$\=\kill
        \> $\scname{P}^q(f, se, r):= \pset{\cquery{v_f}{\sig}{[\kwd{k}; \kwd{range}]}{m}}{ \kwd{find}(se, v_f)= f \land f=\kwd{hash} \land \kwd{range} > 0 \land r = \kwd{range}}$\\
        \>\> $\cup\; \pset{\cquery{v_f}{\sig}{[\kwd{l}; \kwd{k}]}{m}}{ \kwd{find}(se, v_f)= f \land f=\kwd{find\_process}}$\\
        \> $\scname{P}^r(f, se, r):= \pset{\creply{v}{m}}{f=\kwd{hash} \imply 0 \le v < r}.$
    \end{tabbing}
\end{definition}
\noindent Here, the Kripke world of $\scname{P}$ stores the identifier
of the callee, a symbol table used to find the function identifiers
and an auxiliary variable $r$ used to remember the value of
$\kwd{range}$. $\scname{P}^q$ requires that if the callee is
\kwd{hash} then the \kwd{range} must be greater than $0$ and is equal
to $r$. $\scname{P}^r$ requires that if the callee is \kwd{hash} then
the return value $v$ must be less than $r$ (i.e., \kwd{range}).
Inside the safety verification of \code{list.rs}, the requirement of
partial safety simplifies the memory reasoning, as one can assume that
all the memory operations succeed. Otherwise, the partial
safety is satisfied directly.

For \refexmplb{(a.2)}, we must ensure two additional conditions: one
for multi-language interoperation, and the other for the interaction
between safe and unsafe languages.
First, the invocations of \code{buks.c} must conform to the conventions
prescribed by the Owlang interface, which is ensured by $\comprc$. For
example, when a C module calls an Owlang function, we must ensure that
the passed values conform to the expected Owlang types, such as
maintaining compatible struct layouts and field orderings between the
two languages.
Second, at the module boundaries, the C module must be encapsulated by
the ownership interface $\rsown$, which ensures that the memory safety
of the Owlang module is not violated.
%
% \ifdefined\islong
%   Details of the proof are presented in~\secref{sec:buks-proof}.
% \else
%   Details of the proof are presented in the supplementary materials.
% \fi

% However, the domain of $\scname{P}$ and $\rsown$ is $\langi{R}$
% instead of $\langi{C}$. To tackle this, the C module must
% \emph{simulate} an Owlang module at it's boundary by emitting and
% receiving elements in $\langi{R}$ which satisfy $\scname{P}$ and
% $\rsown$. Note that this simulation only applies to external functions
% that are implemented in Owlang.
% %
% The simulation is encoded by $\comprc$ which relates $\langi{R}$ and
% $\langi{C}$. It primarily specifies how Owlang types are converted to
% C types.
% %
% In this work, although the Owlang and C modules are composed at the
% assembly level, we must ensure that their target code adheres to the
% same calling convention (i.e., $\compra$). This is why we require the
% C module to behave like an Owlang module at the source level, so that
% its compiled assembly code naturally conforms to $\compra$, e.g.,
% by~\refexmplb{(d)}.

\myparagraph{Safety Preservation and Safety Checking[\refexmplb{(b.1)} to
  \refexmplb{(b.3)}, and \refexmplb{(c)}]}
We apply~\thmref{thm:own-e2e-safe} to the source partial safety of
\code{list.rs} to prove its total safety at the assembly.
The generic proof structure of \refexmplb{(b.1)}, \refexmplb{(b.2)},
and \refexmplb{(c)} has been presented
in~\figref{fig:owl-comp-safe-proof}, which is independent of specific
examples.
\refexmplb{(b.3)} is obtained by
applying~\thmref{thm:open-safety-preserve} to the end-to-end open
backward simulation of CompCertO. Importantly, no modification to the
existing proofs in CompCertO is required.

\myparagraph{Refinement of Safety Interfaces[\refexmplb{(d)}]}
To enable the target-level composition, we need to ensure that the
composition of $\comprc$ (i.e., the FFI between Owlang and C) and
$\compca$ (i.e., calling convention of CompCertO) correctly reflects
the calling convention $\compra$ of the Owlang compiler. This is
achieved by the refinement of safety interfaces, which derives open
safety from another by strengthening its preconditions or weakening
its postconditions as in program logics.
\begin{definition}
    Given $\scname{P}$ and $\scname{Q}$, $\scname{P}$ is refined by $\scname{Q}$ (denoted as $\screfine{\scname{P}}{\scname{Q}}$) if:
    \[\forall\app w_{\scname{Q}}\app q,\app q\in \scname{Q}^q(w_{\scname{Q}}) \imply \exists\app w_{\scname{P}},\app q\in \scname{P}^q(w_{\scname{P}}) \land (\forall\app r,\app r\in \scname{P}^r(w_{\scname{P}}) \imply r\in \scname{Q}^r(w_{\scname{Q}}))\]
\end{definition}
\noindent We write $\scname{P} \equiv \scname{Q}$ if
$\screfine{\scname{P}}{\scname{Q}}$ and
$\screfine{\scname{Q}}{\scname{P}}$. We can strengthen (weaken) the
interfaces in the open safety:
\begin{theorem}\label{thm:safe-refine}
    Given $L: A \arrli B$, if\; $\screfine{\scname{P}_A'}{\scname{P}_A}$, $\screfine{\scname{Q}_B}{\scname{Q}_B'}$ and $\osafe{L}{\scname{P}_A}{\scname{Q}_B}$ then $\osafe{L}{\scname{P}_A'}{\scname{Q}_B'}$
\end{theorem}
In our example, the proof of \refexmplb{(d)} is derived from the
application of the above theorem to
$\comprc\compcc\compca \equiv \compra$.
Since $\compra$ and $\compca$ are derived via the \emph{transitivity}
of simulation conventions~\cite{direct-refinement}, they directly
relates $\langi{R}$ (respectively, $\langi{C}$) and $\langi{A}$
without exposing the internal details of the compilation. Thanks to
these precise calling conventions, the rewriting from
$\comprc \compcc \compca$ to $\compra$ does not need to
account for the differences in the compilation pipelines between the
Owlang compiler and CompCertO, such as the compilation passes of the
Owlang compiler frontend.

% we use  to rewrite $\comprc\compcc\compca$ into
% $\compra$ by proving that they are equivalent i.e., they refine each
% other.
%
% Since we develop $\compra$ as a direct refinement that directly
% relates $\langi{R}$ and $\langi{A}$, and $\compca$ is the direct
% refinement adopted from CompCertO~\cite{direct-refinement}, the
% rewriting from $\comprc\compcc\compca$ to $\compra$ is
% straightforward, as it requires no reasoning about the internal
% differences between the Owlang compiler and CompCert.

\myparagraph{Safety Composition[\refexmplb{(e)}]}
This step is proved by first applying the compositioanlity of open
safety (\thmref{thm:safety-compose}) to the two safety judgements, and
then applying the safety preservation
(\thmref{thm:open-safety-preserve}) to the adequacy of assembly-level
linking from CompCertO, i.e.,
$\osims{\rid}{\semA{\code{list.s} +
    \code{buks.s}}}{\semA{\code{list.s}} \semlink
  \semA{\code{buks.s}}}$.
Note that the safety interface
$(\scname{P}\uplus \top)\compsymb\rsown\compcc\compra$
in~\figref{fig:proof-struct} is equivalent to
$(\scname{P}\compsymb\rsown\compcc\compra)\uplus
(\top\compsymb\rsown\compcc\compra)$ that is derive directly from the
safety composition.

\myparagraph{Open Safety to Reachability Safety[\refexmplb{(f)}]}
We extend \code{buks.c} by adding a \code{main} function along with
hash map initialization routine (e.g., allocating the array of
buckets). This ensures that the linked program (i.e., \code{list.s} +
\code{buks.s}) is a whole program.  We then apply the soundness of
open safety (\thmref{thm:adequacy}) to establish that this whole
program is reachable safe.

% !TEX root = main.tex
% \vspace{-0.2cm}
\section{Evaluation and Related Work}\label{sec:eval-related}

% Our Coq development took about 15 person-months and 45381 lines of
% code (LOC) on top of CompCertO. We added 1663 LOC for the framework of
% open safety discussed in~\secref{sec:safety}. Although the framework
% is light weight, we spent lots of effort in finding suitable
% definitions for open safety as explained in~\secref{ssec:challenges}.
% %
% Most of the Coq development comes from the verified Owlang compiler and
% the ownership checking as discussed in~\secref{sec:compiler}, which
% contain 25265 and 11387 LOC, respectively.
% %
% Finally, we added 7066 LOC for the running example as introduced
% in~\secref{ssec:running-example} and~\secref{sec:app}.

Our Rocq development took about 15 person-months and 45.4k lines of
code (LOC) on top of CompCertO~\cite{compcerto, direct-refinement}. We
wrote 1.7k LOC for the open safety framework as introduced
in~\secref{sec:safety}.
Although the framework is lightweight---in the sense that it does not
require modifying the existing compiler correctness proofs---we spent
significant effort in identifying a suitable notion of modular safety
that can be preserved through compilation, as explained
in~\secref{sec:ideas}. In particular, we explored defining
reachability-based safety on the modular semantics and attempted to
prove the preservation of VST's safeN predicate, but both approaches
turned out to be unsuccessful.
Most of the Rocq development is spent on the Owlang compiler and
the ownership checking as discussed in~\secref{sec:compiler}, which
contain 25.3k and 11.4k LOC, respectively.
Finally, we added 7.1k LOC for the safety verification of the hash map
example, most of which comes from the manual proof of the source-level
verification, as we need to prove that the invariant is preserved
under each small-step transition. Automating this process or
integrating state-of-the-art verification tools is left as future
work.

\myparagraph{CompCert-based approaches to safety preservation}
Verified Software Toolchain (VST)~\cite{appel11:vst, appel2014program,
  beringer14} extends the end-to-end simulation of
CompCert~\cite{leroy09, Leroy-backend}, to preserve partial
correctness for whole programs from Clight to the assembly.
The partial correctness in VST, i.e., safeN, requires that the
initial state of the module does not get stuck in any of the first $N$
steps (with each external call counting as one).
Therefore, preserving safeN for open modules requires one to show that
for any execution history with $N$ steps in the target, there is a
history with $M$ steps in the source such that they are
simulated. Then one can apply the forward progress property to prove
that the target can take $N+1$ steps.
However, as discussed in~\secref{sec:ideas}, preserving such kind of
``big-step'' safety % partial correctness
through open simulation is in general
not possible.
Another problem for the preservation of safeN is that the
\emph{stuttering} used in the simulation~\cite{stutter-bsim} would
make the construction of $M$ extremely difficult, as the stutter steps
are hidden (i.e., existentially quantified) in the simulation.

% VST-Coq
A recent extension of VST supports compositional verification of Rocq
(Coq) and C programs~\cite{coq-c-popl25}. They use
CertiCoq~\cite{anand2017certicoq} and CompCert to compile Rocq and C
programs, respectively. The safety of C programs is verified in
VST. The safety of Rocq programs is directly verified in Rocq.
% The safety composition is achieved by verified FFI (VeriFFI).
%
However, the verification results in their framework are composed at
the C level via verified FFI (VeriFFI). As discussed above, VST
remains limited in supporting safety preservation for open modules.

% VCC approach
The line of reasearch on verified compositional compilation
(VCC)~\cite{beringer14, stewart15, compcerto, direct-refinement} has
investigated how to establish the semantics equivalence of open
modules under open-world assumptions. These works adopt small-step
open simulations to ensure transitivity, which simplifies compiler
interfaces. However, they primarily focus on proving the refinement
between source and target modules. Building on this foundation, our
work investigates how to preserve safety of open module and to support
the verification of compositional compilers that perform safety
checking, a problem that remains largely unexplored in existing VCC
frameworks.

% \myparagraph{Type-preserving compilation}

\myparagraph{Safety preservation by logical relations}
Prior work on compiler verification for higher-order
language~\cite{hur09, hur11} interprets types into \emph{binary
  logical relations} to establish semantics equivalence between source
and target modules, which states that source and target modules both
diverge or both halt without failure, i.e., it implies safety
preservation.
However, proving the transitivity of logical relations is notoriously
difficult~\cite{hur12, Ahmed06esop}, which makes them challenging to
be applied to realistic compilers.
Moreover, these approaches often require designing a type system for
each intermediate language, which deviates from our goal of reusing
existing proofs in verified compilers.

% TODO: discuss Pilnser
Pilnser~\cite{neis15icfp} is a verified multi-pass compiler for a
higher-order imperative language. The semantics preservation of
Pilnser is defined as \emph{parametric inter-language simulations},
which combines bisimulation and Kripke logical relation~\cite{hur12}
to obtain vertical transitivity. To our knowledge, the safety
preservation for open module in Pilsner remains unexplored. In the
Pilsner paper, the source language is equipped with a type system to
ensure that well-typed programs do not go wrong. However, in order to
preserve this safety property to the target level, one would need
either to design type systems for all intermediate languages and rely
on type-preserving compilation, or to define a form of modular safety
(i.e., safety that accounts for interference from the environment)
that can be preserved by their simulation techniques.

\myparagraph{Multi-language approaches}
% can be removed
%
The multi-language approach~\cite{patterson-pldi22, real-abi} 
grows out of proof-carrying-code~\cite{necula:pcc,necula98} for directly verifying safety of
target modules by using source-level type systems. Specifically, they
interpret source types as terms written in the target language. As a
result, type composition at the source level corresponds to term
composition at the target level, enabling sound multi-language
interoperation. In their approach, the well-typedness of a source term
implies the safety of its compiled result.
Although directly proving safety for the target programs is appealing,
their approach incorporates the entire compiler into the semantics
interpretation. That is, proving the soundness theorem must reason
about the entire compilation process.
It remains to be investigated that if this approach is practical for
supporting multi-pass optimizing verified compilation and
compositional verification involving verifying compilation.

DimSum~\cite{sammler2023dimsum} encodes calling conventions into
modular semantics via \emph{semantic wrappers} to support multi-language
interoperation.
We encode calling conventions into the open safety via $\compcc$
operator (\defref{def:compcc}), where the $\exists$ quantifier in the
pre- and post-conditions (i.e., assume and assert) plays similar roles
as the dual non-determinism~\cite{back2012refinement} used in the
semantic wrappers.
Despite this similarity, it is unclear how to define safety in their
semantic model or prove the preservation of safety by their
refinements, which is a direction worth exploring.

\myparagraph{Behavior refinement}
In behavior refinement, the target module refines the source module if
all behaviors (i.e., event traces) of the target are included in those
of the source. As a result, any safety property satisfied by the
source module will also be satisfied by the target. This allows us to
establish the safety of the compiled code by verifying the safety of
the source module.
Conditional Contextual Refinement~\cite{ccr23} combines
contextual refinement and separation logic. They transform the
refinement that relies on the pre- and post-conditions into a
standard contextual refinement predicate by encoding these conditions
into the module semantics using dual
non-determinism~\cite{back2012refinement}.
Although their approach shows promise in encoding a compiler’s calling
convention (which can be viewed as a form of pre- and post-conditions)
into contextual refinement, it is unclear how to encode such
calling conventions into their underlying separation logic framework.

% , e.g.,
% $\htriple{P}{L_t \sqsubseteq L_s}{Q}$, into a standard contextual refinement predicate $L_t \sqsubseteq \langle L_s \sqsubseteq L_s}{Q}$

% its transitivity

\myparagraph{Verification of other safety properties}
The way we define memory errors is similar to the existing work of
identifying the violation of safety properties, such as thread safety
(i.e., absence of data races)~\cite{cascompcert, RustBelt} and spatial
memory safety~\cite{softbound, checkedC}.
Therefore, we believe that our partial safety approach should work for
ensuring other safety properties. For example, since data races are
undefined behaviors triggered by memory writes and reads, we can extend
the error state predicate to describe such undefined behaviors as
another kind of memory error.

\myparagraph{Linear languages with memory management}
The $\text{L}^3$~\cite{ahmed20073} language is a linear language
equipped with ownership (or capability) types and manual memory
management via explicit drop operations. Our Owlang is an ownership
subset of Rust, which is also a linear language. The type system of
$\text{L}^3$ guarantees that each drop operation is executed if and
only once. Recent work~\cite{wagner2025linearity} further shows that
such linear type systems can ensure free of memory leaks.
In contrast, the memory management in Owlang follows that in the Rust
language, i.e., combining the ownership semantics and RAII (Resource
Acquisition Is Initialization) to insert drop operations to release
memory resource. In this aspect, our main contribution is the formal
verification of the drop elaboration pass in a realistic verified
compiler. One future direction is to verify that the insertion of drop
operations can also guarantee free of memory leaks.

With respect to semantic type soundness for linear languages, the
semantic interpretation of ownership types in $\text{L}^3$ divides the
heap memory into two disjoint parts: one owned by the type and one
representing the rest, much like the points-to assertion in separation
logic.  Similarly, in Owlang, we employ an internally disjointed tree
structure (i.e., some kind of resource algebra) called the memory
footprint (introduced in~\secref{sssec:own-check-verification}) to
interpret ownership types semantically and to prove the correctness of
ownership checking.
Recent work~\cite{patterson-pldi22} investigates safe interoperation
between $\text{L}^3$ and a garbage-collected (GC) language. Their
approach defines the conversion between types in the two languages as
glue code written in the target language, capturing both static and
dynamic semantics. Specifically, defining glue code as a no-op models
static semantics (e.g., conversions between integer types), while
defining it as some code representing the translation from GC-managed
pointers to ownership pointers can model the dynamic semantics of the
conversions between pointer types in two languages.
In Owlang, we define the safety requirement for the language
interoperation as a rely-guarantee condition called $\rsown$ at the
source-level semantics, which concerns only the static semantics. This
kind of rely-guarantee condition can be composed with the calling
conventions derived from the compiler, thereby enabling the
representation of safety assertions for the assembly modules, e.g.,
$\scname{P}\compsymb\rsown\compcc\compra$ for \code{list.s} in
~\figref{fig:proof-struct}.
To support more flexible interoperation—such as between Owlang and a
GC'd language—it may be necessary to define explicit glue code that
captures the dynamic semantics of interoperation, and to prove that
this code satisfies the safety interface at the boundary of the Owlang
side.

\myparagraph{Formalization of Drop Elaboration}
One key feature of Rust is its automatic resource management through
ownership semantics and RAII.  To implement this feature, the Rust
compiler first inserts drop statements for out-of-scope variables
whose types may own resources, and then performs drop elaboration to
adjust these statements so that they are executed only when the
corresponding variables actually own resources at runtime.
A key challenge in verifying drop elaboration lies in defining the
operational semantics for the language before drop
elaboration~\cite{opsem-before-elab}. As discussed in this GitHub
issue, there are two main approaches: (1) augment the semantics with a
dynamic ownership environment (e.g., runtime “magic tables” storing
drop flags); or (2) let drop elaboration itself define the semantics,
i.e., only provide semantics after drop elaboration. While the latter
approach is favored in the Rust community—for example,
Miri~\cite{jung2025miri} defines MIR semantics after drop
elaboration—it is unsuitable for verifying the Owlang (or even Rust)
compiler.
For compiler verification, we must define semantics for the surface
language, where drop operations are not explicitly written but should
still be executed for variables that are reassigned or go out of
scope.  Moreover, both move checking and borrow checking in the Rust
compiler are performed \emph{before} drop elaboration, which means
their correctness must be established with respect to the
surface language semantics rather than the post-elaboration semantics.

To our knowledge, the only formalization of drop elaboration so far
is~\cite{fukala2024formalization}.  Unlike their work, which focuses
on features only relevant to drop elaboration, we aim to verify the
compilation of an ownership-based subset of Rust. This goal introduces
several technical challenges, including proving the soundness of the
initialization analysis used during drop elaboration and verifying the
correctness of inserting stack-allocated drop flags (a particularly
difficult task in the CompCert framework, since it alters the memory
structure).
The formalization in~\cite{fukala2024formalization} introduces a
tree-structure state for initialization analysis to minimize the
number of generated drop flags and thus reduce runtime overhead.  We
also plan to adopt this idea within the Kildall framework to define a
similar structure for optimizing the generated code.

\section{Conclusion and Future Work}\label{sec:conclusion} 

We have presented open safety, a composable definition of safety for
heterogeneous modules, and shown that it can be checked by verifying
compilation and preserved end-to-end by verified compilation. The
effectiveness of our approach is demonstrated on the verified and
verifying compilation of an ownership language, and a hash map example
composed of safe and unsafe code.
There are two directions for the future work. One is to support borrow
checking in our Owlang compiler, from which we can obtain a verified
compiler for a sequential subset of Rust. The other is to connect the
state-of-the-art verification tools (e.g., separation logics and model
checkings) with our open safety to further extend the verification
chain.

\paragraph{Extension of Owlang to a subset of Rust}
We are currently extending Owlang with reference types and verifying a
version of borrow checking based on Polonius~\cite{rust-polonius}.
From this verification, we aim to derive a safety interface for borrow
checking, analogous to $\rsown$ from the ownership checking. This
interface can be viewed as a formalization of safe encapsulation—a
rely-guarantee condition that characterizes the interaction between
safe Rust and the unknown environment.

\paragraph{Connecting open safety with verification tools}
To reduce the verification effort required under open safety, we can
leverage existing verification frameworks such as
Iris~\cite{iris-ground} and VST~\cite{appel11:vst, appel2014program}.
The main challenge lies in bridging the gap between these tools and
our notion of open safety. Taking Iris as an example, the semantic
interpretation of Hoare triples in Iris is defined via the weakest
precondition (wp). We observe that, by definition, wp must be
preserved throughout execution, and states satisfying wp are
guaranteed to make progress~\cite{iris-ground, melocoton,
  iris-compcert}. This naturally aligns with the definition of open
safety. One potential problem may be the handling of step-indexing,
which requires further investigation.

\bibliography{refs}

\ifdefined\islong
\appendix

%%%%%%%%%%%%%%%%%%%%%%%%%% Technical Report Contents %%%%%%%%%%%%%%%%%%%%%%%%%%%%%

\section{Verified Compilation of the Ownership Language}\label{sec:compiler-frontend}

\subsection{The Ownership Language}\label{ssec:owlang-syntax}

% !TEX root = main.tex

% \section{Syntax of Owl}\label{sec:owl}

% We present the syntax of the Owl language in~\figref{fig:owl-syntax}.

\begin{figure}[t]
\begin{align*}
& \text{Identifiers} & id, l \in \app & \text{String}\\
& \text{Places} & p ::=\app & id \stk \deref{p} \stk p.l \stk \dcast{p}{l}\\
& \text{Based Types} & \tau^B ::= \app & \texttt{unit} \stk \texttt{bool} \stk \texttt{i32} \stk \texttt{f32} \stk \dots \\
& \text{Field Lists} & \varphi ::= \app & [\tfields] \\
& \text{Types} & \tau ::=\app & \tau^B \stk \tbox{\tau} \stk \app \tstruct{id}{\varphi} \stk \tenum{id}{\varphi}\\
& & | & \app \tfun\\
& \text{Constants} & c ::= \app & () \stk \texttt{true} \stk \texttt{false} \stk n \stk \dots\\
& \text{Pure Expressions} & e^P ::= \app & c \stk p \stk \texttt{uop}\app e \stk e_1 \app\texttt{bop}\app e_2 \stk \cktag{p}{l}\\
& \text{Move Expressions} & e^M ::= \app & \move{p} \\
& \text{Expressions} & e ::= \app & e^P \stk e^M\\
& \text{Statements} & s ::= \app & \texttt{skip} \stk \sassign{p}{e} \stk \sassignvar{p}{e} \stk \sassign{p}{\texttt{Box}(e)} \stk \sassign{p}{e_f(e_1,...,e_n)}\\
& & | & \app \slet \stk s_1;s_2 \stk \sif \\
& & | & \app \texttt{loop}\app s \stk \texttt{break} \stk \texttt{continue} \stk \sreturn\app p\\
& \text{Variable Declarations} & dcl ::= \app & \vardecls \\
& \text{Internal Functions} & \mathit{fd} ::= \app & \texttt{fn}\app id(dcl_1)\rightarrow \tau \{dcl_2; s\}\quad (dcl_1\!: \text{parameters},\app dcl_2 \!: \text{local variables})\\
& \text{External Functions} & \mathit{fe} ::= \app & \texttt{extern}\app \texttt{fn}\app id(dcl_1)\rightarrow\tau\\
& \text{Functions} & f ::= \app & \mathit{fd} \stk \mathit{fe}\\
& \text{Open Modules} & Mods ::= \app & \{(id_1, f_1),...,(id_n, f_n)\}
\end{align*}
\caption{Syntax of Owlang}
\label{fig:owl-syntax}
\end{figure}

We present the syntax of Owlang in~\figref{fig:owl-syntax}. From
bottom to top of~\figref{fig:owl-syntax}, an Owlang module
($\mathit{Mods}$ in~\figref{fig:owl-syntax}) is composed of a list of
functions that are either internal functions or external
functions. Note that we do not support global variable which is
considered unsafe. In fact, one can use global variables in unsafe
languages (e.g., C) and then compose the unsafe modules with Owlang
modules.
The syntax of internal and external functions is similar to the syntax
of function in Rust. Inside the body of internal functions, we have
assignment statement ($p:=e$), function call statement
($\sassign{p}{e_f(e_1,...,e_n)}$) and other control flow statements,
which are standard in imperative languages.

Owlang supports creating an enum object and then assigning the enum
object to a place $p$ by the statement $\sassignvar{p}{e}$. One
example is the assignment at line 22 of~\figref{sfig:list} (i.e.,
\code{*l = List::Cons(node)}), which creates a \code{List} object
using \code{List::Cons} as the variant and \code{node} as the value of
this variant, and then assigns it to \code{*l}.
To support heap allocation, Owlang has $\sassign{p}{\texttt{Box}(e)}$
statement, whose execution would allocate heap memory to store the
value evaluated from the expression $e$. The assignee $p$ must have
\code{Box} type, which can also be seen as the ownership type in
Owlang. Note that we encode the heap allocation in the syntax of
Owlang instead of calling it from the standard library (i.e., calling
the external function \code{Box::new(\_)} in Rust syntax). This design
choice can simplify the reasoning of the language.

In some imperative languages, such as Clight in CompCert, the l-value
expressions and r-value expressions are not distinguished explicitly
in the language syntax. Instead, we explicitly distinguish them in the
Owlang syntax, where we use Places represent l-value expressions and
Expressions for r-value expressions. The definition of Places is
similar to that in MIR in the rust compiler, which contains local
variable, deference, field access, and downcast ($\dcast{p}{l}$). If a
place is of type enum and represent variant $l$, then it can be
downcast to the body of this variant using $\dcast{p}{l}$. For
example, at line 15 of~\figref{sfig:list}, the match arm
\code{List::Cons(node)} would be desugared (in the phase of parsing of
our Owlang compiler) to \code{node := \move{(*l as Cons)}}, meaning
that the content of the variant \code{Cons} of \code{*l} is moved to
\code{node}.

We classify the expressions into \emph{pure} expressions ($e^P$) and
\emph{move} expressions ($e^M$). Pure expressions are used to express
computation that does not modify the evaluation environment, such as
arithmetic computation by unary operation (\code{uop}) or binary
operation (\code{bop}). The expression $\cktag{p}{l}$ computes that if
the place $p$ (with some enum type) has variant $l$. For example, at
line 13-15 of~\figref{sfig:list}, the match statement would be
desugared into a sequence of if-else statements where each match
branch would be translated to a $\cktag{p}{l}$, such as
\code{List::Nil} to \code{\cktag{*l}{Nil}} and \code{List::Cons} to
\code{\cktag{*l}{Cons}}. 
The move expressions have only one constructor $\move{p}$. One typical
usage of move expression is to transfer the ownership of a place $p_1$
to the place $p_2$ by assignment $\sassign{p_1}{\move{p_2}}$, where
both $p_1$ and $p_2$ have type $\code{Box}(\tau)$ for some $\tau$. The
reason why we distinguish move expressions from pure expressions is
that moving a place should not occur in any pure computation. For
example, we should not write \code{\move{p} + 3}. One way to avoid
mixing move and pure expression is to check it in compilation, but for
simplicity, we explicitly enforce it in the syntax. In our surface
language (the language before parsing), we can mix move and pure
expressions as they will be desugared by the parser.

We now turn to the design of types systems. There are based types
($\tau^B$) including integer, float and others scalar types, whose
inhabitant can be safely copied as the types with Copy trait in Rust.
$\tbox{\tau}$ represents the ownership pointer that points to the heap
memory. $\tstruct{id}{\varphi}$ and $\tenum{id}{\varphi}$ are
structure and enum types, both of which contain an identifier and a
list of fields $\varphi$. For example, the \code{Node} type
in~\figref{sfig:list} is defined as
$\tstruct{\code{Node}}{[(\code{key}, \code{i32}); (\code{val},
  \tbox{\code{i32}}); (\code{next}, \tbox{\code{List}})]}$, and the
\code{List} type in~\figref{sfig:list} is defined as
$\tenum{\code{List}}{[(\code{Nil}, \code{unit}); (\code{Cons},
  \code{Node})]}$. Note that here we do not expand \code{List} in the
definition of \code{Node} type (or expand \code{Node} in the
definition of \code{List} type), which seems different from the
definitions in~\figref{fig:owl-syntax}. This is because \code{List}
and \code{Node} are mutually dependent of each other. In our Rocq
development, as CompCert does, we use a \emph{composite environment}
to map the struct/enum identifiers to their field lists. So when there
is any struct/enum type is used in the definition of other types, we
can just refer to its type identifier rather than its full field
list, e.g., we refer to \code{List} instead of the field list of
\code{List} in the definition of \code{Node}.

% \myparagraph{Semantics}

% To make this section consistent, we should consider how do we
% introduce the ownership semantics in the later section.

% \subsection{The Owl-IR language}

\subsection{Implementation of the Owlang Compiler}

The compilation chain of the Owlang compiler is depicted
in~\figref{fig:owl-compiler}, which consists of three compilation
passes that transform Owlang programs to Clight programs. We
illustrate these three passes using a running example shown
in~\figref{fig:owl-compiler-exm}, where the source Owlang module is
shown in \figref{sfig:source-owl}, the Owl-IR module lowered from the
source module is in~\figref{sfig:owlir1}, and the Owl-IR module after
Drop Elaboration (before Clight Generation) is
in~\figref{sfig:owlir2}.
To improve the readability, the modules shown
in~\figref{fig:owl-compiler-exm} are written in Rust style, but the
readers can also easily relate them to the formalized syntax defined
in~\figref{fig:owl-syntax}. Note that in practice, users do not need
to explicitly write \code{\textbf{move}}, as it can be inferred by the
parser.

%
% The formal syntax of Owlang is provided in appendix~\secref{sec:owl}.
%

The source program (\figref{sfig:source-owl}) defines a structure
\code{Point} containing two fields of type \tbox{\code{i32}}. In the
main function, it initializes a place \code{p} of type \code{Point},
and then it \emph{randomly} assign (i.e., move) one of the fields of
\code{p} to the place \code{a}. We want to use this program to
illustrate the mechanism of \emph{automatic memory management} in
Owlang, i.e., how the resources of ownership pointers are freed, and
how can it be compiled down to the Clight program, where we will see
the necessity of the introduction of \emph{ownership semantics}.

\begin{figure}[t]
\begin{subfigure}{.47\textwidth}
\centering
\begin{lstlisting}[language=Rust, style=colouredRust,mathescape=true,numbers=left,
escapechar=\%]
struct Point {x: Box<i32>, y: Box<i32>}
fn test() {
  let p = Point {x: Box(1), y: Box(2)};
  let a: Box<i32>;
  if rand() { a = move p.x; }
  else { a = move p.y; }
  return; }
\end{lstlisting}
\caption{The Source Owlang Module}
\label{sfig:source-owl}
\end{subfigure}
\begin{subfigure}{.48\textwidth}
\centering
\begin{lstlisting}[language=Rust, style=colouredRust,mathescape=true,numbers=left,
escapechar=\%]
fn test() {
  %\insertstmt{drop(p);}%
  p = Point {x: Box(1), y: Box(2)};
  if rand() { %\insertstmt{drop(a);}% a = move p.x; }
  else { %\insertstmt{drop(a);}% a = move p.y; }
  %\insertstmt{drop(a);}% %\insertstmt{drop(p);}%
  return; }
\end{lstlisting}
\caption{The Owl-IR Module (After Lowering)}
\label{sfig:owlir1}
\end{subfigure}
\begin{subfigure}{.96\textwidth}
\centering
\begin{lstlisting}[language=Rust, style=colouredRust,mathescape=true,numbers=left,
escapechar=\%]
fn test() {
  %\elabstmt{flag\_px = false; flag\_py = false;}%
  %\removestmt{\insertstmt{drop(p);}}% p = Point {x: Box(1), y: Box(2)}; // Owned = {p.x, p.y}, Unowned = {a}
  %\elabstmt{flag\_px = true; flag\_py = true;}%
  if rand() {
    %\removestmt{\insertstmt{drop(a);}}% %\elabstmt{flag\_px = false;}% a = move p.x; } // Owned = {a, p.y}, Unowned = {p.x}
  else {
    %\removestmt{\insertstmt{drop(a);}}% %\elabstmt{flag\_py = false;}% a = move p.y; } // Owned = {a, p.x}, Unowned = {p.y}
  %\insertstmt{drop(a);}%  // Owned = {a, p.x, p.y}, Unowned = {p.x, p.y}
  if %\elabstmt{flag\_px}% { %\elabstmt{drop(p.x);}% } if %\elabstmt{flag\_py}% { %\elabstmt{drop(p.y);}% }
  return; }
\end{lstlisting}
% \vspace{-0.3cm}
\caption{The Owl-IR Module (After Drop Elaboration)}
\label{sfig:owlir2}
\end{subfigure}
% \vspace{-0.3cm}
\caption{A Running Example for the Owlang Compiler}
\label{fig:owl-compiler-exm}
% \vspace{-0.4cm}
\end{figure}

\myparagraph{Lowering}
The Lowering pass (\figref{sfig:owlir1}) makes variable scopes
explicit and inserts \code{drop} statements (shown in blue) at points
where ownership variables (i.e., variables whose types contain
\code{Box}) either go out of scope or are reassigned.

% may be we should write a brief introduction (say that our
% implementation is referred to the RFC) and move this paragraph to
% appendix.

\myparagraph{Drop Elaboration}
The Drop Elaboration pass (\figref{sfig:owlir2}) refines the inserted
\code{drop} statements by analyzing their execution behavior. It
retains \code{drop} statements that will definitely execute, removes
those that are definitely skipped, and inserts dynamic checks---using
auxiliary variables---for those whose execution depends on the control
flow.

To determine which variables may or may not hold ownership of their
values at a given program point, the compiler performs an
\emph{initialization analysis} (init-analysis). This analysis computes
two sets: \code{Owned} and \code{Unowned}, which over-approximate the
variables that may (or may not) be initialized, i.e., hold
ownership. When control flow merges, the analysis joins these sets via
union to conservatively approximate the dynamic ownership status.

At each program point, a \code{drop} statement is retained if the
corresponding variable is not in \code{Unowned} (i.e., it must have
ownership of its value), and removed if it is not in \code{Owned}
(i.e., it must not have ownership of its value). For variables that
fall into the intersection of both sets (e.g., \code{p.x} and
\code{p.y} at line 9 of~\figref{sfig:owlir2}), the compiler generates
auxiliary variables to track their runtime ownership status, ensuring
that \code{drop} operations execute correctly in all branches.

\myparagraph{Clight Generation}
The Clight Generation includes translating the Owlang types to C types
(e.g., translating \code{enum} into tagged union in C) and generating
drop functions for the user-defined ownership types (e.g., the
destructor for the linked list implemented in~\figref{sfig:list}).

% We referred to the Rust compiler to implement these compilation
% passes. The Lowering pass is similar to the MIR
% construction~\cite{mir-construct}. The Drop Elaboration is from this
% RFC~\cite{non-zero-copy}. We implement the init-analysis using the
% data flow analysis framework based on the Kildall's algorithm from
% CompCert.

% For example, it translates \code{Box} pointers to raw pointers and
% \code{enum} types to tagged unions implemented by a struct whose first
% field is the enum discriminant and second field is a C union type
% storing the enum variant. The drop semantics of the user-defined
% ownership types (e.g., \code{Point}) is translated to drop functions
% which drop the values of their fields (e.g., free the memory pointed
% by \code{Point}).

\subsection{Ownership Semantics and Verification of Drop Elaboration.}\label{sssec:verify-owl-compiler}

We focus on the verification of the Drop Elaboration pass and omit the
proofs for other passes. Before Drop Elaboration, whether a
\code{drop(p)} statement is executed depends on dynamically checking
if \code{p} holds ownership of its value--—a check that is encoded
in the semantics of Owl-IR, which we refer to as \emph{ownership
  semantics}.
After this pass, each \code{drop} statement is guaranteed to be
executed unconditionally. This is because Drop Elaboration refines
\code{drop} statements and introduces auxiliary variables to track
whether it holds ownership, allowing their execution to be determined
statically by control flow rather than the dynamic ownership
semantics.

The ownership semantics is represented by a local state called the
\emph{ownership status} in the Owl-IR semantics, denoted as the set
$\ownst$, which tracks variables that currently hold ownership. A
variable is added to $\ownst$ when it is assigned and removed when it
is moved or dropped. For instance, the statement \code{a =
  \textbf{move} b} removes \code{b} from $\ownst$ and adds \code{a} to
$\ownst$, indicating ownership transfer from \code{b} to \code{a}.
In the ownership semantics, executing \code{drop} involves checking
whether the variable is in $\ownst$ to decide whether its value should
be dropped (e.g., deallocation of its memory).

The key to verifying Drop Elaboration is proving the soundness of the
initialization analysis (init-analysis). Recall that init-analysis is
used to statically approximate the ownership information of variables,
whereas in the source program, ownership semantics captures this
information dynamically.
The proof follows the standard approach of abstract interpretation. We
show that the init-analysis soundly over-approximates the ownership
semantics by establishing a subset relation between the concrete
domain (i.e., the ownership status $\ownst$) and the abstract domain
(i.e., the \code{Owned} and \code{Unowned} sets
in~\figref{sfig:owlir2}). We then prove that this relation is
preserved throughout the execution, ensuring that the elaborated
\code{drop} statements faithfully simulate the dynamic \code{drop}
behavior in the source program.

% For the correctness of the Owl compiler, we define an end-to-end
% simulation conventions $\compra$ and prove that it is equivalent to
% the simulation convention composed from each pass:
% %
% \begin{lemma}\label{lem:ra-equiv}
%     $\compra \equiv \rid \compsymb \rinjp \compsymb (\rinjp \compsymb \comprc) \compsymb \compca$
% \end{lemma}
% %
% The key of this proof is to show that the simulation conventions used
% in $\compca$ is commutative with $\comprc$ so that the $\rinjp$ used
% in the Owl compiler frontend and CompCert can be composed. Finally, we
% prove the end-to-end backward simulation of the Owl compiler by
% composing the forward simulation from each pass
% of~\figref{fig:owl-compiler} and applying~\lemref{lem:ra-equiv}:

%%%%%%%%%%%%%%%%%%%%%%%%%% End of Technical Report Contents %%%%%%%%%%%%%%%%%%%%%%%%%%%%%

\fi

\end{document}